\documentclass[manuscript]{aastex}
\usepackage{txfonts}
\usepackage{graphicx}

\slugcomment{Accepted  to ApJ}

\shorttitle{Gas emissions in Planck cold dust clumps\\
---A Survey of the J=1-0 Transitions of $^{12}$CO, $^{13}$CO, and C$^{18}$O
}

\shortauthors{Wu et al.}

\begin{document}

\title{Gas emissions in Planck cold dust clumps\\
---A Survey of the J=1-0 Transitions of $^{12}$CO, $^{13}$CO, and C$^{18}$O}

\author{Yuefang Wu\altaffilmark{1}, Tie Liu\altaffilmark{1}, Fanyi Meng\altaffilmark{2}, Di Li\altaffilmark{3,}\altaffilmark{4,}\altaffilmark{5}, Sheng-Li Qin\altaffilmark{6},
Bing-Gang Ju\altaffilmark{7,}\altaffilmark{8}}

\altaffiltext{1}{Department of Astronomy, Peking University, 100871,
Beijing, China; ywu@pku.edu.cn} \altaffiltext{2}{Yuan Pei school,
Peking University, 100871, Beijing China}\altaffiltext{3}{ National
Astronomical Observatories, CAS, Chaoyang Dist., Datun Rd, A20,
Beijing 100012, China}\altaffiltext{4}{Space Science Institute,
Boulder, CO, USA}\altaffiltext{5}{Jet Propulsion Laboratory,
California Institute of Technology, Pasadena, CA}

\altaffiltext{6}{I. Physikalisches Institut, Universit\"{a}t zu
K\"{o}ln, Z\"{u}lpicher Str. 77, 50937}\altaffiltext{7}{Purple
Mountain Observatory, Qinghai Station, 817000, Delingha, China
}\altaffiltext{8}{Key Laboratory for Radio Astronomy, CAS}

\begin{abstract}

A survey toward 674 Planck cold clumps of the Early Cold Core
Catalogue (ECC) in the J=1-0 transitions of $^{12}$CO, $^{13}$CO and
C$^{18}$O has been carried out using the PMO 13.7 m telescope. 673
clumps were detected with the $^{12}$CO and $^{13}$CO, and $68\%$ of
the samples have C$^{18}$O emission. Additional velocity components
were also identified.
A close consistency of the three line peak velocities was revealed
for the first time. Kinematic distances are given out for all the
velocity components and half of the clumps are located within 0.5
and 1.5 kpc. Excitation temperatures range from 4 to 27 K, slightly
larger than those of $T_d$. Line width analysis shows that the
majority of ECC clumps are low mass clumps. Column densities
N$_{H_{2}}$ span from 10$^{20}$ to 4.5$\times10^{22}$ cm$^{-2}$ with
an average value of (4.4$\pm$3.6)$\times10^{21}$ cm$^{-2}$.
N$_{H_{2}}$ cumulative fraction distribution deviates from the
lognormal distribution, which is attributed to optical depth. The
average abundance ratio of the $^{13}$CO to C$^{18}$O in these
clumps is 7.0$\pm$3.8, higher than the terrestrial value. Dust and
gas are well coupled in 95\% of the clumps. Blue profile, red
profile and line asymmetry in total was found in less than 10\% of
the clumps, generally indicating star formation is not developed
yet. Ten clumps were mapped. Twelve velocity components and 22 cores
were obtained. Their morphologies include extended diffuse, dense
isolated, cometary and filament, of which the last is the majority.
20 cores are starless.
Only 7 cores seem to be in gravitationally bound state. Planck cold
clumps are the most quiescent among the samples of weak-red IRAS,
infrared dark clouds, UC~H{\sc ii} region candidates, EGOs and
methanol maser sources, suggesting that Planck cold clumps have
expanded the horizon of cold Astronomy.

\end{abstract}

\keywords{ISM: clouds - ISM: structure-ISM: kinematics and dynamics - Stars: formation - Stars: protostars}

\clearpage

\section{Introduction}
Large samples significantly improve our understanding of star
formation. At the beginning of star formation studies in the early
70s, the Palomar Sky Survey (PSS) plates provided astronomers
optical selected nebulae as targets of star forming regions. The
catalogued Shapeless HII regions \citep{shar59} from PSS served as
sources for investigating gas and dust properties of molecular cloud
complex \citep{evans77,Har77}. Nearby dark cores as well as cloud
fragments were from Lynds dark nebula \citep{str75,sne80,clar81}.
After visual inspection, 70 small opaque spots were chosen for the
surveys of low-mass cores with the $^{13}$CO, C$^{18}$O and NH$_{3}$
respectively \citep{my83a,my83b}. A number of sources resulted in
these earliest observations are still primary examples in low mass
star formation so far. However, "optical dark" selection method is
limited for probing star forming in the deep of molecular clouds.
\cite{bal83} chose infrared sources for detecting high velocity
outflows in high-mass star formation regions. Particularly IRAS
point sources afforded plenty samples for high mass star formation
regions. Based on the similar shapes of the far infrared flux
distribution of all embedded O type stars, the IRAS color-color
criteria were used to choose UC HII region candidates \citep{wood89}
and further refined by molecular line studies \citep{ce92,wat97}.
Precursors of UC HII regions were also obtained from luminous IRAS
sources and used for a number of surveys to examine the early
characteristics of high mass star formation
\citep{mol96,sr02,beu02,wu06}. In recent years, earlier samples for
high mass star formation come from infrared dark clouds surveyed by
MSX. These are extinction features against the bright mid-infrared
background of the Galaxy \citep{egan98}. A number of starless
massive cores were detected, which are with narrower line widths and
lower rotation temperatures than both UC HII region precursors and
UCHII regions but with similar masses \citep{ra06,sr05}. However,
MSX is limited to $\vert$b$\vert$$\leq6^{\arcdeg}$ of our Galaxy.

Now Planck Surveys provide a wealth of early sources which are cold and with unprecedented complete space distribution. The
cold core Catalogue of Planck Objects (C3PO) includes 10783 sources which are mainly cold clumps, intermediate structures of the
fragmentation scenario. Their temperatures and densities range from 7 to 17 K and 30 to 10$^{5}$ cm$^{-3}$ respectively, derived
from the fluxes in the three highest frequency Planck bands (353, 545, 857 GHz) and the 3000 GHz of IRAS band \citep{ade11a}. This
enables us to probe the characteristics of the prestellar phase or starless clumps. The all sky nature of the Planck cold clump sample is particularly useful for studying the global properties of Galactic star formation. Follow up study with high resolution
observations by Herschel Satellite revealed extended regions of cold dust with colour temperatures down to 11 K. The results show
different evolutionary stages ranging from a quiescent, cold filament to clumps with star formation activities \citep{ju10}.

These known properties of Planck cold clumps currently were revealed
with various bands of continuum emissions and only a set of 8 C3PO
sources selected from different environment \citep{ade11b} were
investigated with molecular lines. Examination with molecular lines
is another essential aspect for understanding the properties of the
Planck cold clumps. Molecular line studies of the Planck clumps,
which are an unbiased sample of cold dust clumps in the Milky Way,
will provide clues to probe a number of critical questions about
clumps and star formation: What are the morphologies, physical
parameters and their variations in the Galaxy? What are the initial
conditions of star formation, which we do not yet know in our
current state of knowledge? What are the dynamic factors in a
variety of the clumps and is there any premonition of collapse? Is
there any depletion and what are cold chemistry phenomena? In which
environment can the cold clump exist? What is the highest Galactic
latitude where stars can form? CO is the most common tool to probe
molecular regions. Although there is "dark gas" which is undetected
in the available CO and HI surveys \citep{ade11c}, CO is still a
basic probe to address the above questions.

In this paper we report a survey of Planck cold clumps with J=1-0
lines of the $^{12}$CO and its major isotopes $^{13}$CO and
C$^{18}$O using the telescope of Purple Mountain Observatory at
Qinghai province in western China. Our target sources are chosen
from the Early Cold Core Catalogue (ECC), which are the most
reliable detections of C3PO clumps. So far we have surveyed 674
sources with single point observations and mapped a subset of ten
clumps in different locations. In the next section the observations
are described. In section 3 we present the results. The discussions
are in section 4 and a summary is given in section 5.

\section{Observations}
Using the 13.7 m telescope of Qinghai Station of Purple Mountain
Observatory (PMO) at De Ling Ha, the survey towards the ECC clumps
with the lines of $^{12}$CO (1-0) and $^{13}$CO (1-0) was carried
out from January to May, 2011. The half-power beam width (HPBW) at
the 115 GHz frequency band is 56$\arcsec$. The pointing and tracking
accuracies are better than 5$\arcsec$. The main beam efficiency is
$\sim$50\%. A newly installed SIS superconducting receiver with nine
beam array was employed as the front end. $^{12}$CO (1-0) line was
observed at upper sideband (USB) and the other two lines were
observed simultaneously at lower sideband (LSB). The typical system
temperature (T$_{sys}$) in SSB mode is around 110 K and varies about
10\% for each beam. An FFT spectrometer was used as the back end,
which is with a total bandwidth of 1 GHz and 16384 channels. The
equivalent velocity resolution is 0.16 km~s$^{-1}$ for $^{12}$CO
(1-0) line and 0.17 km~s$^{-1}$ for $^{13}$CO (1-0) and C$^{18}$O
(1-0) lines respectively. The Position switch (PS) mode
was adopted for single point observations. The off position for each "off" source
was carefully chosen from the area within 3$\arcdeg$ from the "on" source, which has no or extremely weak CO emission
based on the CO survey of the Milky Way \citep{dame87,dame01}. For those regions without CO data, we chose the off postions based on the IRAS 100 $\mu$m images. The integration time at
the on/off position is 30 s.

We also mapped ten sources which are with rather strong CO emission. On-The-Fly (OTF) observing mode was applied
for mapping observations. The antenna continuously scanned a region
of 22$\arcmin\times22\arcmin$ centered on the Planck cold clumps with
a scan speed of 20$\arcsec$~s$^{-1}$. The rms noise level was 0.2 K
in main beam antenna temperature T$^{*}_{A}$ for $^{12}$CO (1-0), and 0.1 K for $^{13}$CO (1-0)
and C$^{18}$O (1-0). Since the edges of the OTF maps are rather
noisy, only the central 14$\arcmin\times14\arcmin$ regions are
analyzed. The OTF data were then converted to 3-D cube data with a
grid spacing of 30$\arcsec$. The IRAM software package GILDAS was
used for the data reduction \citep{gui00}.

\section{Results}
To make the observation with a sufficiently high elevation for the 13.7
m telescope at De Ling Ha, our samples were chosen from ECC with
Dec.$\geq-20\arcdeg$. There are 674 sources from ECC which satisfies
this criterion. The name and coordiantes of
the 674 clumps are listed in Table 1 (Column
1-7). One clump marked with $"**"$ is without detection, which needs to further check. There are 36 clumps without suitable background reference positions for
the CO emission observations, which were marked with $"*"$ after
their names and excluded in further analysis. Clumps with Latitude
larger than 25$\arcdeg$ are referred as high latitude clumps. There are 41 high latitude clumps. According to
\cite{dame87}, the distributions of the detected clumps are assigned to different molecular complexes. The regions and corresponding
Longitude and Latitude scopes are shown in columns 1-3 of Table 2. The fourth column of Table 2 gives number of the clumps in each complex.
 673 clumps were detected to have
the $^{12}$CO and $^{13}$CO emission, and 68\% of the clumps have the
C$^{18}$O emission. 108 additional velocity components were identified
in both the $^{12}$CO and $^{13}$CO emission, 50\% of which have
C$^{18}$O emission. We made a Gaussian function fitting for each
distinguishable velocity component of the spectral lines. The results of
emissions and their physical parameters are presented in the following.

\subsection{Emission components and characteristics of line profiles}
Most of the observed clumps have single velocity component in CO emission.
Among the 673 sources, 123 and 16 have double or possible double velocity components
respectively, and 18 and 10 have three or possible three
velocity components. Multiple components or blend emission were detected
in 52 cold sources. The spectral line of each distinguishable
component was fitted with a Gaussian function. Centroid velocity
V$_{lsr}$, antenna temperature T$_{A}^{*}$ and full width at half
maximum $\Delta \textrm{V}$ (FWHM) were obtained for all the 3
CO lines, which are given at Table 3. Centroid velocity of $^{13}$CO was taken as the systemic velocity of each velocity components and listed
in the second column of Table 4. Dense core masses are related with molecular line widths \citep{wang09}. We adopted an average value of $^{13}$CO line width 1.3 km~s$^{-1}$ of the nearby dark cores \citep{my83b}as a criterion to distinguish
high-mass and low-mass clumps. If a clump
contains one or more velocity components with the $^{13}$CO line width larger
than 1.3 km~s$^{-1}$, it is considered as a candidate of high mass
clumps (group H), otherwise, it is taken as a candidate for low mass
clumps (group L). In some of the observed clumps, the spectral lines
of the emission components depart from Gaussian profile, which probably reflect the state of gas
or originates from systematic motions and star
formation activities. We classify the different non-Gaussian
profiles with the following characters:

BA$^{+}$: blue profile; RA$^{+}$: red profile; BA: blue asymmetric; RA: red asymmetric;
De: possible depletion;  BW: blue wing; RW: red wing; W: wings; P: pedestal; Dob: Double components;
T: Three components; M.B: Multiple or Blended components.

The normalized velocity difference $\delta
V=(V_{thick}-V_{thin})/\delta V_{thin}$ is applied to identify the
blue- and red- profiles \citep{mar97}, where V$_{thick}$ is the peak
velocity of $^{12}$CO (1-0), V$_{thin}$ and $\delta V_{thin}$ are
the systemic velocity of optically thin lines and the line width of
optically thin lines, respectively. We use $^{13}$CO (1-0) and
C$^{18}$O (1-0) as the optically thin lines. If $\delta V<-0.25$,
the line is classified as "blue profile", and if $\delta V>0.25$,
the line is classified as "red profile". The sources with blue or
red profiles are listed in Table 5. We define $\delta
V(12)=(V_{thick}-V12)/\delta V_{thin}$ to investigate line
asymmetries, where V12 is the center velocity of $^{12}$CO (1-0).
The sources with asymmetric profiles are listed in Table 6. The
spectrum characteristics are denoted in the last Column of Table 3.
Figure 1 presents the examples of the spectra of the J=1-0 lines of
the $^{12}$CO, $^{13}$CO and C$^{18}$O with red, green and blue
colors respectively, including single, double and three components
and the characteristics of the spectral profiles.

\subsection{Observed parameters}
The histograms of the line center (peak) velocity difference between
the lines of $^{12}$CO and $^{13}$CO (V12-V13) and those of
$^{13}$CO and C$^{18}$O (V13-V18) lines are shown in Figure 2 a$)$
and b$)$. The distributions (blue solid histograms) of both the
V$_{12}$-V$_{13}$ and V$_{13}$-V$_{18}$ are quite symmetric around
the zero, which are fitted with a normal distribution (red curve).
However, their distributions are much narrower than a standard
normal distribution (red curve), but have a sharp peak at zero (see
Figure 2 a$)$ and b$)$).  94\% of the clumps have V$_{12}$-V$_{13}$
less than 3 $\sigma$ of the velocity resolution given by the
spectrometer, and 98\% of the clumps have V$_{13}$-V$_{18}$ less
than 3 $\sigma$ of the velocity resolution. The velocity
correlations between lines of $^{12}$CO-$^{13}$CO and
$^{13}$CO-$^{18}$CO are shown in Figure 2 c$)$ and d$)$.
Y=(1.002$\pm$0.001)X+(0.005$\pm$0.016) for V$_{12}$ (Y) to V$_{13}$
(X) and Y=(0.992$\pm$0.005)X+(0.140$\pm$0.057) for V13 (Y) to V18
(X) show very good correlations and the correlation coefficient is
100\%. This is the first time to show the good agreement of the
center velocities of the $^{12}$CO and its main isotopes $^{13}$CO
and C$^{18}$O lines with such a large sample. These results show
that the peak velocities of the three CO lines agree very well, and
demonstrate that they originate from the same emission regions.
Comparisons of Fig. 2 a$)$ and b$)$, Fig.2 c$)$ and d$)$ show that
the agreement between V$_{lsr}$ of the $^{13}$CO and C$^{18}$O lines
is better than that of the $^{12}$CO and $^{13}$CO lines, which may
suggest that the $^{12}$CO is easily affected by dynamic factors in
the clump or its environment.

Since there are fewer sources with detected the C$^{18}$O lines than those
with the $^{13}$CO lines, we adopt the center velocity of the $^{13}$CO as
the systemic velocity of the clumps in the following analysis. For
the components without the $^{13}$CO (1-0) emission, the center
velocities of the $^{12}$CO were adopted as the systemic velocity.

Main beam antenna temperatures of the three transitions for all the components
detected were obtained. T$_{12}$ ranges from 0.5 - 12 K. The three
clumps G209.28-19.62 (12 K), G192.54-11.56 (9.2 K)
and G207.35-19.82 (8.8 K) have highest temperature, which are all
located in the Orion region. Generally, the antenna temperatures of $^{13}$CO
and C$^{18}$O are also high in the $^{12}$CO emission strong clumps.
For the above three clumps, T$_{13}$ and T$_{18}$ are 6.8, 1.4; 5.0,
0.6; and 3.1, 0.4 K respectively. The ratio of T$_{12}$/T$_{13}$ is
between 1.7-1.8, showing that the
$^{12}$CO line emissions have large opacity. The ratio of
T$_{13}$/T$_{18}$ ranges between 4.5-8.3, with an average value of
6.0 which is close to the terrestrial value. However the ratio may
be different in different regions, for example in the Ophiuchs
complex, the antenna temperatures of the three transitions are 6.7, 4.4 and 2.5 K for
G003.73+18.30 and 5.9, 4.4 and 1.4 K for G006.41+20.56. The
ratio of T$_{13}$/T$_{18}$ are 1.8 and 3.1 for the two clumps
respectively, much less than 5.5, suggesting where the $^{13}$CO lines seem optical thick. For all the
detected components, the histograms of T$_{12}$, the ratios of
T$_{12}$/T$_{13}$ and T$_{13}$/T$_{18}$ are present in Figure 3
a$)$, b$)$ and c$)$ respectively. The peak of T$_{12}$ is around 3
K. For the ratio of T$_{12}$/T$_{13}$, the mean value is 2.2 with a
standard deviation of 1.4, showing the $^{12}$CO (1-0) line
emissions of the cold clumps are optical thick generally. The mean
value of the ratio of T$_{13}$/T$_{18}$ is 3.9 with a standard
deviation of 1.7.

We found all the distributions of the observed parameters as well as
the derived parameters seem to skew to the right with a long tail at
high value side. We try to depict their distributions with a
lognormal distribution. The probability density function (PDF) of a
log-normal distribution is:
\begin{equation}
f_{X}(x;\mu,\sigma)=\frac{1}{x\sigma\sqrt{2\pi}}e^{-\frac{(\textrm{ln}x-\mu)^{2}}{2\sigma^{2}}},x>0
\end{equation}
where x is value of the variable, the parameters denoted $\mu$ and
$\sigma$, are the mean and standard deviation, respectively, of the
variable's natural logarithm. The Kolmogorov-Smirnov test (K-S test)
is applied to identify whether the parameters follow a lognormal
distribution. The decision to reject or accept the null hypothesis
is based on comparing the P-value with the desired significance
level, which is 0.05 in this paper. If the P-values from K-S test is
larger than 0.05, the parameter should follow the reference
distribution. The statistics and K-S test results of the parameters
are summarized in Table 7.

There is a similar characteristic of the distributions of T$_{12}$,
T$_{12}$/T$_{13}$. They are both with a power-law like long tail but
can not be well described with a lognormal distribution. The tail of
the distribution of T$_{13}$/T$_{18}$ is less than those of T$_{12}$
and T$_{12}$/T$_{13}$, suggesting T$_{12}$ may be the more sensitive
physical element than T$_{13}$ and T$_{18}$ for star formation. The
distribution of T$_{13}$/T$_{18}$ has a lognormal shape with P-value
of K-S test as high as 0.234.

From Table 3 one can see the line widths of the emission components
are narrow generally. Most of the clumps have line widths smaller
than 1.3 km~s$^{-1}$. There are 162 high mass clump candidates, among
them 68 have two or more velocity components.  The clumps at high latitude
all are low-mass clump candidates with single component except
G159.23-34.49 which is in high-mass group and with two components.
Figure 4 a$)$ presents the distributions of the FWHM of all the
three lines: the mean values and the standard deviations of the
three line widths are 2.0$\pm$1.3, 1.3$\pm$0.8 and 0.8$\pm$0.7
km~s$^{-1}$ respectively. The shapes of all the distributions are
similar to each other and are lognormal distributed.

\subsection{Derived physical parameters}
The excitation temperature derived from radiation transfer equation
is:
\begin{equation}
T_{r}=\frac{T_{a}^{*}}{\eta_{b}}=\frac{h\nu}{k}\left[\frac{1}{exp(h\nu/kT_{ex})-1}-\frac{1}{exp(h\nu/kT_{bg})-1}\right]\times\left[1-exp(-\tau)\right]f
\end{equation}
here $T_{r}$ is the brightness temperature corrected with beam
efficiency $\eta_{b}$. Assuming $^{12}$CO emission is optically thick ($\tau\gg1$) and the filling factor f=1, the excitation temperature $T_{ex}$ can be
straightforwardly obtained. The T$_{ex}$ is given in column 6 of
Table 4. We assume the excitation temperatures of $^{13}$CO and C$^{18}$O are the same as that of the $^{12}$CO (1-0) in the following analysis. The values range from 3.9 K to 27 K, wider than the dust
temperature range 7 - 17 K \citep{ade11a}. The mean value with the
standard deviation is 10$\pm$2.6 K, which is smaller than the mean value of the dust temperature (12.8$\pm$1.6 K) based on
aperture photometry with a local background subtraction by Herschel photometric observations towards 71 Planck cold fields \citep{ju12}, indicating the gas maybe heated by the dust. There are 12 clumps with
T$_{ex}>$17 K. All these hottest clumps are located in the Orion and
Taurus regions, showing high excitation temperature may be related to star
formation conditions. 93 components with T$_{ex}<$ 7 K are referred
to as coldest clumps among which G093.66+04.66 (3.9 K) is in the 1st
Quadrant, G098.10+15.83 (5.0 K) and G112.63+20.80 (5.8 K) are in the
Cepheus. All other coldest clumps are the weaker components of the
clumps with double or three emission components. The T$_{ex}$
histogram and its lognormal fit are shown in Figure 5. It's
distribution can be well fitted by a lognormal shape with P-value of
K-S test as large as 0.227. But the tail of the distribution seems
much smaller than that of line widths and velocity dispersions. In
the following analysis T$_{ex}$ was taken as gas kinetic temperature
assumed the local thermodynamic equilibrium (LTE) holds.

Both the optical depths $\tau_{13}$ and $\tau_{18}$ at the emission peak of the
corresponding lines were calculated with equation (2) assuming the filling factor f=1, which are listed in column 7
and 9 in Table 4. $\tau_{13}$ ranges from 0.1 to 4.1. Clumps with
largest $\tau_{13}$ include : G026.45-08.02 (3.2) and G058.07+03.29
(3.2) of the 1st Quadrant; G164.94-08.57 (4.1) and G179.10-06.27
(3.9) in Taurus and G182.04+00.41 (3.8) located at Anti-center. For the ten
clumps with the smallest optical depth (0.1), all of them are the minor
components of the double or more component sources except
G121.88-08.76. Figure 6 a$)$ shows the histogram of $\tau_{13}$ and
its lognormal fitting. The mean value is 0.93$\pm0.56$.

The column density of a molecular line can be obtained with the
theory of radiation transfer and molecular excitation as following \citep{gar91}:
\begin{equation}
N=\frac{3k}{8\pi^{3}B\mu^{2}}\frac{exp[hBJ(J+1)/kT_{ex}]}{(J+1)}\times\frac{(T_{ex}+hB/3k)}{[1-exp(-h\nu/kT_{ex})]}\int\tau_{V}dV
\end{equation}
where B, $\mu$, J are the rotational constant, permanent dipole
moment, and the rotational quantum number of the lower state of the
molecular transition. Column densities of the $^{13}$CO and
C$^{18}$O molecules were calculated and given in Column 8 and 10 of
Table 4. The maximum C$^{18}$O column densities of the Planck clumps
are 1.1$\times$10$^{16}$ cm$^{-2}$ and 1.3$\times$10$^{16}$
cm$^{-2}$ in G028.56-00.24 and G160.51-16.84 respectively and the
smallest ones are 1.0$\times$10$^{14}$ cm$^{-2}$ in G26.93-20.68 and
G102.72-25.96. The column densities of the C$^{18}$O span a wider
range than those of the nearby dark clouds \citep{my83a} which are
from 3$\times$10$^{14}$ to 2.7$\times10^{15}$ cm$^{-2}$. The
different column density ranges between our sample and the nearby
dark clouds may be attributed to the larger space distribution of
Planck cold clumps than the nearby dark clouds.

The $^{13}$CO column densities were calculated for 782 velocity
components. For the 437 components with both C$^{18}$O and $^{13}$CO
column densities, the abundance ratio of X$_{13}$/X$_{18}$ were
calculated and listed in Column 13 of Table 4. The histogram and the
lognormal distribution of X$_{13}$/X$_{18}$ fitting is presented in
Figure 6 b$)$. The distribution shape of the ratio X$_{13}$/X$_{18}$
is similar to that of $\tau_{13}$. The mean value of
X$_{13}$/X$_{18}$ is 7.0$\pm$3.8, higher than that of the
terrestrial ratio 5.5. The distribution of X$_{13}$/X$_{18}$ can be
well depicted by a lognormal distribution with P-value of K-S test
as high as 0.388.

Molecular hydrogen column densities of each velocity component were derived
according to the column densities of the $^{13}$CO. The fractional
abundance of [H$_{2}]$/[$^{13}$CO]=89$\times$10$^{4}$ was adopted. We
also calculated the N$_{H_{2}}$ for the components without $^{13}$CO
emission by assuming $^{12}$CO (1-0) emission to be optically thin,
the excitation temperature of 10 K and
[H$_{2}]$/[$^{12}$CO]=10$^{4}$, but these components are
excluded in further statistics . Figure 6 c$)$ presents the plots of
the N$_{H_{2}}$ histogram and its lognormal fitting. It spans from
10$^{20}$ to 4.5$\times10^{22}$ cm$^{-2}$, which is larger than that
of the 8 sources on the average \citep{ade11b}. From Herschel follow-up observations, the peak column densities
of 71 clumps range from 4$\times$10$^{20}$ to 7.4$\times10^{22}$ \citep{ju12}, which are slightly larger than the column densities obtained in this work.
The four clumps with largest $^{13}$CO column density are G209.28-19.62, G028.56-00.24,
G033.70-00.01 and G158.37-20.72. The N$_{H_{2}}$ of G209.28-19.62 located
in the Orion complex is as high as
4.5$\times$10$^{22}$ cm$^{-2}$. G028.56-00.24 and G033.70-00.01 with
N$_{H_{2}}$ of 3.4 and 2.8$\times$10$^{22}$ cm$^{-2}$ are located in the 1st
Quadrant. G158.37-20.72 with N$_{H_{2}}$ of 2.3$\times$10$^{22}$
cm$^{-2}$ is in Taurus, which is at 7$\arcmin$ north of NGC 1333
\citep{str76} and with a very red young star SSV 13
\citep{harvey98}. The $^{12}$CO lines of all these clumps have large line widths of
2.57-9.35 km~s$^{-1}$, showing the turbulence support for
these clumps. The column density distribution (Figure 6
c$)$ can not be fitted with a unique lognormal distribution, but
exhibits a power-law like tail, similar to that identified from
active star formation regions \citep{ka09}, indicating that some Planck
cold clumps are locate in active star forming regions.

The one dimensional non-thermal ($\sigma_{NT}$ ) and thermal ($\sigma_{Therm}$)  velocity dispersions can be estimated as following:
\begin{equation}
\sigma_{NT} = \left[\sigma_{^{13}CO}^{2}-\frac{kT_{ex}}{m_{^{13}CO}}\right]^{\frac{1}{2}}
\end{equation}
\begin{equation}
\sigma_{Therm} = \left[\frac{kT_{ex}}{m_{H}\mu}\right]^{\frac{1}{2}}
\end{equation}
where $\sigma_{^{13}CO}=\frac{\Delta V_{13}}{8\textrm{ln}(2)}$ and $T_{ex}$
are the one dimensional velocity dispersion of the $^{13}$CO (1-0) and
excitation temperature, respectively. k is Boltzmann's constant,
$m_{^{13}CO}$ is the mass of $^{13}CO$, $m_{H}$ is the mass of
atomic hydrogen, and $\mu$=2.72 is the mean molecular weight of the
gas. Then the three-dimensional velocity dispersion $\sigma_{3D}$
can be estimated as:
\begin{equation}
\sigma_{3D} = \sqrt{3(\sigma_{Therm}^{2}+\sigma_{NT}^{2})}
\end{equation}
The three dimensional and non-thermal velocities dispersions
shown in Figure 4 b$)$ and c$)$ also present similar distribution as
line widths. As shown in Table 7, the P-values of K-S test for
lognormal hypothesis for line widths and velocity dispersions are
all much larger than 0.05, indicating their distributions can be
remarkably well described by a lognormal fit. The lognormal
behaviors of volume or column density in molecular clouds were
frequently reported in recent observations
\citep{rid06,fro07,goo09}, which are often interpreted as a
consequence of supersonic turbulence in the observed clouds
\citep{va94}. From our results, the effect of supersonic turbulence
in the clouds should be more likely reflected in the lognormal
behaviors of the distributions of line widths and velocity
dispersions than in the distributions of column density. Figure 4
d$)$ plots the distribution for the ratios of $\sigma_{NT}$ to
$\sigma_{Therm}$. One can find most of the clumps have
$\sigma_{NT}$ larger than $\sigma_{Therm}$, indicating that turbulent
motions dominate in these clouds. The shape of their ratio
distribution is similar to the distributions of the FWHM of the
three transitions, the $\sigma_{3D}$ and $\sigma_{NT}$. But the tail part
is narrower than the others, which may suggest that the thermal
motions may not be as sensitive as non-thermal motions for revealing
star formation activity.

\subsection{Mapping results}
In the ten mapped clumps, 12 velocity components were detected. Six of them
have FWHM of the $^{13}$CO (1-0) larger than 1.3 km~s$^{-1}$ and the
other 4 belong to group L. The CO line profiles of the clump
G089.64-06.59 , one of the velocity components of G157.60-12.17,
G179.29+04.20 and G196.21-15.50 show blue asymmetric signature, The integrated
intensities are shown in the contour maps in Figure 7.
The morphologies of these clumps are different. G001.38+20.94 and
G180.92+04.53 show both diffuse features and dense clump. G108.85-00.80, G157.60-12.17b, and G196.21-15.50 exhibit a filamentary
shape with a chain of dense clumps. G161.43-35.59 and G194.80-03.41 appear as a
dumbbell also in filamentary structure. Four cores are detected in G006.96+00.89a and G161.43-35.59 respectively.
G049.06-04.18 is an isolate clump while G089.64-06.59 shows a cometary-like structure. Filamentary structures are the majority among these clumps, which is consistent with the
Herschel follow-up observation results \citep{ju12}.
Juvela et al. (2012) also found the filaments often fragment into sub-structures. Such
behavior is revealed in the gas emission too as in G006.96+00.89a shown in
Figure 7..

There are 22 cores identified with 2 dimensional gaussian fits to
the ten mapped sources. The physical parameters of these cores are
presented in Table 8. The positions of the cores are shown as offset
relative to the reference positions in column 4. Deconvolved size a
and b (column 5) is the major and minor angular sizes of the cores
and measured from the contours at half of the maximum intensity of
the $^{13}$CO (1-0) lines. The radius $R=\frac{\sqrt{ab}}{2}D$,
where D is the distance, in column 6 distributes from several tenth
to about 3 pc. Column 7-12 list the clump parameters: T$_{ex}$,
N$_{H_{2}}$, velocity dispersions and volume density
n$_{H_{2}}$=N$_{H_{2}}$/2R respectively. Then the core mass can be
derived as M=$\frac{4}{3}\pi \cdot R^{3}\cdot n_{H_{2}}\cdot
m_{H_{2}}\cdot \mu_{g}$, where $m_{H_{2}}$ is the mass of a hydrogen
molecule and $\mu_{g}$=1.36 is the mean atomic weight of the gas.
Core mass calculated with LTE assumption is given in column 13.
Column 14 and 15 are virial mass and the Jeans mass respectively
(see next Section). The final two columns are the group and the
location of the clumps. The column density of the cores ranges from
1.3 to 7.7 $\times$10$^{21}$ cm$^{-2}$. The LTE masses of the cores
range from 9 to 1.5x10$^4$ M$_{\sun}$ with a median value of 110
M$_{\sun}$. In the Herschel follow-up observations, the 26
identified filaments have column density ranging from 0.9 to 19.2
$\times$10$^{21}$ cm$^{-2}$ and masses ranging from 3.5 to
3.2x10$^3$M$_{\sun}$ with a median value of 120 M$_{\sun}$
\citep{ju12}, which are consistent with the case of the cores in
this work.

\section{Discussion}

\subsection{The line center velocities}
The well coincidence of three transitions observed towards the Planck clumps is not usual in
star formation regions. Line center velocities of different molecular
species could be significant offset from the systematic velocity in
active star forming regions. In the six NH$_{3}$ clumps of
G084.81-01.09, the deviations between the $^{12}$CO and $^{13}$CO are
all larger than 1 km~s$^{-1}$ \citep{zhang11}. V$_{lsr}$ of
the $^{12}$CO deviated from that of $^{13}$CO or C$^{18}$O are also seen
in infared dark clouds. In 61 infrared dark clouds, there are 10\% sources
with V$_{lsr}$ deviation of $\gtrsim$ 1 km~s$^{-1}$ from the $^{12}$CO
and the C$^{18}$O lines \citep{du08}. The rather large discrepancy
between the V$_{lsr}$ of $^{12}$CO and $^{13}$CO can be also seen in the
sub-millimeter clumps \citep{qin08}. Line center velocity difference
of various molecular species may origin from molecular layers with
different temperature or trace different  kinematical gas layers
\citep{Ber97,mu11}. The deviation between V$_{13}$ and V$_{18}$ of
the Planck clumps also tends to be smaller than those of
\cite{my83a}. All these suggest that the cold clumps are the quietest
molecular regions found so far as a whole.

\subsection{Distances of the clumps:}
Distance is essential to investigate the spatial distribution and
physical conditions of the clumps. \cite{ade11d} have estimated
distance for 2619 C3PO clumps using various extinction signatures.
They also found that there are 127 Planck cold clumps associated
with IRDCs which have a kinematic distance from \cite{si06}. In our
sample there are only 2 clumps closed to the sources of \cite{si06}
at the 0 latitude Galactic layer. There are certainly a part of the
ECC clumps have distance estimated with extinction methods. We
estimated the kinematic distances of the clumps using the Vlsr of
the clumps which could be a comparison for those with known
distance. We adopt the rotation curve of \cite{cle85} R$_{\sun}$=8.5
kpc and $\Theta_{\sun}$=220 km~s$^{-1}$ in our calculation. To see
the possible physical relation of the components in clumps with
double and three peaks, the distance of each component was
calculated. For the clumps within the inside of the solar circle
there are two solutions. The clumps are perhaps located at the front
of Galactic bulk of the diffuse background, since extinction is
rising along a line of sight that crosses a dust clump
\citep{ade11a}. So the near value of the distances was adopted.
Among our sample there is a number of clumps located in molecular
complex with known distance. Since these complexes are with rather
large areas and clumps resident inside are with different V$_{lsr}$,
the kinematic distances of the clumps within one complex are quite
different. Therefore for every clump within the same complex their
distances are also given out as the kinematic distances. However, in
a case that a clump has ambiguity on the distance, the one close to
the known distance of the complex was adopted.

The histogram of the kinematic distances is plotted in Figure 8.
Distances were obtained for 741 $^{13}$CO components. 51\% of the
components are with distances within 0.5 and 1.5 kpc. The mean value
is 1.57 kpc, smaller than those associated with infrared dark clouds
by \cite{si06}. The reason may be due to cloud properties, of which
most of our clumps or components belong to low mass group and the
infrared dark clouds of \cite{si06} are at the first quadrant of the
mid-plane and with distances between 0.7 to 7.8 kpc \citep{si06}.

\subsection{Physical Parameter distributions in the Galaxy}
\subsubsection{Excitation temperature and the ratio of line strength of $^{12}$CO and $^{13}$CO}
Figure 9 presents the distribution of the T$_{ex}$ and the ratio of
T$_{12}$/T$_{13}$. Fig. 9 a$)$, c$)$  are for radial changes and
b$)$, d$)$ for the altitude from the Galactic plane. The excitation
temperature is higher than 10 K from 4 to 8 kpc. The early observations also showed the high
excitation of the gas emission at the 3-7 kpc molecular ring
(Goldsmith 1987 and the references there in). Around 8 kpc, T$_{ex}$
is also larger than 10 K, which may be related to the emission of the
giant molecular clouds near the Sagittarius arm. The ratio of
T$_{12}$ to T$_{13}$ is high at R$\sim$5 kpc, then deceases with R
and has a valley between 5 - 8 kpc, then have a lowest value at $\sim$6 kpc, suggesting the
brightness temperature of $^{13}$CO (1-0) is relatively high in this
region.

Fig. 9 b$)$ shows that excitation temperature changes with the
altitude. There are two peaks, 11 and 13 K at Z$\sim$350 pc and
$\sim$470 pc respectively. The changes of the ratio of
T$_{12}$/T$_{13}$ shown in Fig. 9 d$)$ are rather monotone and reach
the low value at Z$\sim$470 pc, suggesting the brightness temperature
of $^{13}$CO is relatively high at this altitude.

\subsubsection{Velocity dispersion}
Variations of the velocity dispersion with the distance from the
Galactic center, and the altitude from the Galactic plane are
investigated. The radial variations of $\sigma_{3D}$, $\sigma_{NT}$
and the ratio of $\sigma_{NT}/\sigma_{Therm}$ are plotted in Figure
10 a$)$, c$)$ and e$)$ respectively. The variation of the
$\sigma_{3D}$ and $\sigma_{NT}$ as well as the ratio of
$\sigma_{NT}/\sigma_{Therm}$ with R are about the same and they
reached the maximum at R$\sim$5 kpc, which suggest that the dynamic
process is most violent at the 5 kpc Galactic ring. From 6 kpc, the
$\sigma_{3D}$, $\sigma_{NT}$ and the ratio of
$\sigma_{NT}/\sigma_{Therm}$ seem linearly increase with R,
indicating turbulence becomes more violent in the outer part of the
Galaxy.

Figure 10 b$)$, d$)$, f$)$  present changes of the velocity
dispersions $\sigma_{3D}$, $\sigma_{NT}$ and the ratio of
$\sigma_{NT}/\sigma_{Therm}$ with the altitude. One can see that
they all decrease with the increasing of the altitude from the
Galactic disk to highness 475-525 pc, showing the closer to the
galactic plane, the stronger of the turbulent process.  At
Z$\sim$680 pc all of them reached a minor peak. We found the clumps
at this minor peak are distributed around (l$\sim$174$\arcdeg$,
b$\sim$17$\arcdeg$) or (l$\sim$4$\arcdeg$, b$\sim$17$\arcdeg$), this
minor peak may be concerned with the emission regions of Taurus and
$\rho$ Oph. From panel e$)$ and f$)$, one can see non-thermal motion
dominates the line broadening. This is the first time to obtain an
evidence for non-thermal line broadening from survey of the $^{13}$CO (1-0) lines.

One can see that there are some differences among the radial
variations of the velocity dispersion and excitation temperature.
The maximum of T$_{ex}$ -R variation is at rather high values around
4- 8 kpc and reaches maximum at 6 kpc. The T$_{ex}$ variation is
milder than that of $\sigma_{NT}$ suggesting that the gas heating
and cooling occur in a wider spatial region than the turbulence.

\subsubsection{$^{13}$CO opacity, X$_{13}$/X$_{18}$ and H$_{2}$ column density}
Figure 11 a$)$ shows the radial variation of the optical depth of
the $^{13}$CO (1-0) lines. The smallest value is at the 5 kpc
ring. Between 5.5 and 8 kpc there is a high feature, then
decreases till 14 kpc. One reason for its low valley is T$_{ex}$ is
rather high around 5 kpc (see Fig. 9 a$)$). Besides, its emission is
relatively low comparing with that of the $^{12}$CO. For example,
G017.22-01.47 at R= 4.90 kpc is $\tau_{13}$ =0.3, T$_{ex}$
=10.1 K and T$_{13}$=0.42 K; G033.70-00.01, R=5.05 kpc,
$\tau_{13}$ =0.5, T$_{ex}$ =9.2 K and T$_{13}$=1.12 K;
G028.56-00.24, R= 5.28 kpc, $\tau_{13}$= 0.4, T$_{ex}$ =10.9 K,
T$_{13}$=1.13 K.

The ratio of X$_{13}$ to X$_{18}$ presented in Figure 11 c$)$ is
rather low between 5-7 kpc and its corresponding values range from
$\sim$6 to 7, still higher than the terrestrial value. At 8 kpc and
$>$10 kpc the value is near 8.

Figure 11 e$)$ shows the radial variation of the column density of
hydrogen molecules. Clearly it presents an enhancement at 5 kpc
where the most dense and massive star formation regions within our
Galaxy are located. Then it is almost at the similar level till
outer region except at 9 kpc where there is a minor low valley.
Owing to the small $\tau_{13}$ around 5 kpc (see Fig. 11 a$)$ the
column density is mainly affected by the velocity dispersion
$\sigma_{3D}$ or $\sigma_{NT}$ shown in Figure 10. To confirm the
Galactic distribution of the column density, the radial distribution
of the flux density at 857 GHz dust emission detected by the
Planck was plotted at Figure 11 g$)$. The variations agree with that of
the column density very well. At 9 kpc the 857 GHz flux is a
little higher showing another dense structure (Goldsmith 1987 and
the reference therein).

The variation of $\tau_{13}$ with altitude is shown in Figure 11
b$)$. It exhibits a high feature between 350-550 pc and reaches its
maximum at 450 pc. The change of the ratio of X$_{13}$/X$_{18}$ seen
in Fig. 11 d$)$ seems to be opposite to that of $\tau_{13}$ with its
lowest point at Z=450 pc. Fig. 11 f$)$ presents the variation of the
molecular hydrogen column density with Z. At Z= 300 and 500 pc, the
values are higher than in the other regions. There is a low valley at Z=450.
Combining the altitude variation of $\tau_{13}$ and the velocity
dispersion of Fig. 10 b$)$, d$)$, f$)$ where $\sigma_{NT}$ is at low
values, again showing that non-thermal line width is the major
factor to determine the gas column density. Between 350-550 pc the
flux at 857 GHz is higher too, which are consistent with the
variation of N$_{H_{2}}$ as a whole. These results revealed that the
column density reaches the maximum at R=5 kpc, a low valley at Z=450
pc and mainly caused by non-thermal velocity dispersion, which
are also not reported before.

\subsubsection{Parameters of the clumps in different molecular complexes}
For the 12 complexes included in our sample, a statistical
analysis of the physical parameters was made. The corresponding
average values are presented in Table 2. They display different
trends: The famous star formation regions including Ophiuchs, Orion,
Oph-Sgr and Taurus harbor 253 observed clumps. They have the
highest excitation temperatures and column densities. The average
$^{13}$CO FWHM of these clumps are less than 1.5 km~s$^{-1}$, even in
Orion it is only 1.29 km~s$^{-1}$, suggesting that low mass clumps are the
dominant sources in the Planck cold clumps. Their non-thermal
velocity dispersion is almost two times of the thermal one except in
Ophiuchs where $\sigma_{thermal}\sim\sigma_{NT}$. Cephens harbors 87
observed clumps. The FWHM of $^{13}$CO J=1-0 line is between Orion
and the above mentioned star formation regions. $\sigma_{NT}$ is
also the dominant factor for the line broadening. A common character
can be seen that all the 4 Quadrants and the anti-center regions have
FWHM of $^{13}$CO J=1-0 line $\gtrsim$ 1.5 km~s$^{-1}$. All of them
belong to the high mass group. The $\sigma_{NT}$ is about 4 times of
the $\sigma_{thermal}$, indicating that these regions have stronger
dynamic processes than other star formation regions.

\subsection{Line profiles}
There are 15 clumps having absorption dip at the line center, which is rather
symmetric relative to the V$_{lsr}$. Eight of them show the dip in
all the three transitions. They may be the candidates
for the $^{12}$CO depletion. In the other 7 clumps, only $^{12}$CO
line has the center dips which may originate from self-absorption.
Mapping is needed to further examine the properties of the line
center dips.
18 and 15 cores were identified have blue and red profiles (see
Table 5). Blue profiles are a kind of typical feature of molecular
clump collapse \citep{zhou93}. While the red profile could originate
from expansion of the clump or outflow motion. The ratios of the
clump with the blue and red profiles to the total clump numbers of
clumps are small, which are 2.7\%  and 2.2\% respectively. The blue
excess $E=(N_{b}-N_{r})/N_{t}$ is 0.004, here $N_{b}$, $N_{r}$ and
$N_{t}$ are the clump numbers with the detected blue and red
profiles and the surveyed sample \citep{mar97}. E of the Class -I, 0
and I clumps is 0.31, 0.30 and 0.30 shown in HCN (3-2) line
respectively \citep{ev03} while it is 0.15-0.17 for UC HII region
precursors and 0.58 for UCHII regions detected with HCO$^{+}$(1-0)
lines \citep{wu07,fu05}. Nevertheless a sample of 27 Orion starless
clumps, 9 have blue profile and 10 have red one, implying that the
blue excess E is -0.04 \citep{ve08}. The ratios of blue and red
profiles to these cores are 33\% and 37\%, greatly exceeding those
of our sample, suggesting that star formation activities occur more
frequently in the Orion cores than in the Planck cold clumps. The
small ratio of the blue and red profiles means that most of the
Planck cold clumps do not have systematic star forming motion yet.
We also identified 19 and 13 cores with blue and red line asymmetry respectively. Different from blue and red profile, line asymmetry reflects whole gas motion of the core, which may results in interaction of the core and its environment.

The high velocity wings are rare in Planck cold clumps. Among the
surveyed clumps, only 3 clumps were detected with blue wing and 6
with red one, 8 with both blue and red wings and 5 with pedestal
feature, showing rather rare star formation feed back activities.

\subsection{Conditions of the clumps at high latitude clumps}
There are 41 clumps located at latitude higher than 25$\arcdeg$. Six
clumps have two velocity components and one has three. Five of the
clumps belong to the high mass group. T$_{ex}$ is intermediate among
the 12 complexes (see Table 2). The column density is
3$\times$10$^{21}$ cm$^{-2}$ on the average.  $\sigma_{NT}$ tends to
be smaller among the 12 complexes but still larger than those in the
Ophiechs and Oph-Sgr. Core G089.03-41.28  is with blue profile while cores G182.54-25.34 and G210.67-36.77 are with red profile. Their altitudes are 2.4, 0.64 and 0.15 kpc respectively, showing that
star formation signatures also exist in clumps at high latitude. Here the
kinematic distance was adopted.

Since the diffuse emission was found over all the Galactic sky
\citep{hau84}, many high latitude clouds such as infrared cirrus were also
detected \citep{low84}. \cite{hei93} made a survey for
16$\arcdeg\leq$b$\leq44\arcdeg$, 117$\arcdeg\leq$l$\leq160\arcdeg$
in the 2nd quadrant. They found that the clouds with the CO emission
are 13\% of the survey sample. \cite{ya03} also carried out a CO survey within the latitude
-30$\arcdeg$-(43$\arcdeg$). They identified 110 $^{12}$CO clouds
with a total of mass of 1200 M$_{\sun}$, in which all of the clouds are not dense enough to form
stars. The conditions of our 41 Planck samples are more closer to
star forming statues. Additionally the latitudes of the Planck clumps
exceed the above samples. These results suggest that the ECC clumps are a
good guide to investigate initial conditions or search for
star formation.

\subsection{States of the ten mapped clumps:}
The different morphologies of the contour maps of the ten mapped
clumps show that the Planck clumps contain a rather long
evolutionary sequence, which includes diffuse and elongating
regions, filament structure or cometary shape, multiple cores as
well as isolated core.

Assuming a core is a gravitationally bound isothermal sphere with
uniform density and is supported solely by random motions, the
virial mass M$_{vir}$ can be calculated following \cite{ung00}:
\begin{equation}
\frac{M_{vir}}{M_{\sun}}=2.10\times10^{2}(\frac{R}{pc})(\frac{\Delta V}{km~s^{-1}})
\end{equation}
where R is the radius of the clump and $\Delta V$ is the line width of the $^{13}$CO (1-0).
The virial masses are listed in the column 14th of Table 8.

In molecular clouds, many factors including thermal pressure,
turbulence, and magnetic field support the gas against gravity
collapse. The Jeans mass, which takes into account of thermal and
turbulent support, can be expressed as \citep{hen08}:
\begin{equation}
M_{J}\approx1.0a_{J}(\frac{T_{eff}}{10~K})^{3/2}(\frac{\mu}{2.33})^{-1/2}(\frac{n}{10^4~cm^{-3}})^{-1/2}M_{\sun}
\end{equation}
where $a_{J}$ is a dimensionless parameter of order unity which
takes into account the geometrical factor, $\mu=2.72$ is the mean
molecular weight, $n=\frac{N_{H_{2}}}{2R}$ is the volume density of
H$_{2}$ and $T_{eff}=\frac{C_{s,eff}^{2} \mu m_{H}}{k}$ is the
effective kinematic temperature. The effective sound speed
$C_{s,eff}$ including turbulent support can be calculated as:
\begin{equation}
C_{s,eff}=[(\sigma_{NT})^{2}+(\sigma_{Therm})^{2}]^{1/2}
\end{equation}
The calculated Jeans masses are listed in the 15th column of Table 8.

There are 7 cores with M$_{LTE}$ larger than M$_{vir}$ and
M$_{J}$, which maybe under collapse. There are also 7 cores having M$_{LTE}$ agree M$_{vir}$ and
M$_{J}$ within a factor of three. Considering the uncertainties in
mass estimation, these cores may be in magical states. The remaining cores seem to be
gravitationally unstable. One should keep in mind, these ten clumps are not a
representative sample for the whole ECC, but include the morphologies of the
majority of the Planck clumps. Most of the ECC
clumps show diffuse molecular emission or harbor gravitationally stable dense cores (Liu, Wu \& Zhang, in preparation).

The mapped clumps are noted individually as the following:

G001.38+20.94: It is located in $\rho$ Oph. The gas emission is
diffuse and with size $>$1 pc. The density is lower than 10$^{3}$
cm$^{-3}$ and mass $>$750 M$_{\sun}$. The excitation temperature is
rather high (14 K). The average $\sigma_{3D}$ is only 0.67
km~s$^{-1}$. Actually it is located at (0.5,-1.0) with respect
to L43B, the mixture of
isolated globules and complex \citep{ben89}. Maybe it is in the
transition between diffuse ISM and dense molecular cloud.

G006.96+00.89: It is located in the 4th quadrant. Two velocity
components with V$_{lsr}$ 9.33 and 41.67 km~s$^{-1}$ were detected
in the clump and both have FWHM larger than 1.3 km~s$^{-1}$. The two
velocity components have 4 and 1 cores respectively. No astronomical
object was found associated with this clump so far. G006.96+00.89a
with 4.33 km~s$^{-1}$ velocity component appears elongated from SE
to NW and has a chain of at least four cores. The other component
has an isolated core near the mapping center, and the cores are not
very dense with n lower than 10$^{3}$ cm$^{-3}$.

G049.06-04.18: It is an isolated clump and located in the first
quadrant. The mass calculated with LTE is close to the Jeans and
virial mass. It is CB 198 and contains IRAS 19342+1213 (J2000=19 36
37.8 +12 19 59) located at (8$\arcsec$, 43$\arcsec$) of the clump
\citep{Go06}.

G089.64-06.59: It is a clump of the first quadrant. Its gas clump
tends to be cometary. The starless clump CB 232 AMM 1 is located at
(20$\arcsec$, -36$\arcsec$) of the clump. It harbors an infrared
source IRAS 21352+4307 \citep{hua99}. At about 15$\arcsec$ eastern
there is a near infrared source YC1-I. The LTE mass and the Jean
mass, virial mass all are close to each other.

G108.85-00.80: It belongs to the 2nd quadrant. There are no
associated objects or known cloud was found for this clump. It shows a filamentary structure and is
compact. The $^{13}$CO (1-0) line width is 2.7
km~s$^{-1}$ and is a typical high mass clump. The LTE mass is larger
than both the Jeans and virial mass. It seems very likely to be in
gravitational collapse.

G157.60-12.17: It locates in the Taurus complex and has two components
belonging to L group. Contours of their integrated intensity show the first
component is rather diffuse and the second one contains two
cores. TGU 1064 is located at 65$\arcsec$ east and south
\citep{do05}.

G161.43-35.59: It is a high latitude clump and belongs to L group. It
contains at least 4 cores. No associated object was found.

G180.92+04.53: It is at the side of Anticenter of the Galaxy and
belongs to the H group.

G194.80-03.41. The dumbbell gas emission region elongates in
north-west direction. Two large clumps are connected and each
contains at least two cores. There was no associated object found.
All the cores should be starless. The clump masses are all larger
than the corresponding Jean and virial mass, suggesting they are at
the gravitational bound states. TGU H1364 P8 is about 2$\arcmin$
away \citep{do05}.

G196.2-15.5: This clump is with blue asymmetry line (see Table 6). It is associated with L1595 and a reflection nebular
VDB 40 \citep{ma86}. Three cores were found in this filamentary structure.
All belong to Group L. Except the M$_{J} <$ M$_{LTE}$,
for the core 1, M$_{LTE}$ is less than M$_{J}$ and M$_{vir}$,
but all these masses are close to each other.

The morphology, structure and physical parameters of the small set
of the ten mapped clumps show that Planck cold clumps cover
different phases which may be 1:  in a transition phase from diffuse
ISM to cloud; 2: in a state close to gravitational bound. Among the
10 clumps, 4 are in filamentary or elongated shape which show that
filamentary clumps may be the majority in C3PO clumps. 3:  an
isolated core or multiple cores; 4: starless cores, 5: in a state
that harbor infrared sources. In total 22 cores were found and 20
are starless.

\subsection{Gas and dust coupling}
The range of the clump kinetic temperature is from 4-27 K, wider
than those of dust temperatures (7-17 K) \citep{ade11a}. However
there are only 12 clumps with T$_{k}>$17 K. The $\sim$98\% clumps
have T$_{k}\leq$ 17 K, showing both the dust and gas are cold and
couple well. Figure 12 a) shows a comparison of the T$_{k}$ and
T$_{d}$ in both "Total" and "Cold" (see Fig. 12 a) ). Most of the
clumps are with T$_{d}>$T$_{k}$, indicating that gas could be heated
by dust in these regions \citep{gol74}. For the clumps with
T$_{d}<$T$_{k}$, the gas may be due to ongoing protostellar process.
For example, for the 27 clumps with T$_{k}>$16 k$ >$T$_{d}$, 6 are
in or close to Ophiuchs, 8 in Taurus and 10 in Orion. The column
densities deduced from dust emission and CO lines were plotted in
Figure 12 b). One can see that the range of the values from dust is
about 3 orders in total and that from CO lines is about 2.5 orders,
sightly narrower. However both the column densities concentrate on
10$^{21}$-10$^{22}$ cm$^{-2}$. According to \cite{har98} such column
density is just about the critical value of the cloud collapse.

\subsection{Evolutional phases - A comparison to different star formation samples}
To investigate the physical conditions and examine the possibility of stars
forming in the cold clumps, we compare the line widths of $^{13}$CO
(1-0) and column densities with the following CO molecular line surveys towards different kind of targets:

a$)$ Methanol maser sources \citep{liu10};

b$)$ Candidates of UC HII region chosen with the IRAS colour index and flux limit \citep{wu01,wood89};

c$)$ Candidates of extremely young stellar objects chosen with
redder color IRAS index and smaller flux density than those of UC
HII regions \citep{wang09};

d$)$ Infrared dark clouds (IRDCs) \citep{si06}.

e$)$ Extended green objects (EGOs) identified from the Spitzer GLIMPSE survey \citep{chen10}

Figure 13 a) plots a cumulative fraction of the FWHM of $^{13}$CO
(1-0). It shows that the methanol maser sources have the
largest FWHM and the smallest slope. And the IRDCs and EGOs
have similar shape with the methanol maser sources but with a
slightly larger slope. When the FWHM is less than 3 km~s$^{-1}$,
the slope of IRDCs is almost the same as those of the UC HII
candidates and the redder-weaker IRAS sources. When FWHM becomes
larger than 3 km~s$^{-1}$, the changes of the the UC HII candidates
and the redder-weaker IRAS sources are much steeper than IRDCs, EGOs and
the methanol maser sources. The slope of the fraction of the
redder-weaker IRAS is large and its maximum value is at 6
km~s$^{-1}$. The variation of cumulative fraction function of FWHM
for the Planck cold clumps is the narrowest. The FWHM of the Planck cold clumps
are the smallest comparing with the other samples. For all the
samples used in the comparison with Planck cold clumps, their FWHM
are almost larger than 2 km~s$^{-1}$ while the cumulative fraction
at FWHM$>$2 km~s$^{-1}$ of the cold clumps is less than 10\%.

The comparison of column densities with the samples of the above a), c), d) and e)
is presented in Figure 13 b). The cumulative fraction distribution also
produces the smallest column density range for the Planck cold clumps. IRDCs
have nearly the same shape with Planck cold clumps when column density
below 10$^{21}$ cm$^{-2}$, but are similar with the redder-weaker
IRAS sources at high densities, indicating IRDCs may be at a transition
phase between Planck cold clumps and redder-weaker IRAS sources. The
methanal maser sources and EGOs have the largest column densities, indicating active star formations
in them.

These results show that the Planck cold clumps are quiescent and have
smallest column densities among these star formation samples on the
whole. Most of them seem to be in transition from clouds to dense
clumps.

\section{Summary}
Aiming to understand gas properties of the Planck cold dust clumps we
have carried out a survey for ECC clumps with J=1-0 lines of the
$^{12}$CO, $^{13}$CO and C$^{18}$O. Using the 13.7 m telescope of
PMO, 674 clumps were observed. Their observed parameters V$_{lsr}$,
FWHM and antenna temperature were presented. Distances and physical
parameters T$_{ex}$, velocity dispersions and N$_{H_{2}}$ were
derived, and their spatial distribution, regional difference and environmental effect were
investigated. Ten clumps were mapped and their morphologies and
properties were analyzed. Evolutionary states of the cold clumps were
discussed when comparing with different star formation samples. Our
main findings are as following:

1. With the survey of the 673 clumps, 782 $^{13}$CO emission
components were identified. 437 components have the detection of all
the three transitions. Line center velocity differences for V$_{12}$-V$_{13}$ of
94\% clumps and for V$_{13}$-V$_{18}$ of 98\% clumps less than 3 $\sigma$ of the velocity resolution.
The correlations between the V$_{12}$ and V$_{13}$,
V$_{13}$ and V$_{18}$ have $\sim$100\% confidence. This is the first
time to confirm the agreement of the velocities of the $^{12}$CO and its
major isotopes $^{13}$CO and C$^{18}$O lines at such a large
sample. It suggests that the Planck clumps are quite, cold and
uniform as a whole and have no significant
differences in dynamical or thermal
material layer structures.

2. For each of the identified components, kinematic distances,
galactocentric distance and altitude from the galactic plane were
derived. The distance ranges from 0.1 to 21.6 kpc and 82\% of them
are located within 2 kpc and 51\% of the components have distances
within 0.5 and 1.5 kpc.

3. The mean value of $^{12}$CO antenna temperature T$_{12}$ is
3.08$\pm$1.37 K. Mean ratio of T$_{12}$/T$_{13}$ is 2.15$\pm$1.35 and
T$_{13}$/T$_{18}$ is 3.88$\pm$1.71. Excitation temperature ranges from
3.9-27.0 K. 98\% of the clumps have T$_{ex}$ (or T$_{k}$ under
LTE assumption ) smaller than 17 K. A comparison of the T$_{k}$ and
T$_{d}$ shows that most of the clumps are with T$_{d}>$T$_{k}$
suggesting gas could be heated via collision with dust and dust and
gas are coupled well in 95\% of the clumps. There are clumps with
T$_{k}>$17 K, which are located in star formation regions such as
the Orion and Taurus, suggesting that the high T$_{k}$ is related to star
forming activities.

4. The mean FWHM of the three tansitions are 2.03$\pm$1.28,
1.27$\pm$0.77, 0.76$\pm$0.75 respectively. Adopting 1.3 km~s$^{-1}$
of the FWHM of the $^{13}$CO (1-0) lines as a criterion of high or low
mass clumps, there are $\sim$75\% belonging to L Group, suggesting that the majority of the ECC
clumps are low mass clumps. For the $^{13}$CO lines
three dimension velocity dispersion $\sigma_{3D}$, non-thermal
velocity dispersion $\sigma_{NT}$ were calculated and it is found
that the non-thermal motion is dominant line broadening.

5.  Column densities of the $^{13}$CO and C$^{18}$O were derived. They
span from 10$^{14}$-10$^{16}$ cm$^{-2}$, which are about critical
value for collapse. The ratio of [X$_{13}$]/[X$_{18}$] has mean
value 7.0$\pm$3.8, higher than the terrestrial value. Hydrogen
molecule column densities were obtained for 782 $^{13}$CO
components, which are from 10$^{20}$ to 4.5$\times10^{22}$
cm$^{-2}$. Column densities derived from CO observations cover
narrower range than those deduced from dust emission.

6.  Histograms for antenna temperature T$_{12}$, ratio of
T$_{12}$/T$_{13}$, T$_{13}$/T$_{18}$, excitation temperature
T$_{ex}$, and the velocity dispersions ($\sigma_{3D}$,
$\sigma_{NT}$ and ratio of $\sigma_{NT}/\sigma_{therm}$) as well as
$\tau_{13}$, [X$_{13}$]/[X$_{18}$] and N$_{H_{2}}$) were fitted with
lognormal distribution. The distributions of line widths of the
three transitions, T$_{13}$/T$_{18}$, T$_{ex}$, $\sigma_{3D}$,
$\sigma_{NT}$, $\sigma_{NT}/\sigma_{therm}$ and
[X$_{13}$]/[X$_{18}$] can be well depicted purely by a lognormal
distribution. We found that the
deviation of the N$_{H_{2}}$ cumulative fraction distribution from
the lognormal distribution origins from that of optical depth. The
long tail of the distribution of the column density is consistent
with those found in active star forming regions, indicating some of
the Planck cold clumps reside in star forming regions. We suggest the
distributions of line widths or velocity dispersions are more likely
to reflect the effect of supersonic turbulence in clouds.

7.  Physical parameters variations in Galactic space were
investigated. As a variation with radial the hydrogen molecule
column density reaches the maximum at R$\sim$5 kpc, which is consistent
with the 857 GHz flux density. However $\tau_{13}$ decreases to a low valley
while the $\sigma_{NT}$ or $\sigma_{3D}$ reach the peak. The results suggest that the velocity
dispersion of the molecular line is the dominant factor in determining
the column density. Velocity dispersions decrease with the Altitude
increase and reach the lowest point at Z=450 pc. Although
$\tau_{13}$ is at its highest value here, the column density is at a
low valley, further conforming that
non-thermal line broadening plays a major role in excting the molecular
transition.

8. Line profile characteristics including possible depletion dip,
blue and red profiles as well as high velocity wings were found in
part of the clumps, indicating the star forming activities. However
the number of these clumps less than 10\% of the whole sample, indicating that star formation is not
active yet in Planck cold clumps.

9. Parameters in different molecular complexes in our Galaxy show different
scores. Clumps in molecular complex Ophiuchs, Taurus and Orion show
high excitation temperatures while those in 1st-4th quadrants have
larger velocity dispersions.

10. Ten mapped cold dust clumps have very different gas emission
morphologies, showing that filaments and elongated structures are in
the majority. 22 gas cores were identified with size tenth to 5 pc,
density from 10$^{2}$-10$^{3}$ cm$^{-3}$ and mass from dozens to
thousands of M$_{\sun}$. Only 7 cores seem to be in gravitational
bound state.

11. Planck cold clumps are the most quiescent ones among the samples
of weak-red IRAS, UC HII candidates, IRDCs, EGOs and methanol maser
sources, suggesting that the Planck cold clumps are at the very
early phase in cloud evolution.

This work is a preliminary investigation for Planck cold clumps.
Mapping observations are
 necessary for obtaining properties of the Planck clumps.
Various molecular species, especially dense molecular tracers are
needed for examining dense clumps and star formation processes in
the ECC clumps. Mapping with HI line in ECC clumps is useful to
investigate the transition from diffuse to dense ISM.

\section*{Acknowledgment}
\begin{acknowledgements}
We are grateful to the staff at the Qinghai Station of PMO for their
assistance during the observations. Thanks for the Key Laboratory
for Radio Astronomy, CAS for partly support the telescope operating.
This work was partly supported by China Ministry of Science and
Technology under State Key Development Program for Basic Research
(2012CB821800).
\end{acknowledgements}

\begin{figure}
\includegraphics[angle=0,scale=0.8]{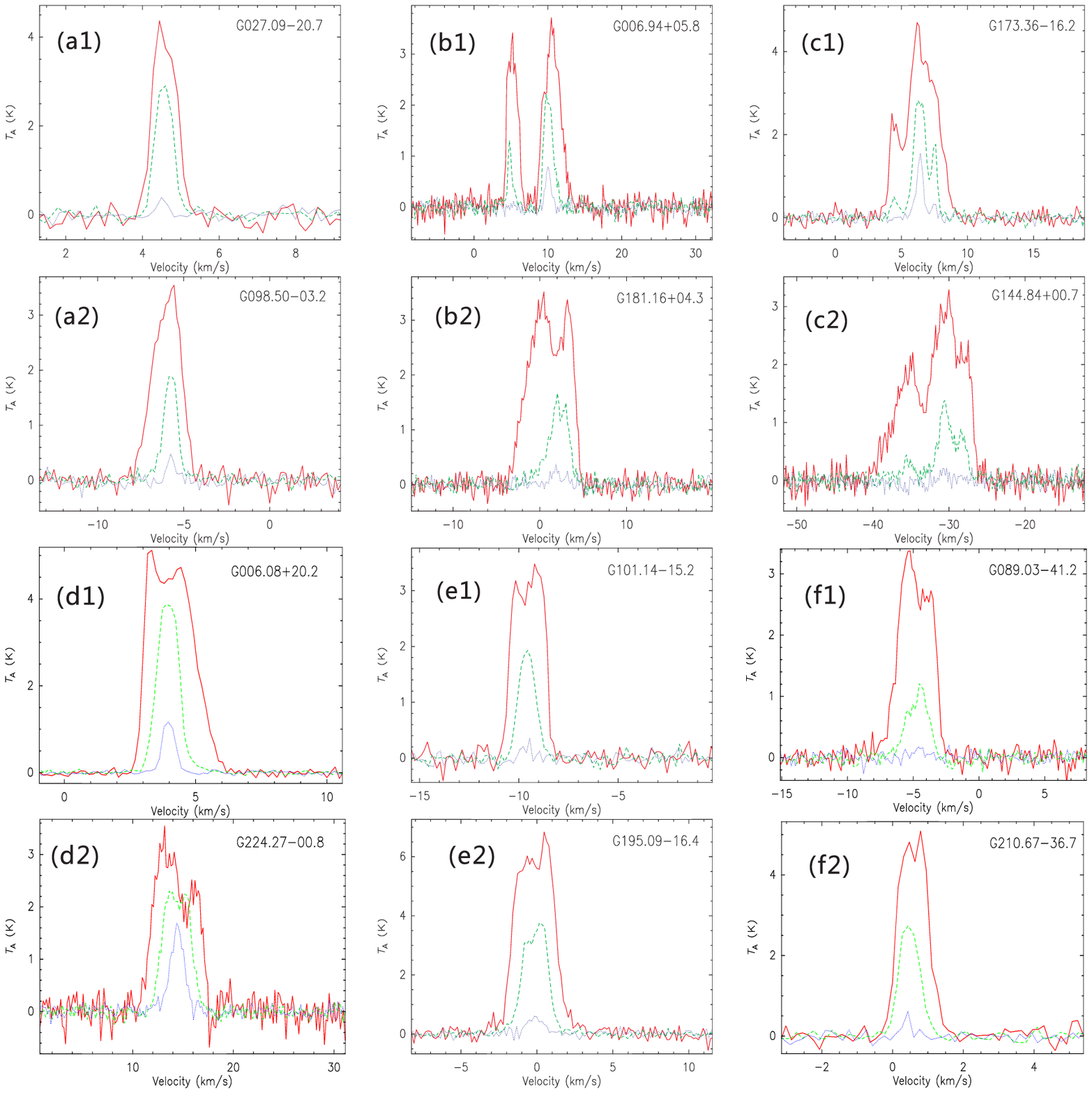}
\caption{Examples of J=1-0 lines of 12CO, 13CO and C18O spectra with different profiles. The classes are as following: a, b, c denote single, double and three velocity component spectrum respectively.
d,e for blue- and red- profile; "1"and "2" for low and high mass group respectively. f1 and f2 for blue and red profile of high latitude clumps. g,h,i, for blue-, red-asymmetry and center dip respectively; j, k, and l for blue wing, red wing and pedestal respectively, "1"and "2" denote the same as above.
}
\end{figure}

\setcounter{figure}{0}

\begin{figure}
\includegraphics[angle=0,scale=0.8]{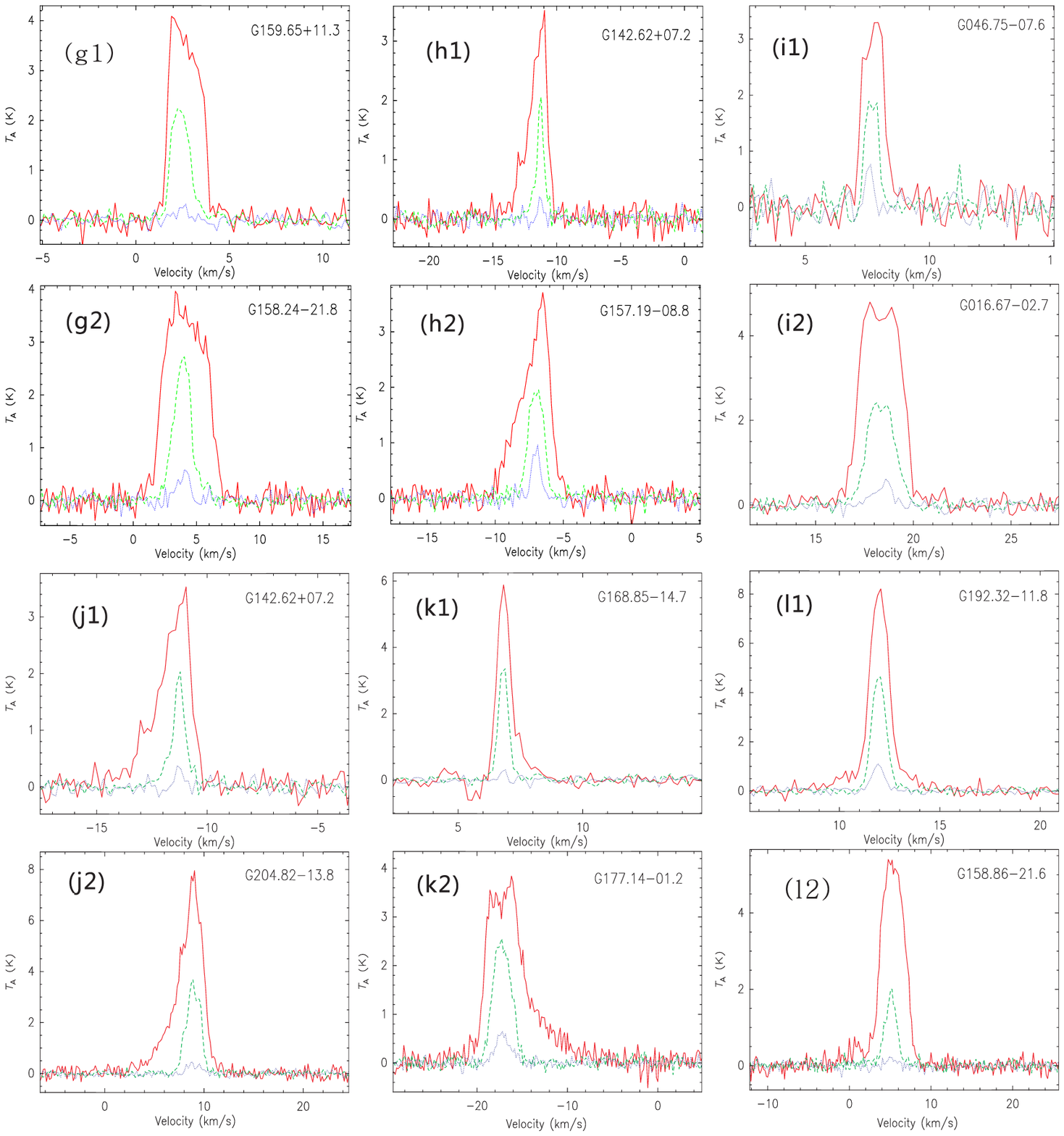}
\caption{continued}
\end{figure}

b

\begin{figure}
\begin{minipage}[c]{0.5\textwidth}
  \centering
  \includegraphics[width=70mm,height=55mm,angle=0]{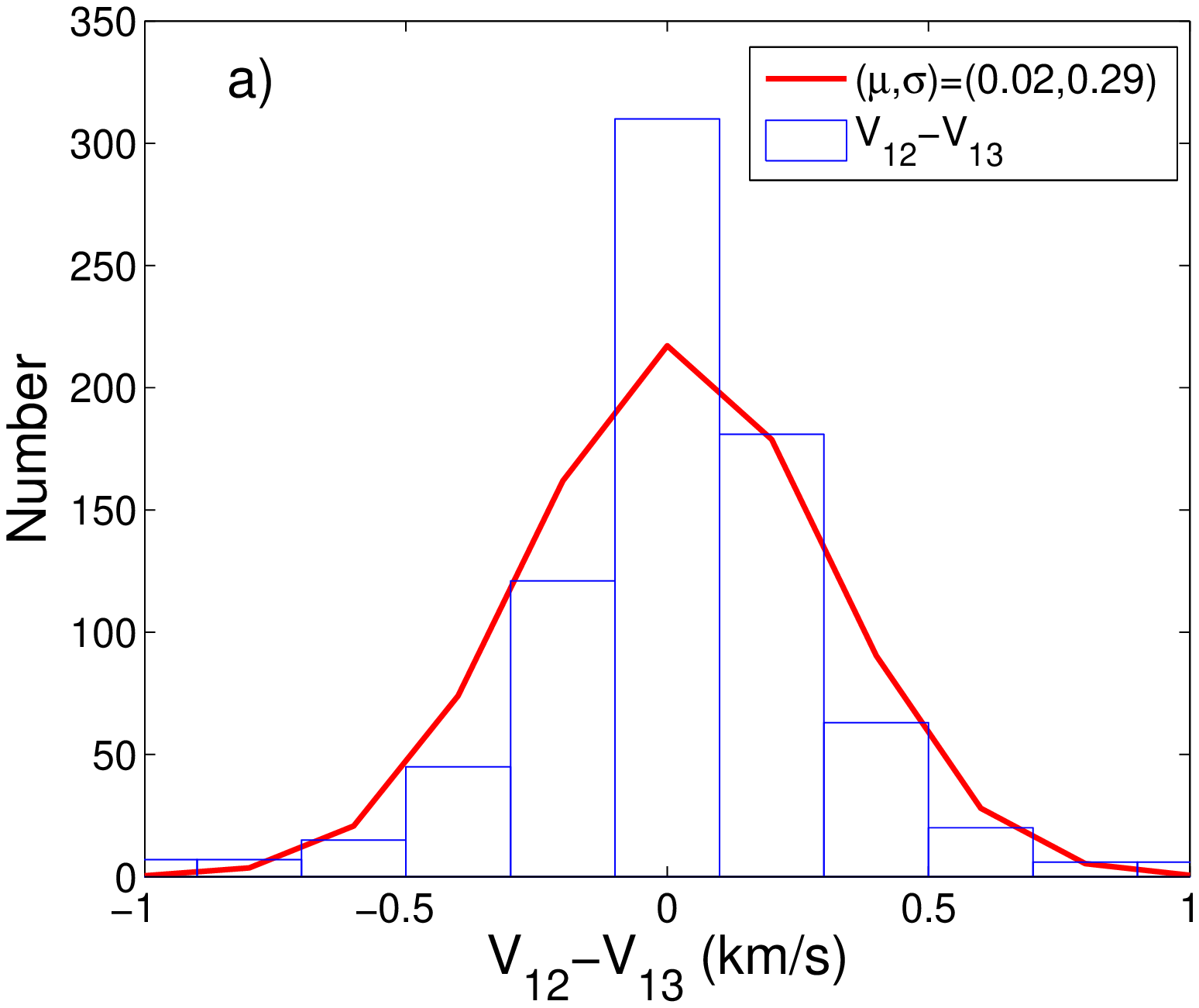}
\end{minipage}
\begin{minipage}[c]{0.5\textwidth}
  \centering
  \includegraphics[width=70mm,height=55mm,angle=0]{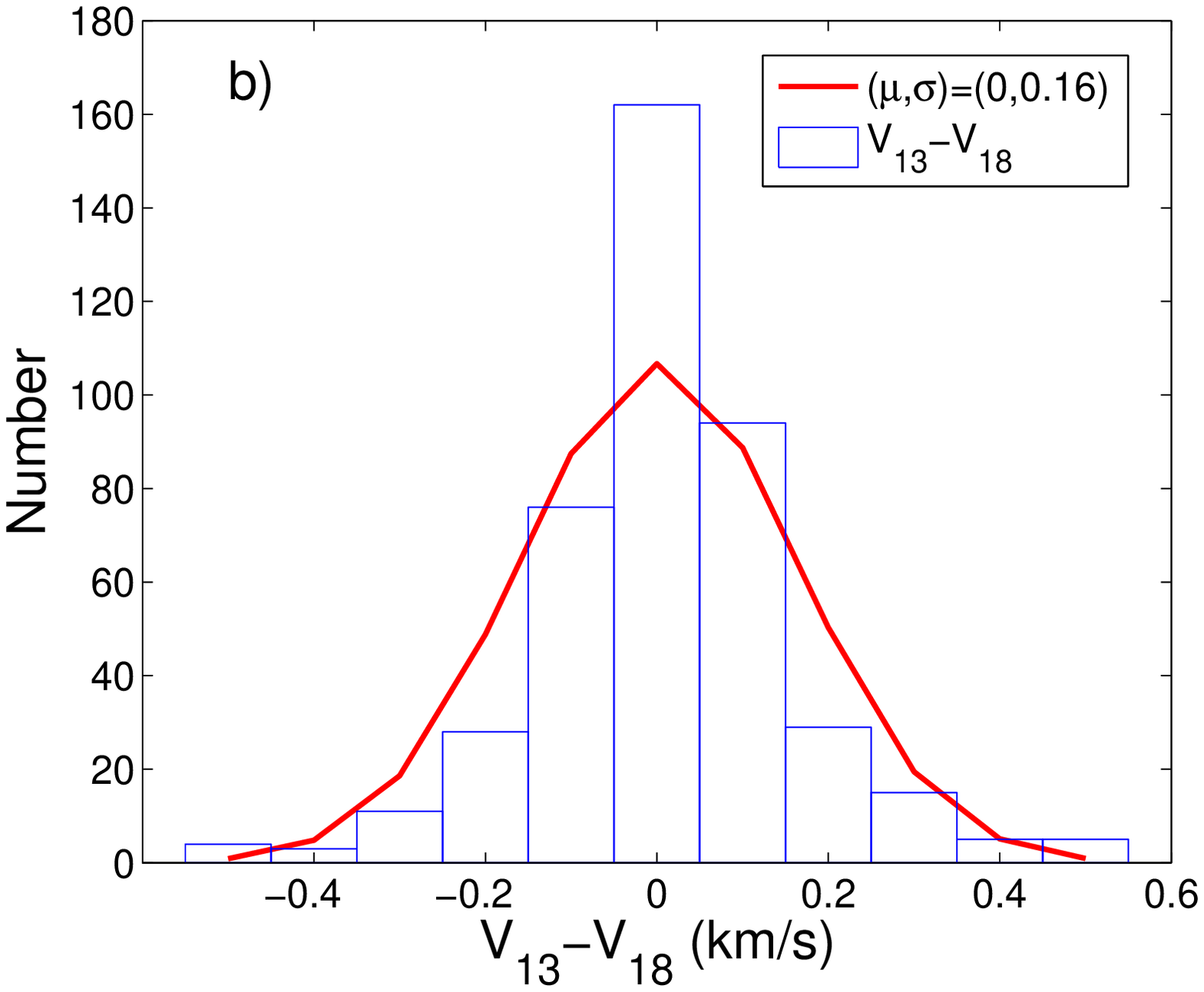}
\end{minipage}
\begin{minipage}[c]{0.5\textwidth}
  \centering
  \includegraphics[width=70mm,height=55mm,angle=0]{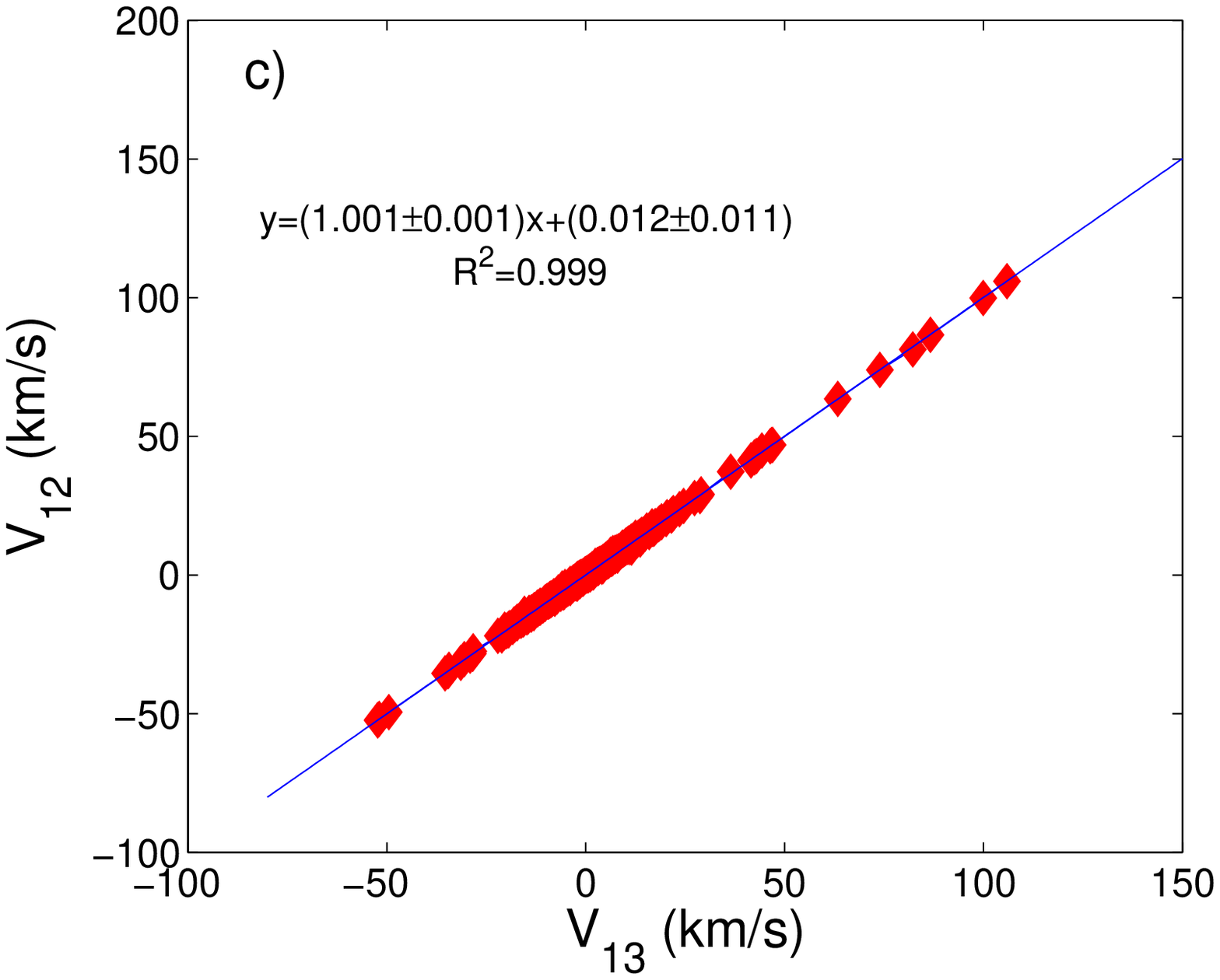}
\end{minipage}
\begin{minipage}[c]{0.5\textwidth}
  \centering
  \includegraphics[width=70mm,height=55mm,angle=0]{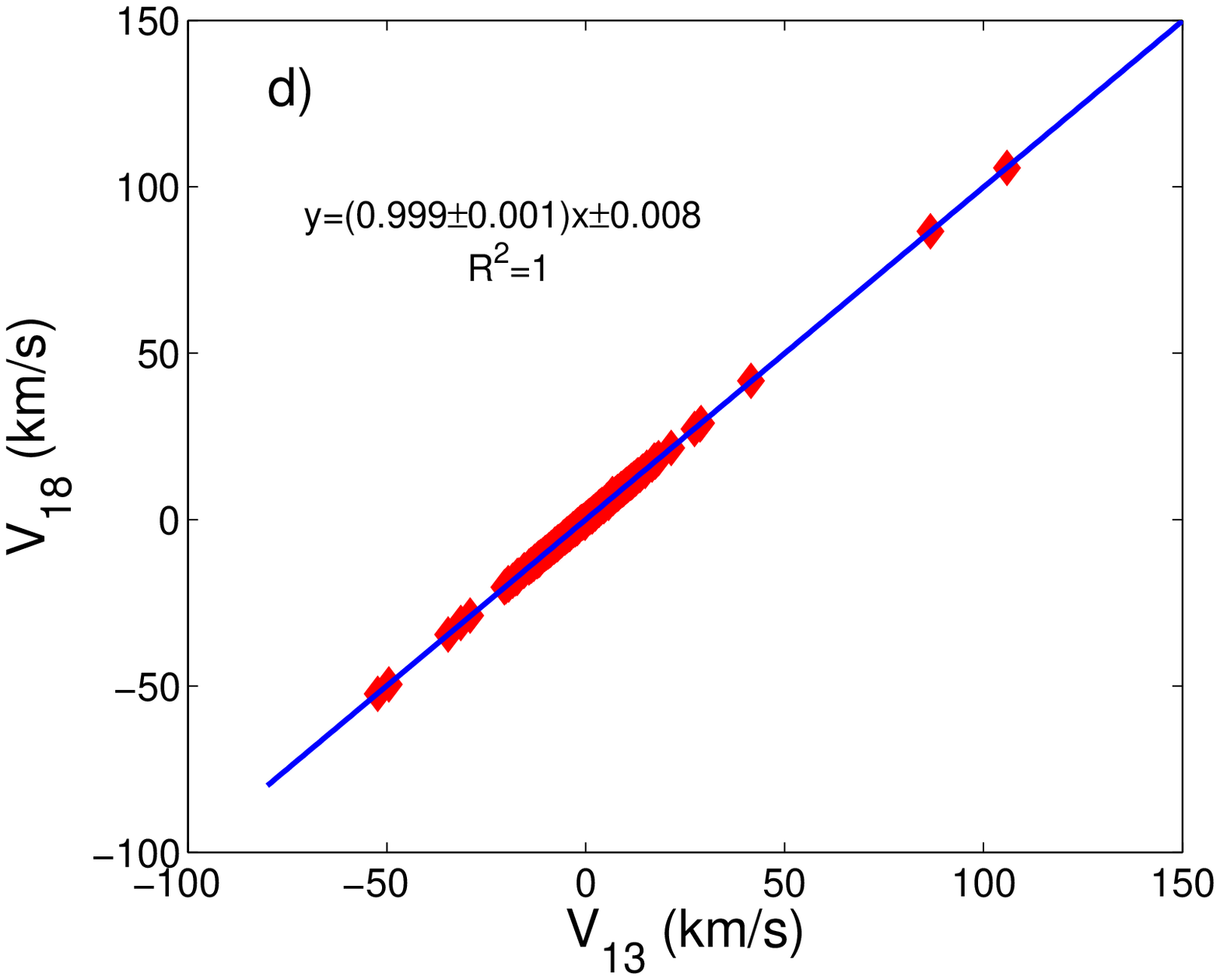}
\end{minipage}
\caption{Line center velocities of the J=1-0 lines of $^{12}$CO, $^{13}$CO and C$^{18}$O: a) and b) the histogram and the normal distribution fits for the difference between V12 and V13, V13 and V18 respectively. The mean $\mu$ and standard deviation $\sigma$ of the normal distributions are presented in the upper-right boxes; c) and d): plots for V12 vs. V13 and V13 :vs. V18 respectively.}
\end{figure}

\begin{figure}
\begin{minipage}[c]{0.33\textwidth}
  \centering
  \includegraphics[width=55mm,height=50mm,angle=0]{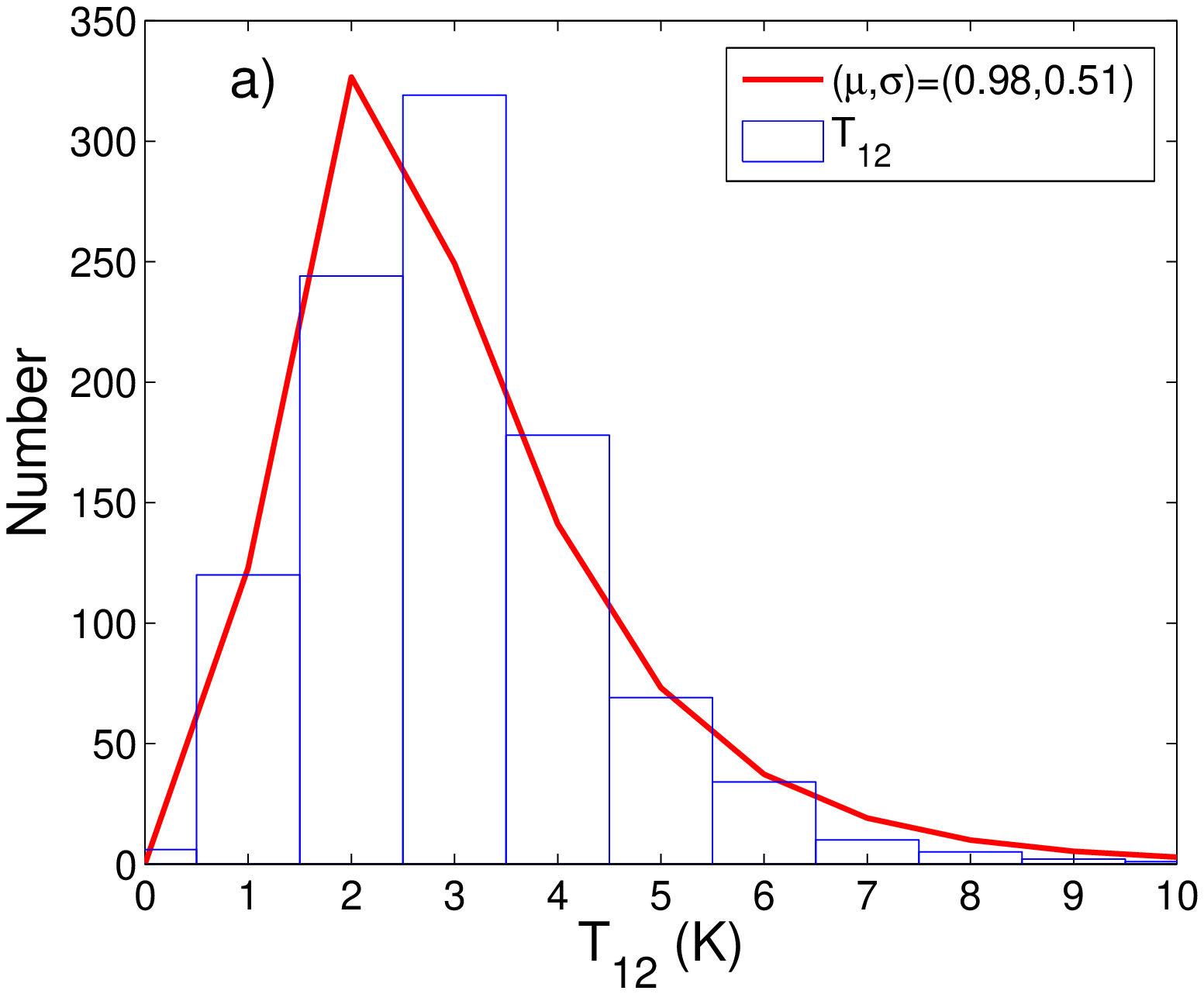}
\end{minipage}
\begin{minipage}[c]{0.33\textwidth}
  \centering
  \includegraphics[width=55mm,height=50mm,angle=0]{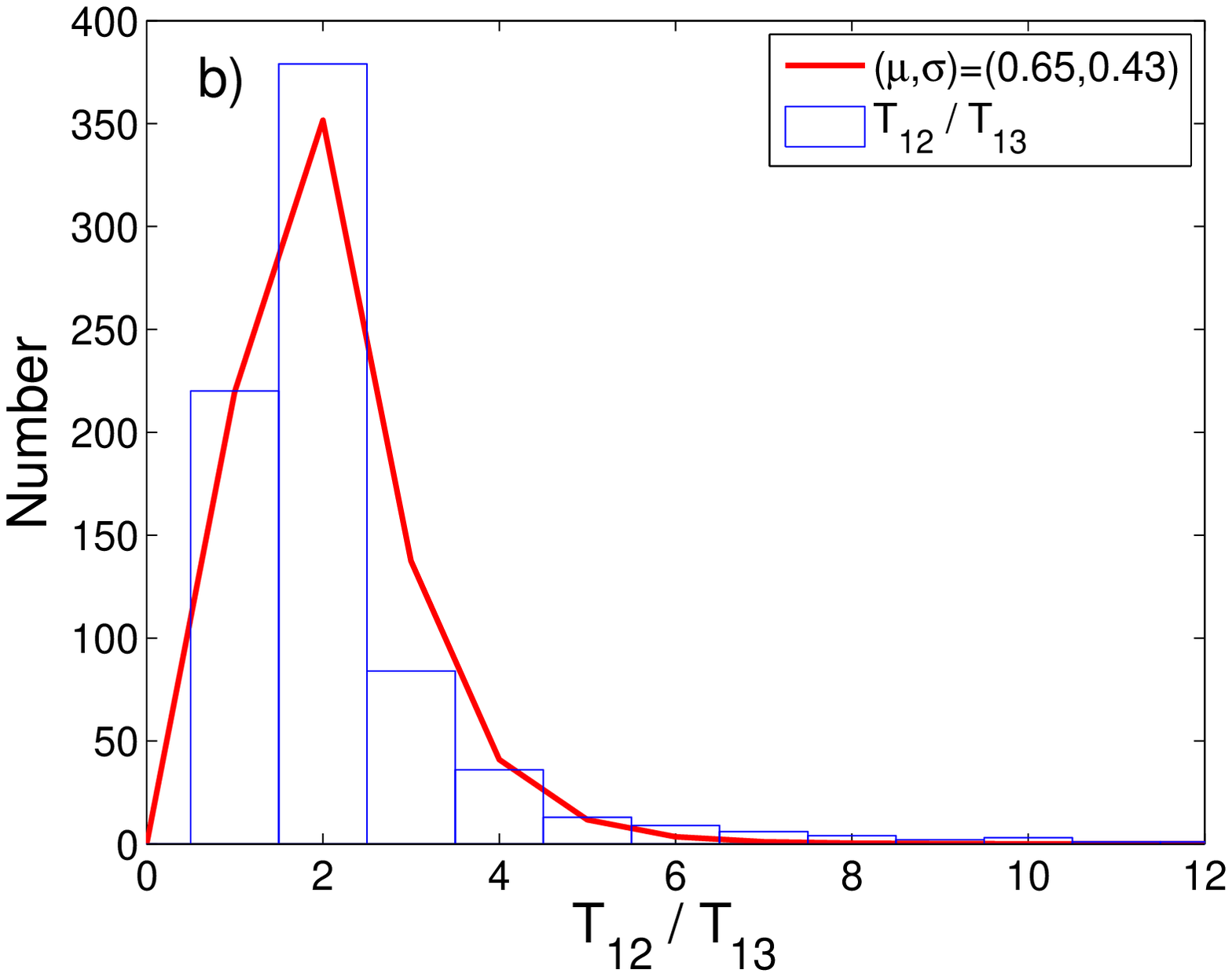}
\end{minipage}
\begin{minipage}[c]{0.33\textwidth}
  \centering
  \includegraphics[width=55mm,height=50mm,angle=0]{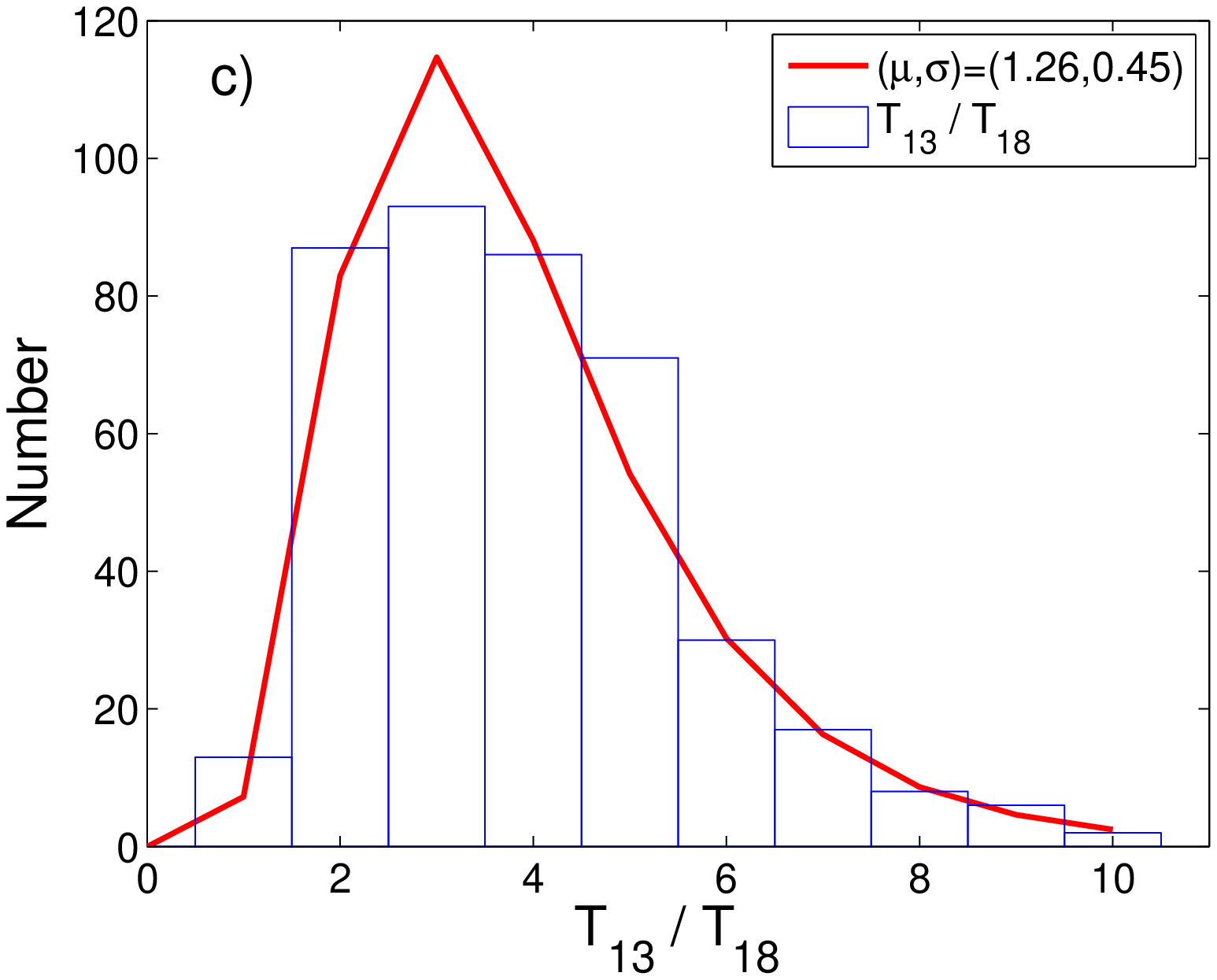}
\end{minipage}
\caption{The frequency distributions of antenna temperature T12, the ratio of T12/T13 and T13/T18.}
\end{figure}

\begin{figure}
\begin{minipage}[c]{0.5\textwidth}
  \centering
  \includegraphics[width=70mm,height=55mm,angle=0]{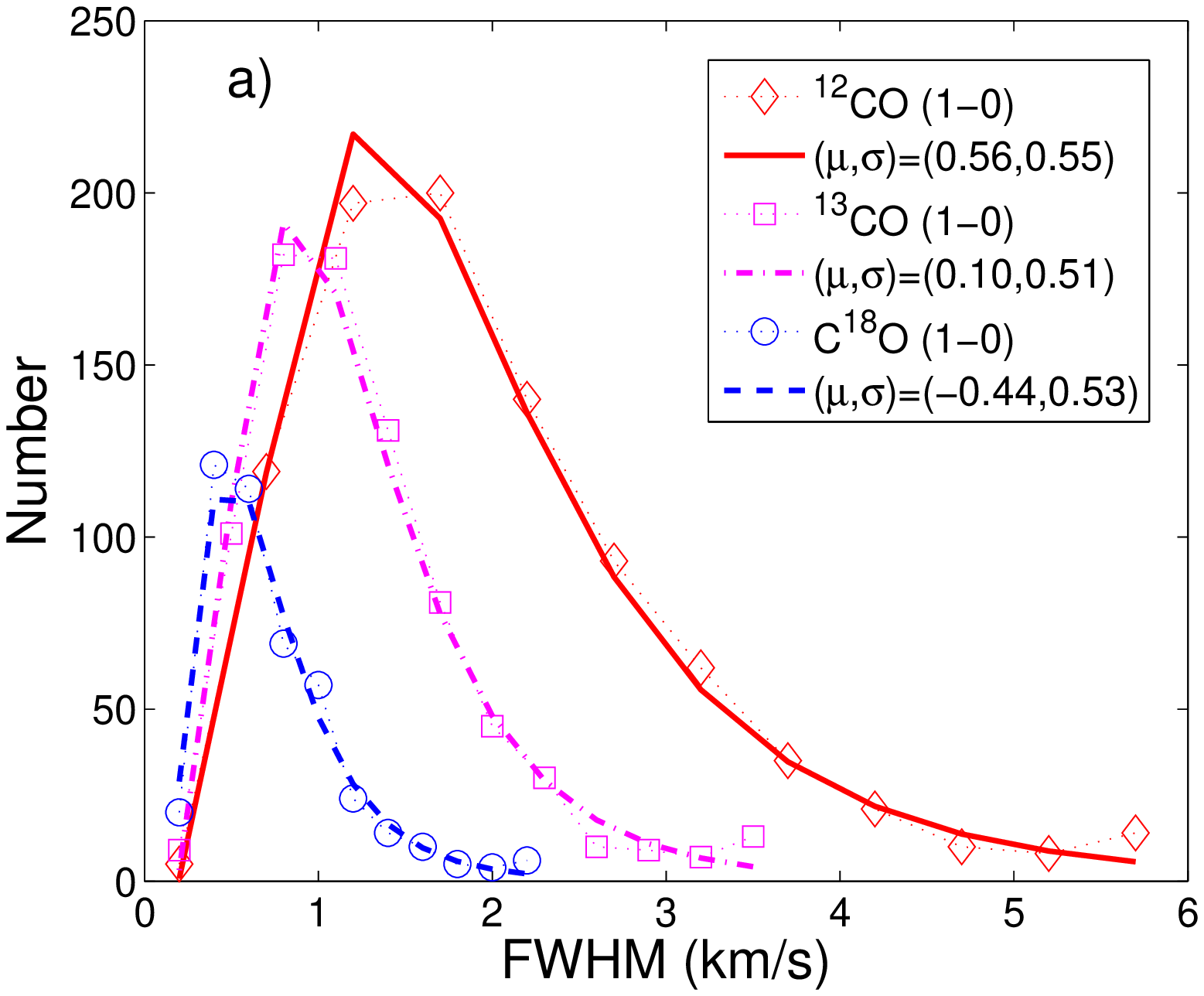}
\end{minipage}
\begin{minipage}[c]{0.5\textwidth}
  \centering
  \includegraphics[width=70mm,height=55mm,angle=0]{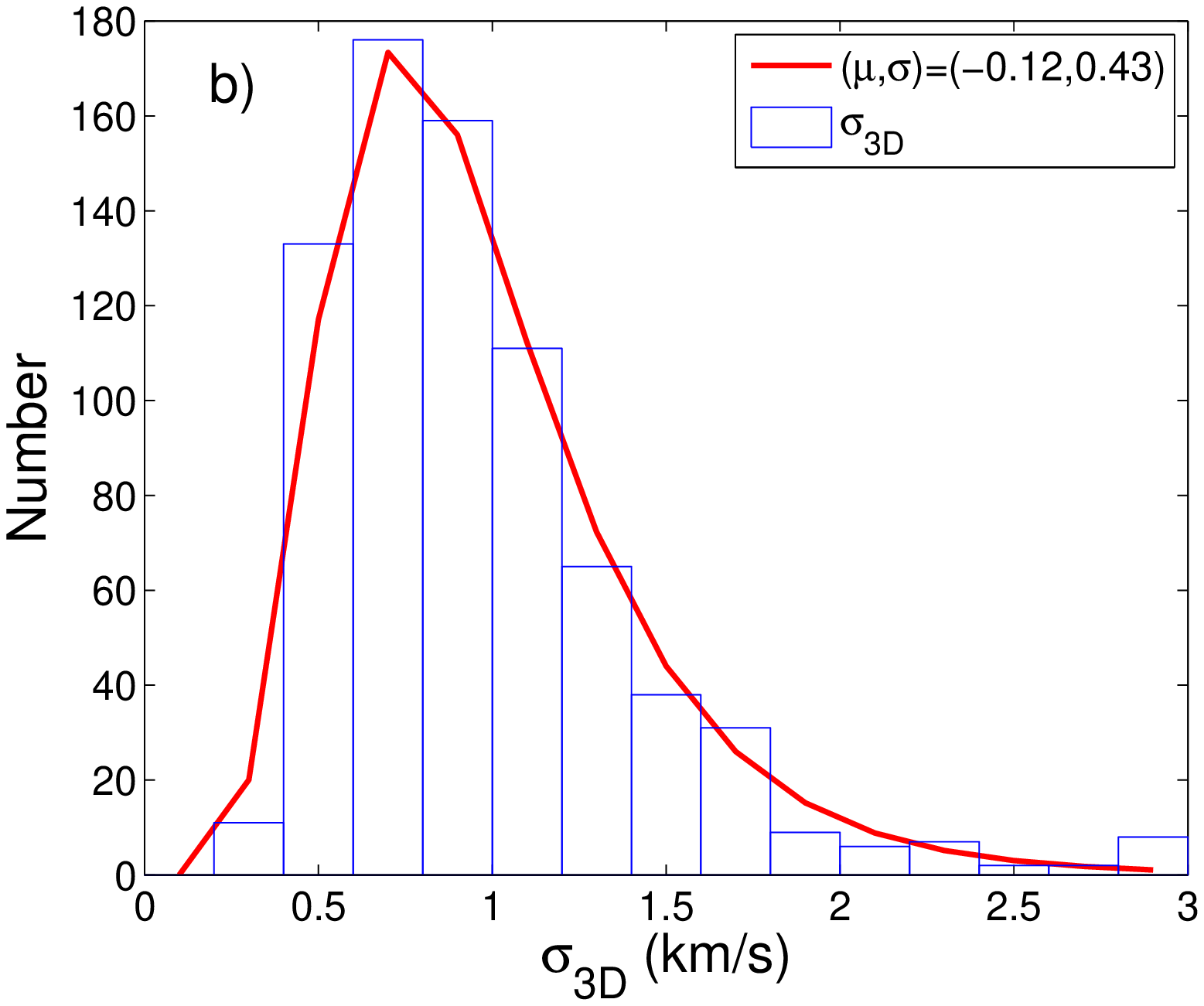}
\end{minipage}
\begin{minipage}[c]{0.5\textwidth}
  \centering
  \includegraphics[width=70mm,height=55mm,angle=0]{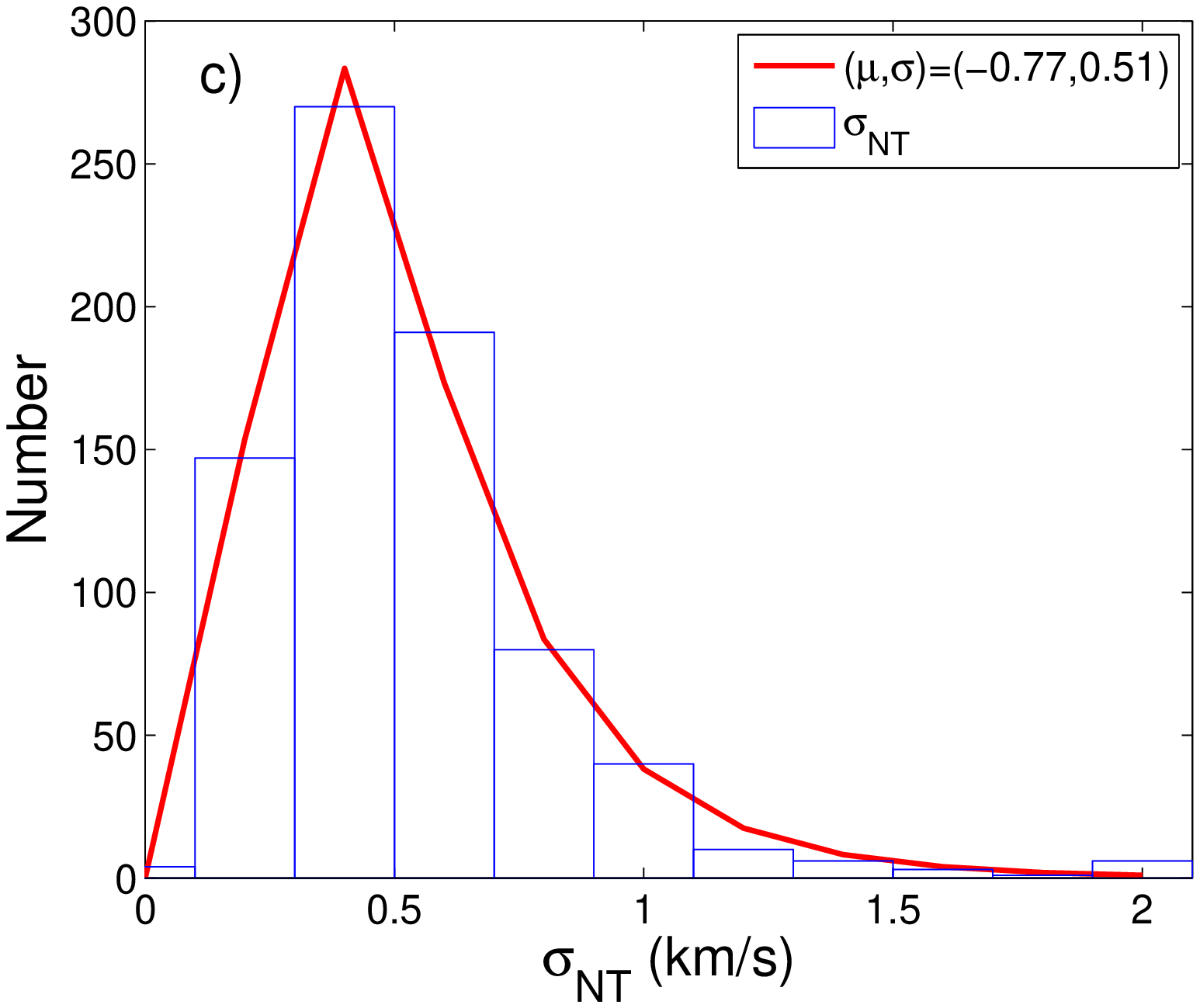}
\end{minipage}
\begin{minipage}[c]{0.5\textwidth}
  \centering
  \includegraphics[width=70mm,height=55mm,angle=0]{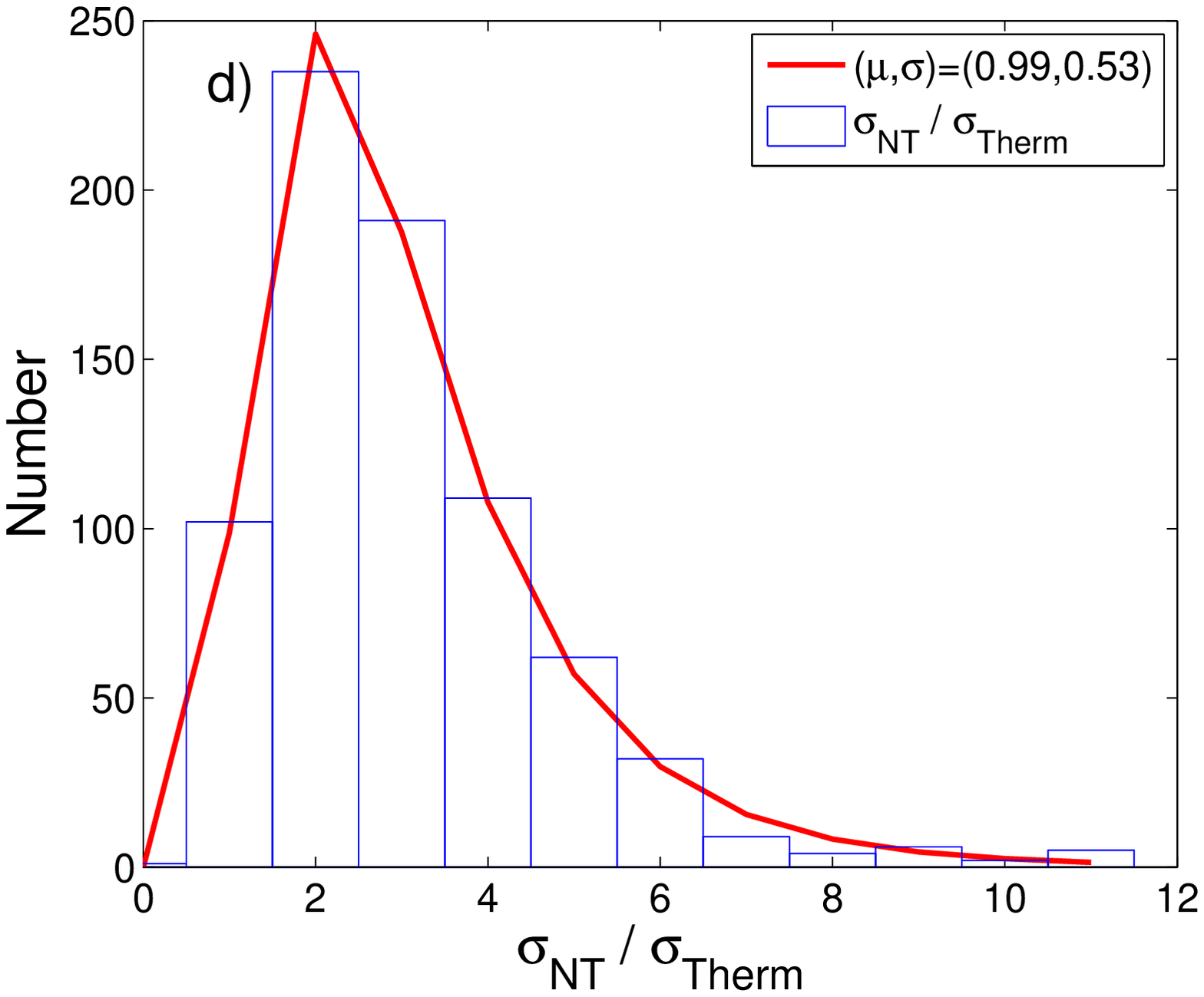}
\end{minipage}
\caption{The frequency distributions of line FWHM and velocity dispersion: a). Distributions and lognormal fitting of FWHM of the CO, $^{13}$CO and C$^{18}$O lines; b) Histogram and the lognormal fitting of the $^{13}$CO line 3D velocity dispersion; c) Velocity dispersion of non-thermal motion 13CO lines;  d) Histogram and lognormal fitting of ratio of the $^{13}$CO line non-thermal and thermal motion. The mean $\mu$ and standard deviation $\sigma$ of the lognormal distribution fits are presented in the upper-right boxes in each panel.}
\end{figure}

\begin{figure}
\includegraphics[angle=0,scale=.50]{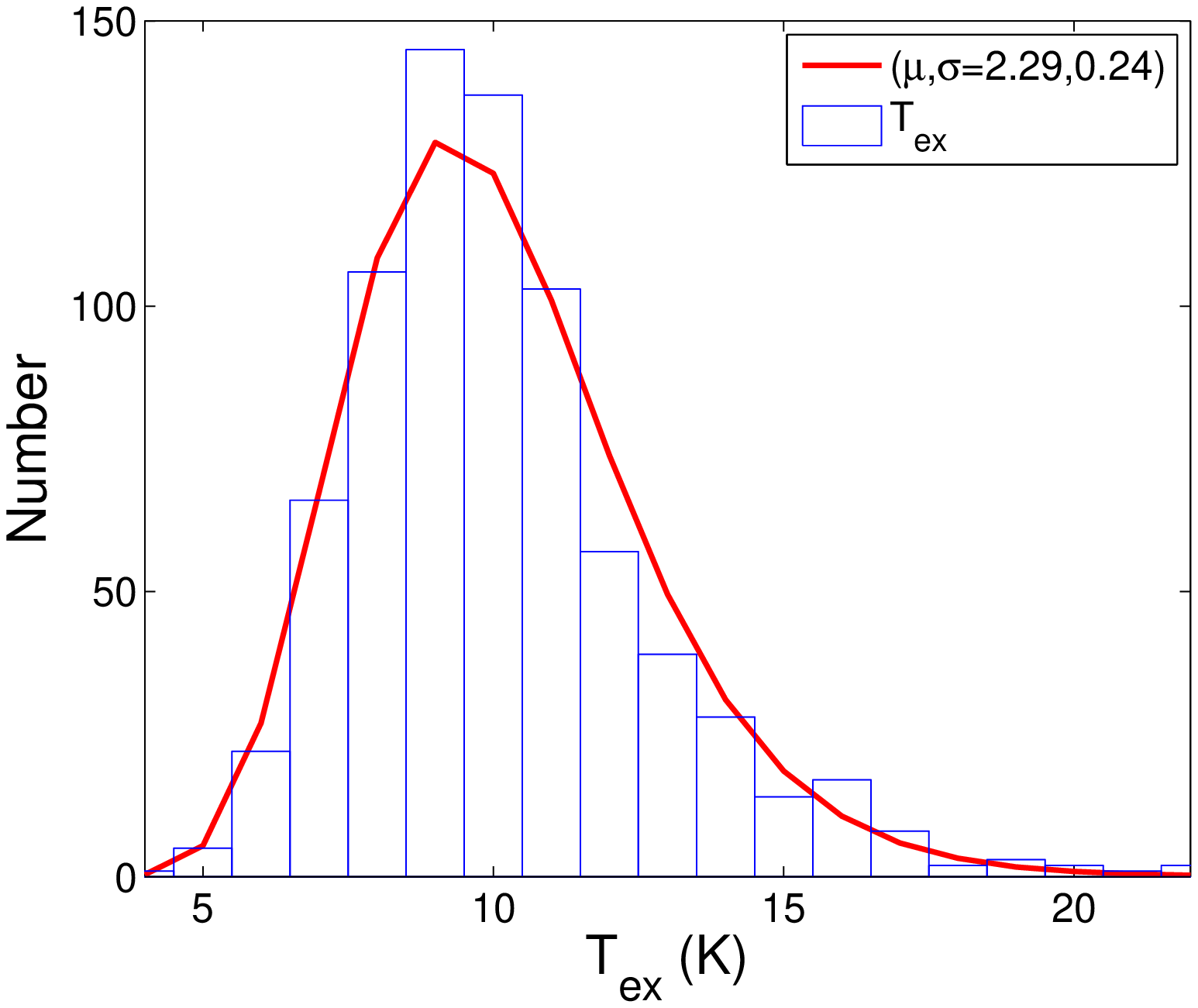}
\caption{The histogram and the lognormal-PDF fitting of excitation temperatures }
\end{figure}

\begin{figure}
\begin{minipage}[c]{0.33\textwidth}
  \centering
  \includegraphics[width=55mm,height=50mm,angle=0]{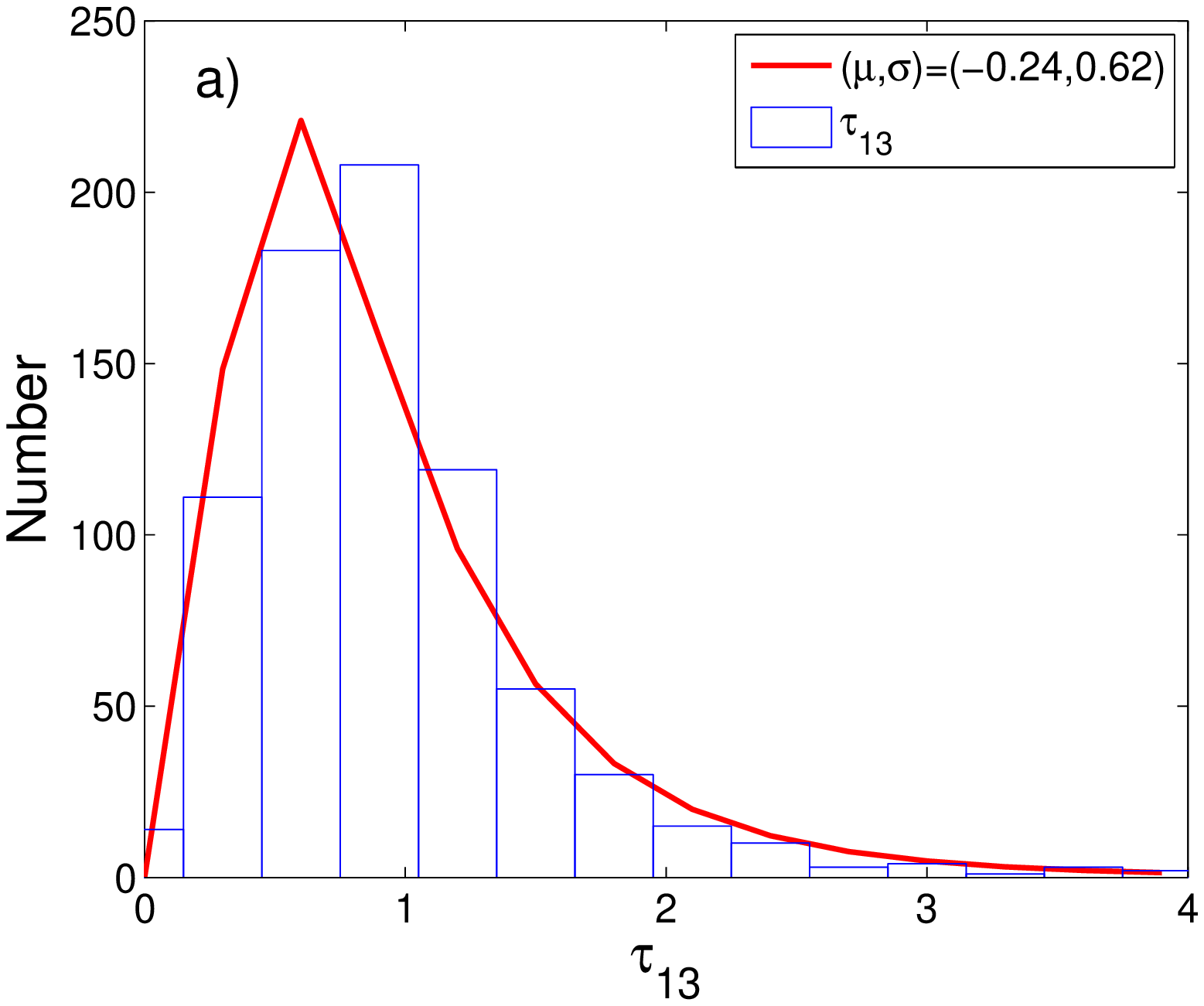}
\end{minipage}
\begin{minipage}[c]{0.33\textwidth}
  \centering
  \includegraphics[width=55mm,height=50mm,angle=0]{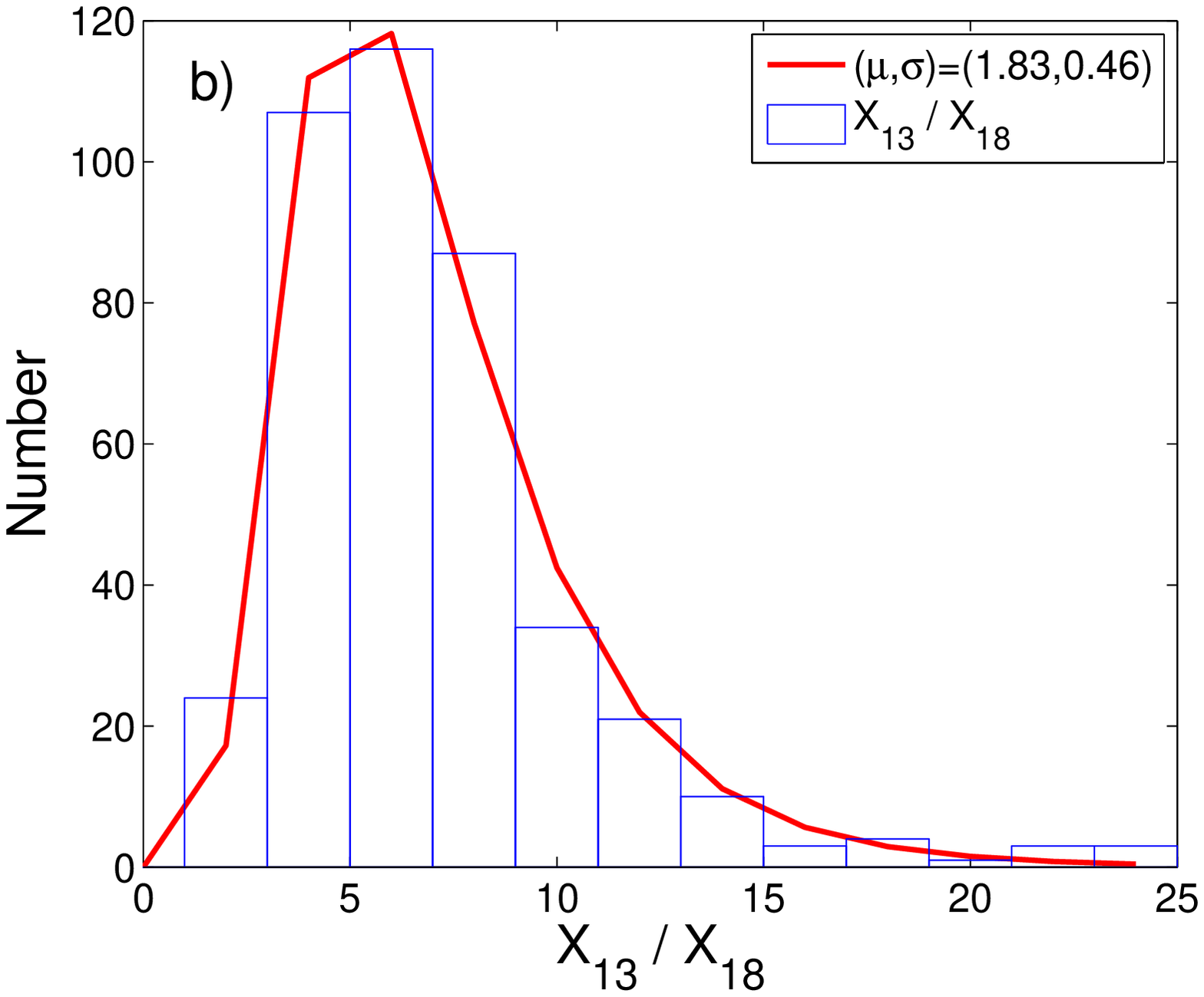}
\end{minipage}
\begin{minipage}[c]{0.33\textwidth}
  \centering
  \includegraphics[width=55mm,height=50mm,angle=0]{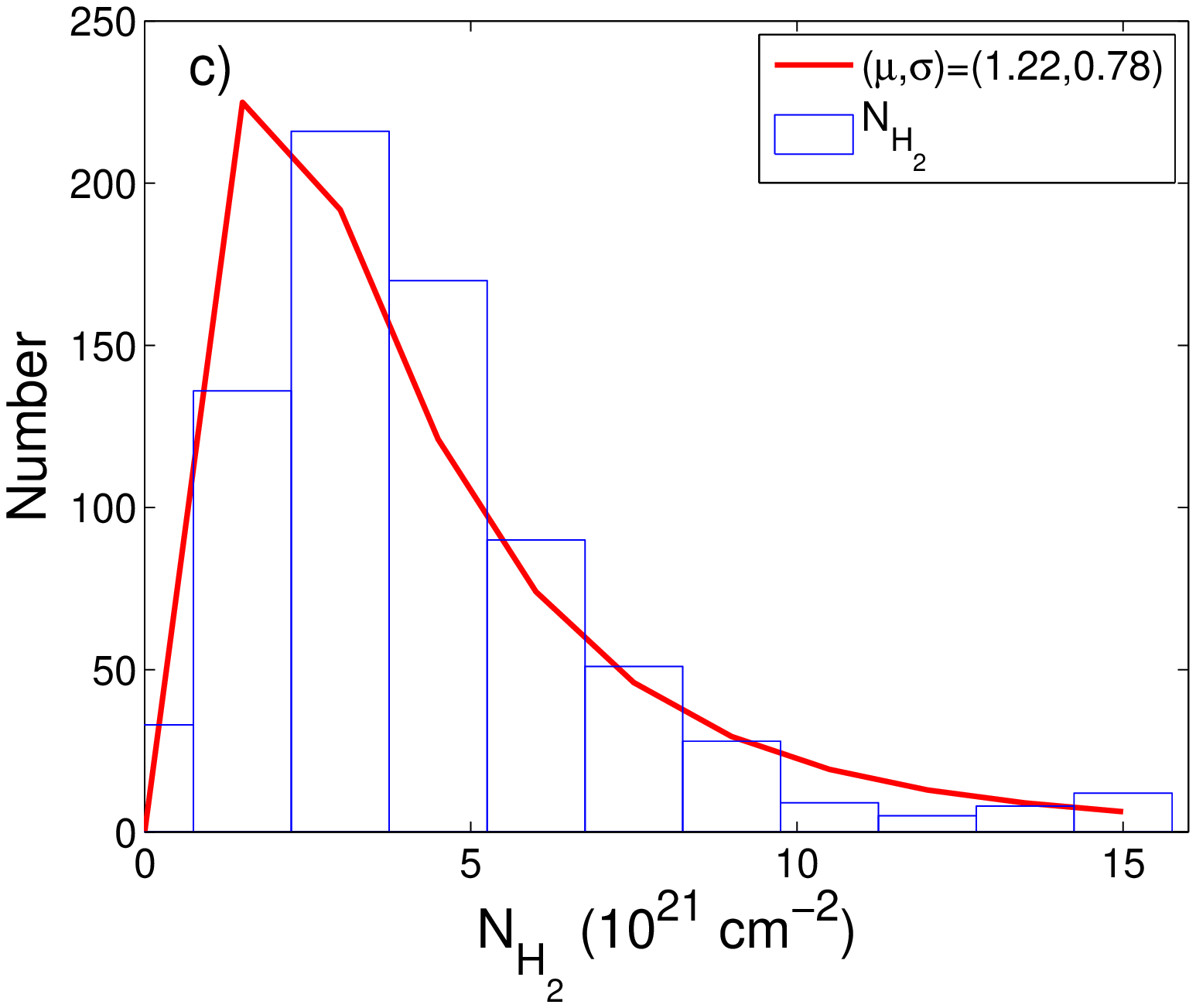}
\end{minipage}
\caption{Histograms of $\tau_{13}$, ratio [X\_13/X\_18] and the column density N$_{H_{2}}$ and their lognormal-PDF fitting}
\end{figure}

\clearpage

\begin{figure}
\begin{minipage}[c]{0.33\textwidth}
  \centering
  \includegraphics[width=40mm,height=50mm,angle=-90]{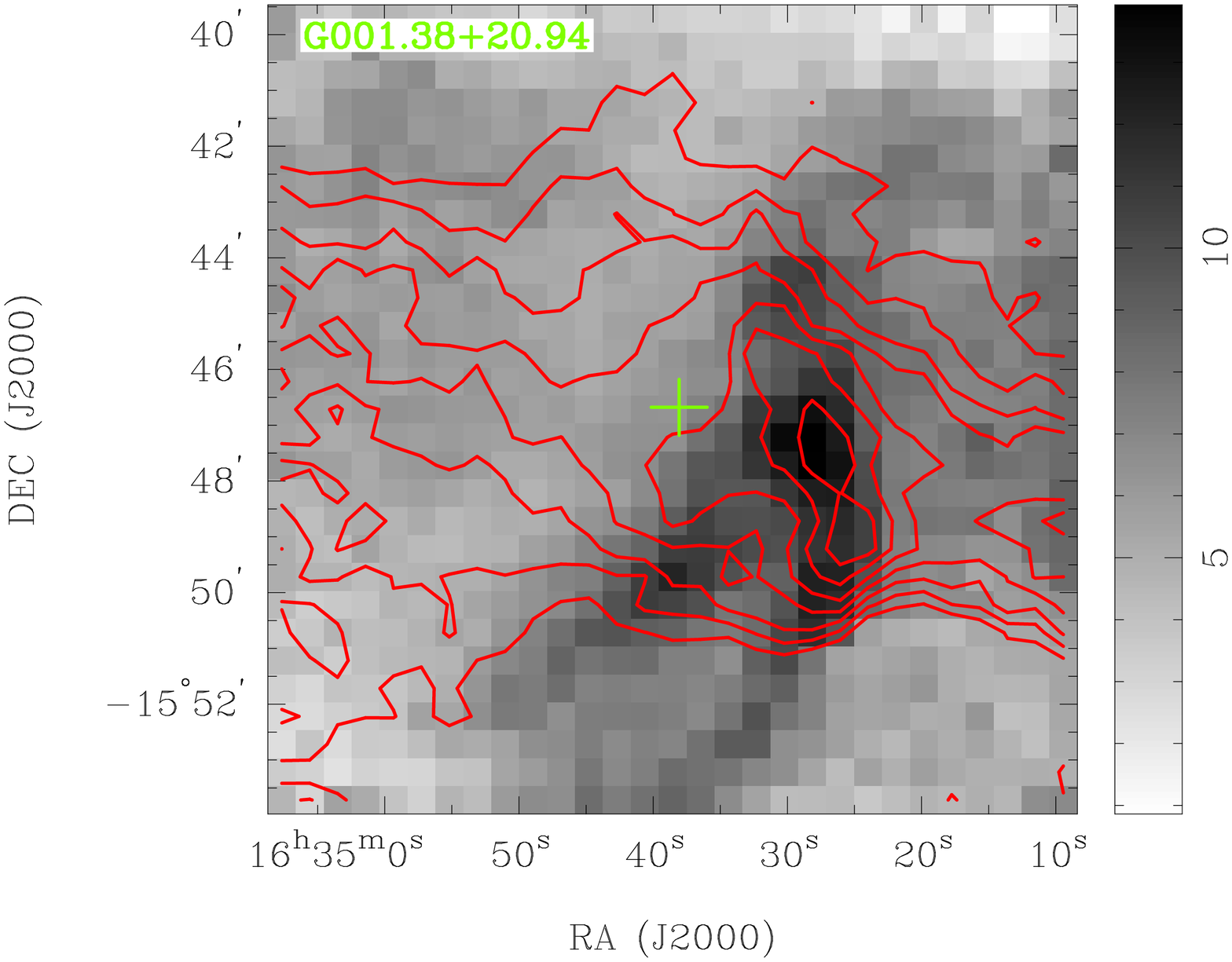}
\end{minipage}
\begin{minipage}[c]{0.33\textwidth}
  \centering
  \includegraphics[width=40mm,height=50mm,angle=-90]{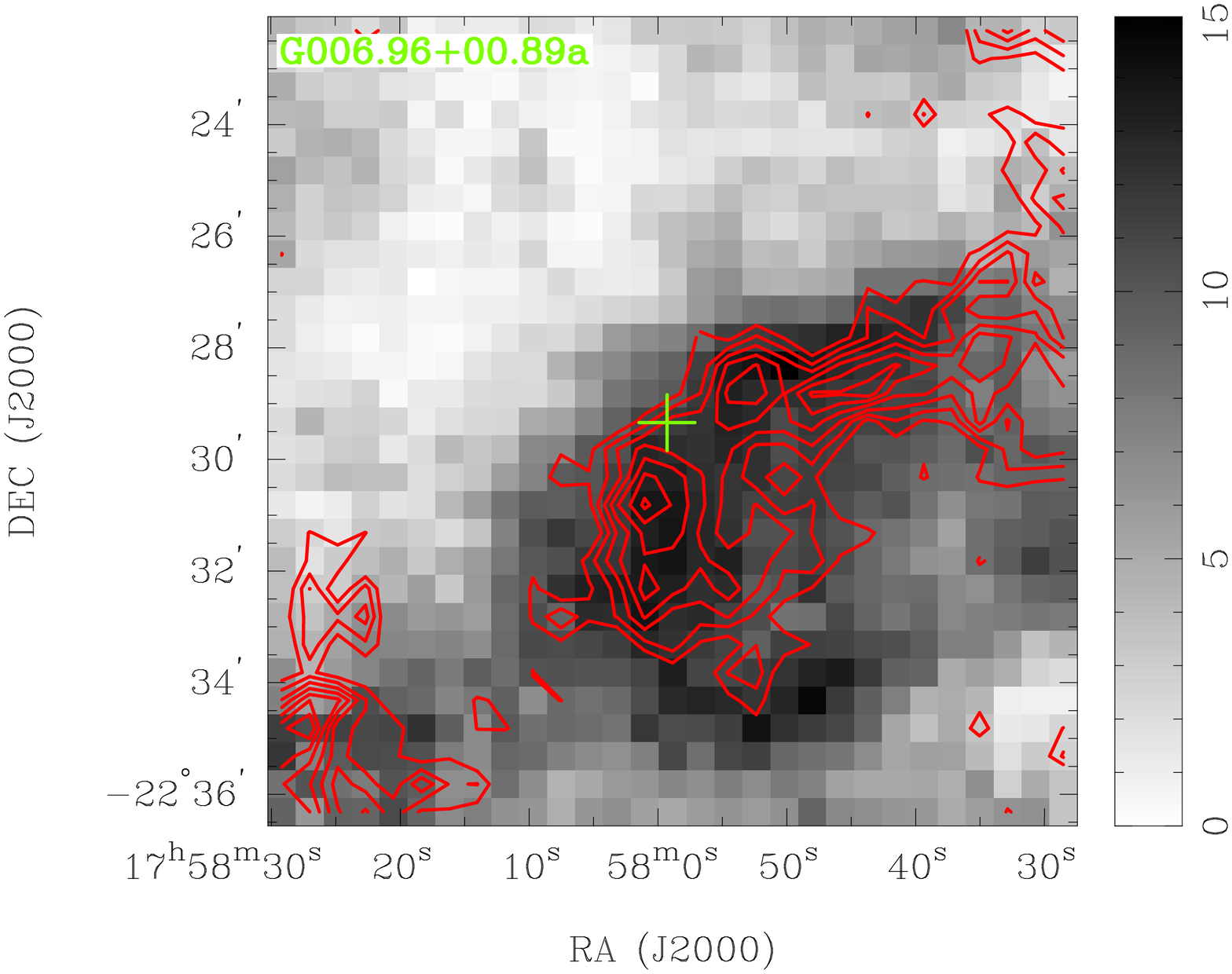}
\end{minipage}
\begin{minipage}[c]{0.33\textwidth}
  \centering
  \includegraphics[width=40mm,height=50mm,angle=-90]{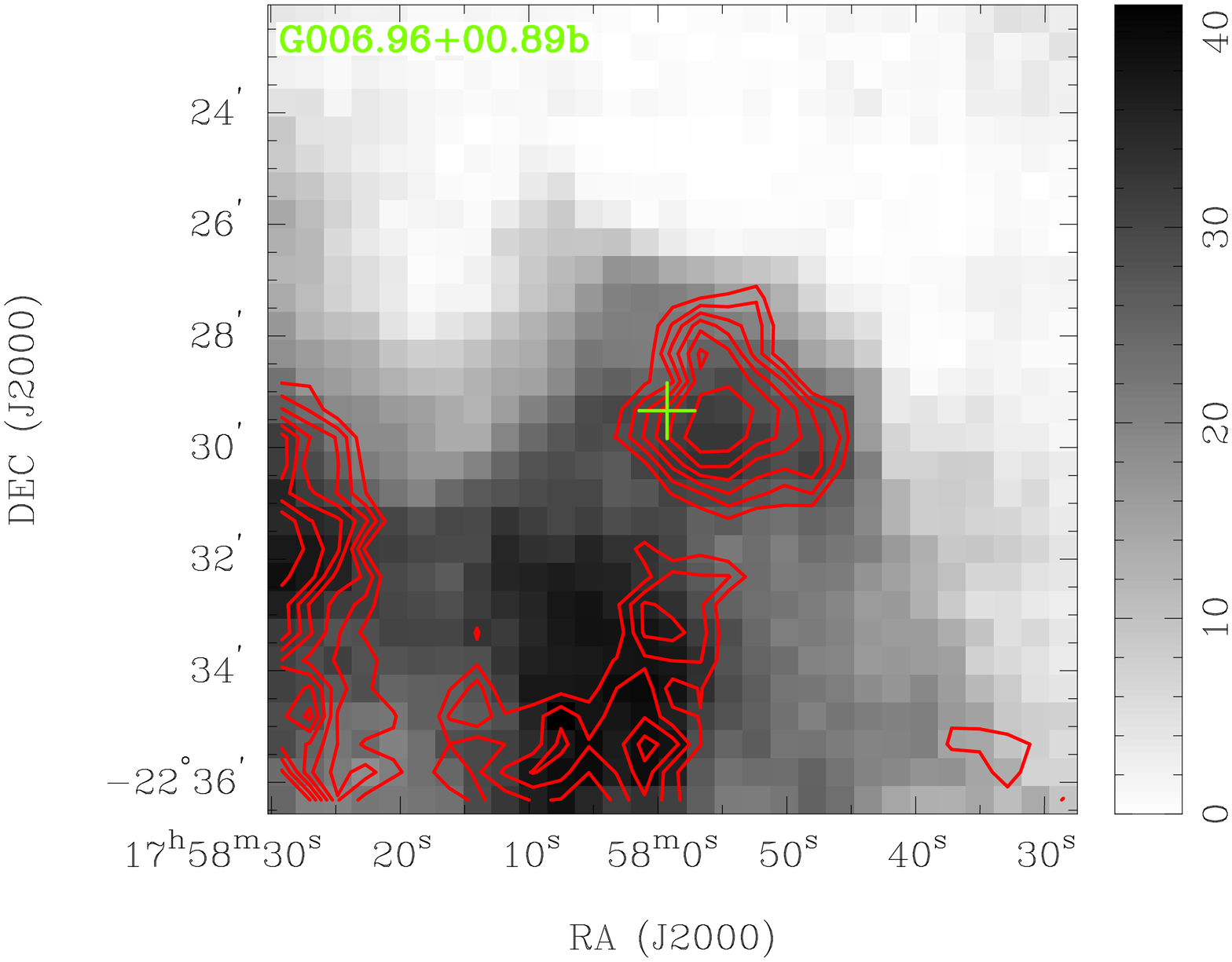}
\end{minipage}
\begin{minipage}[c]{0.33\textwidth}
  \centering
  \includegraphics[width=40mm,height=50mm,angle=-90]{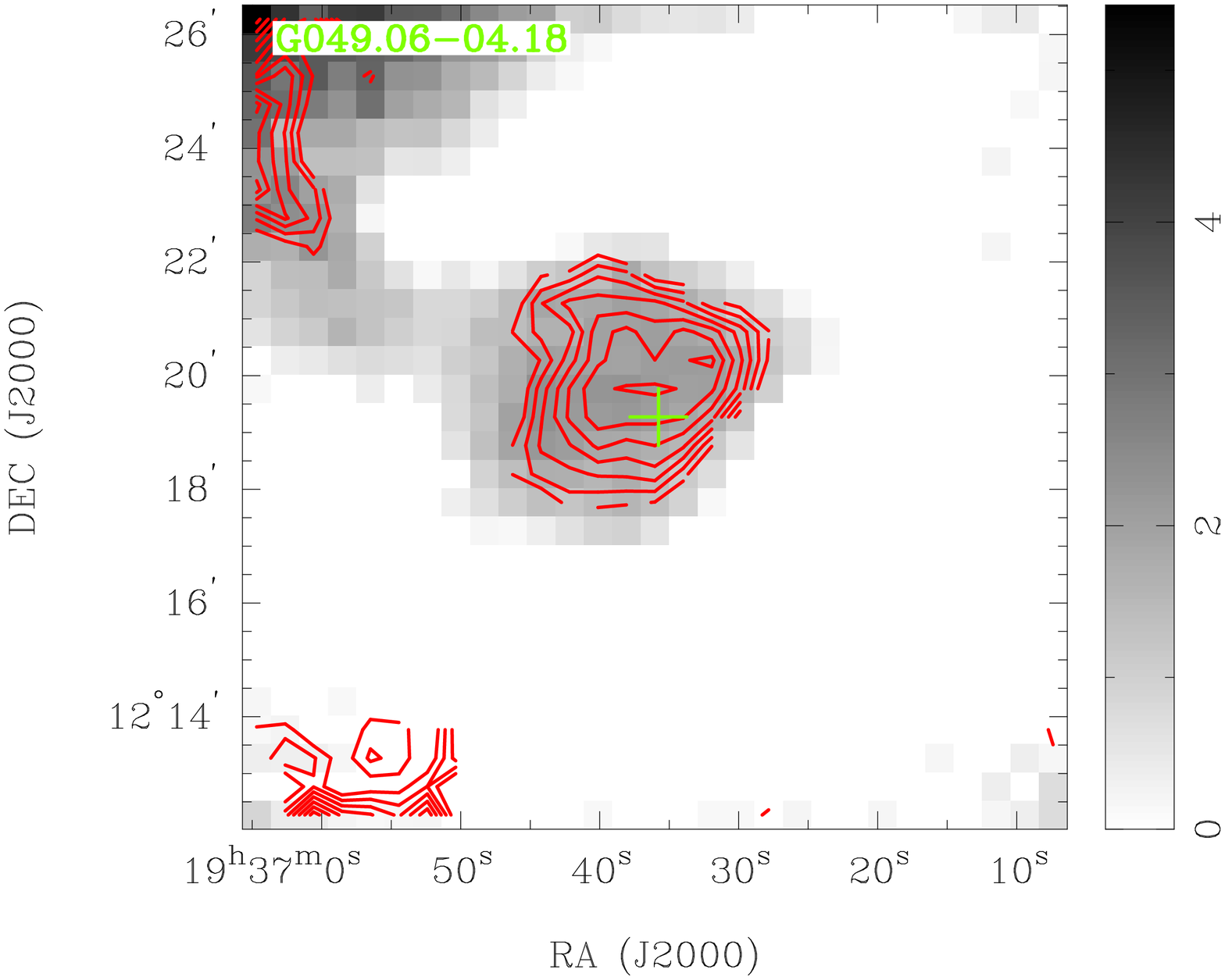}
\end{minipage}
\begin{minipage}[c]{0.33\textwidth}
  \centering
  \includegraphics[width=40mm,height=50mm,angle=-90]{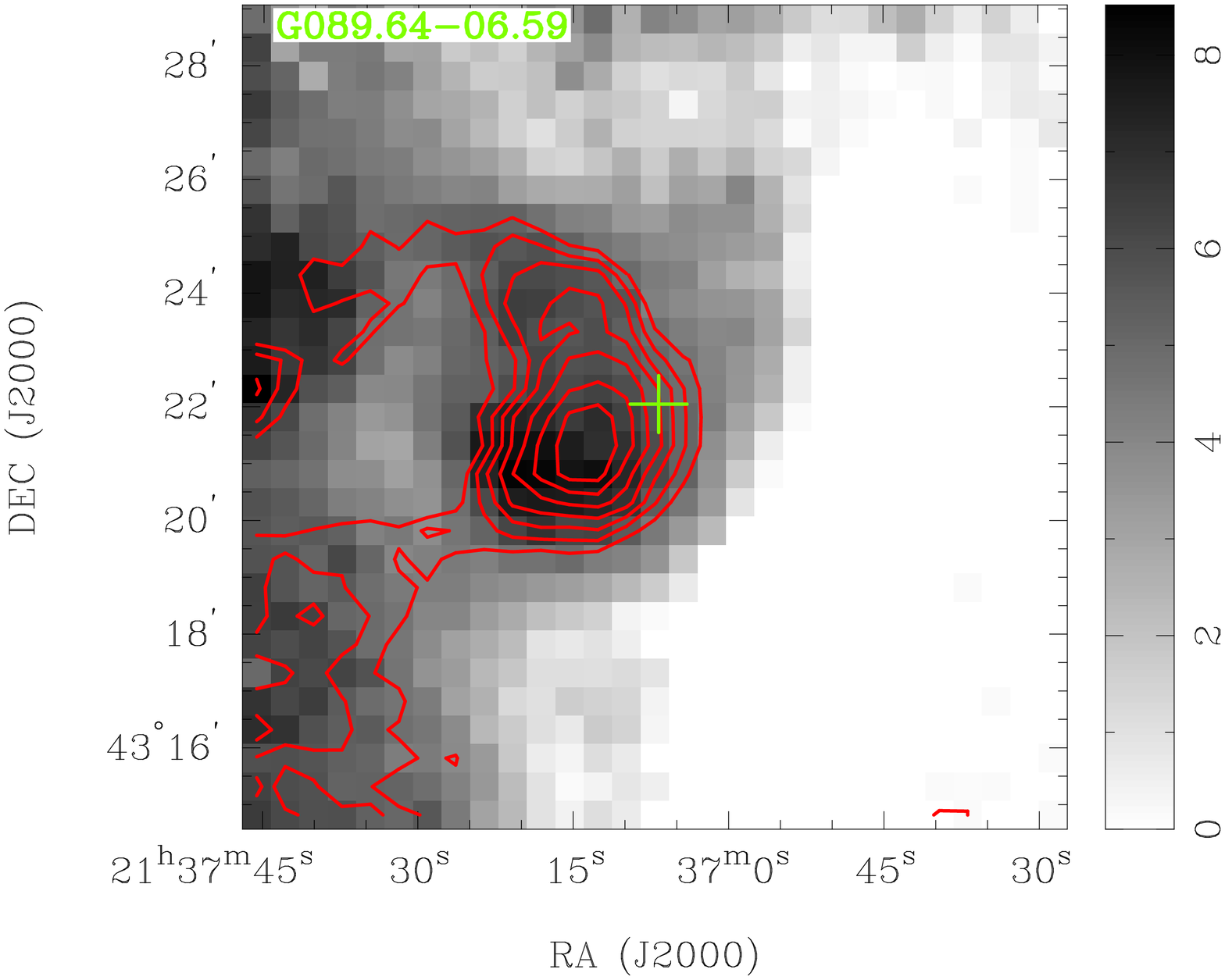}
\end{minipage}
\begin{minipage}[c]{0.33\textwidth}
  \centering
  \includegraphics[width=40mm,height=50mm,angle=-90]{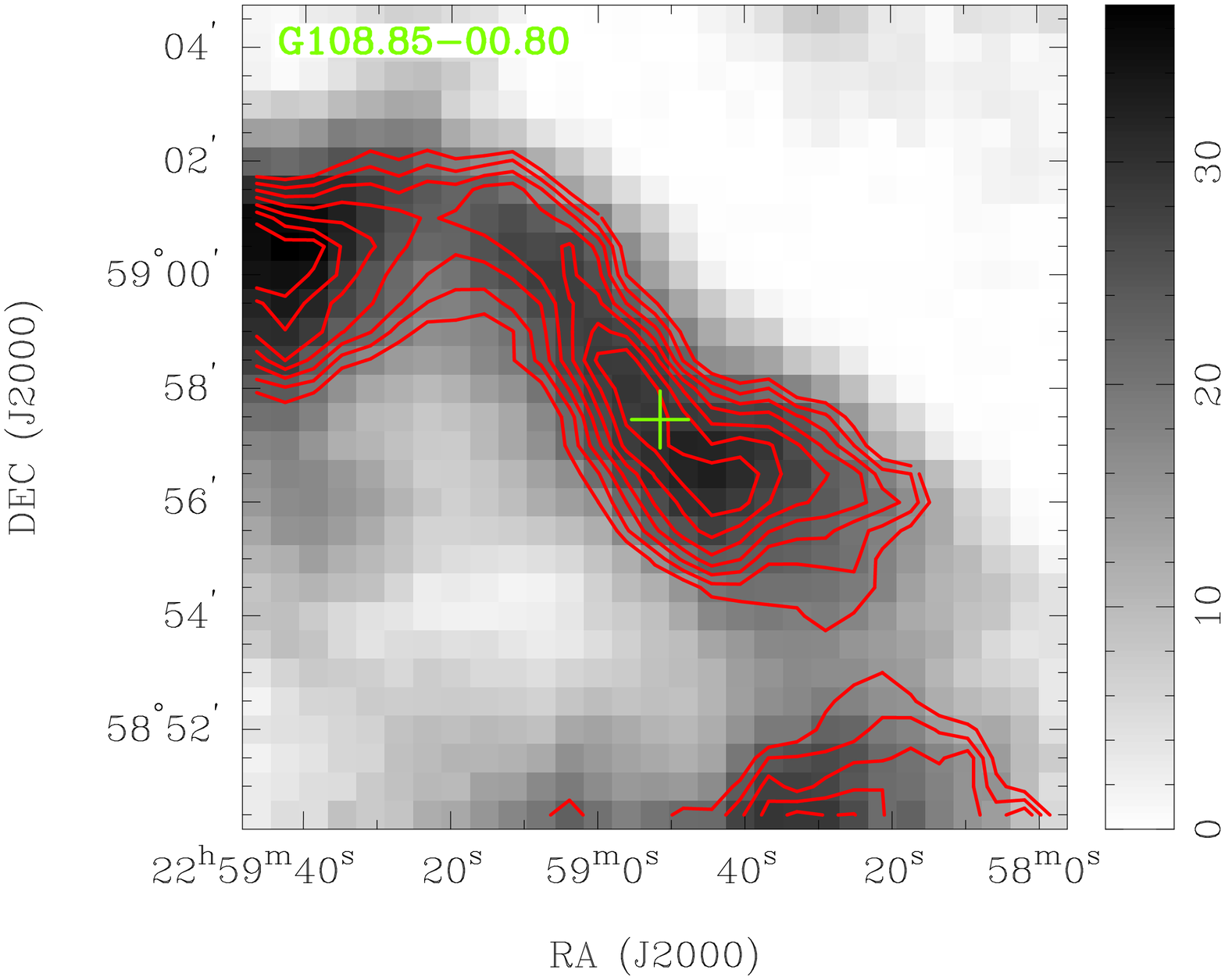}
\end{minipage}
\begin{minipage}[c]{0.33\textwidth}
  \centering
  \includegraphics[width=40mm,height=50mm,angle=-90]{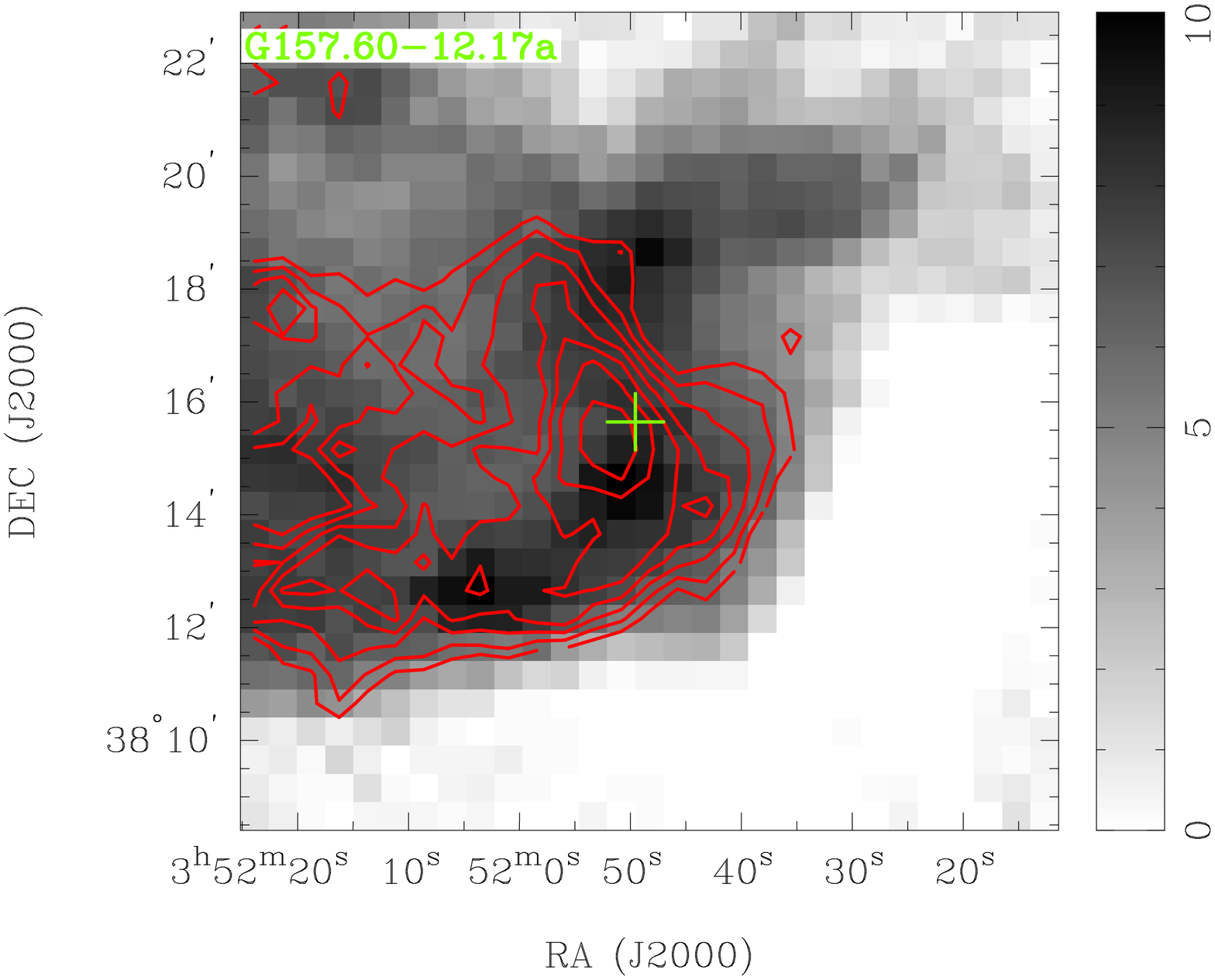}
\end{minipage}
\begin{minipage}[c]{0.33\textwidth}
  \centering
  \includegraphics[width=40mm,height=50mm,angle=-90]{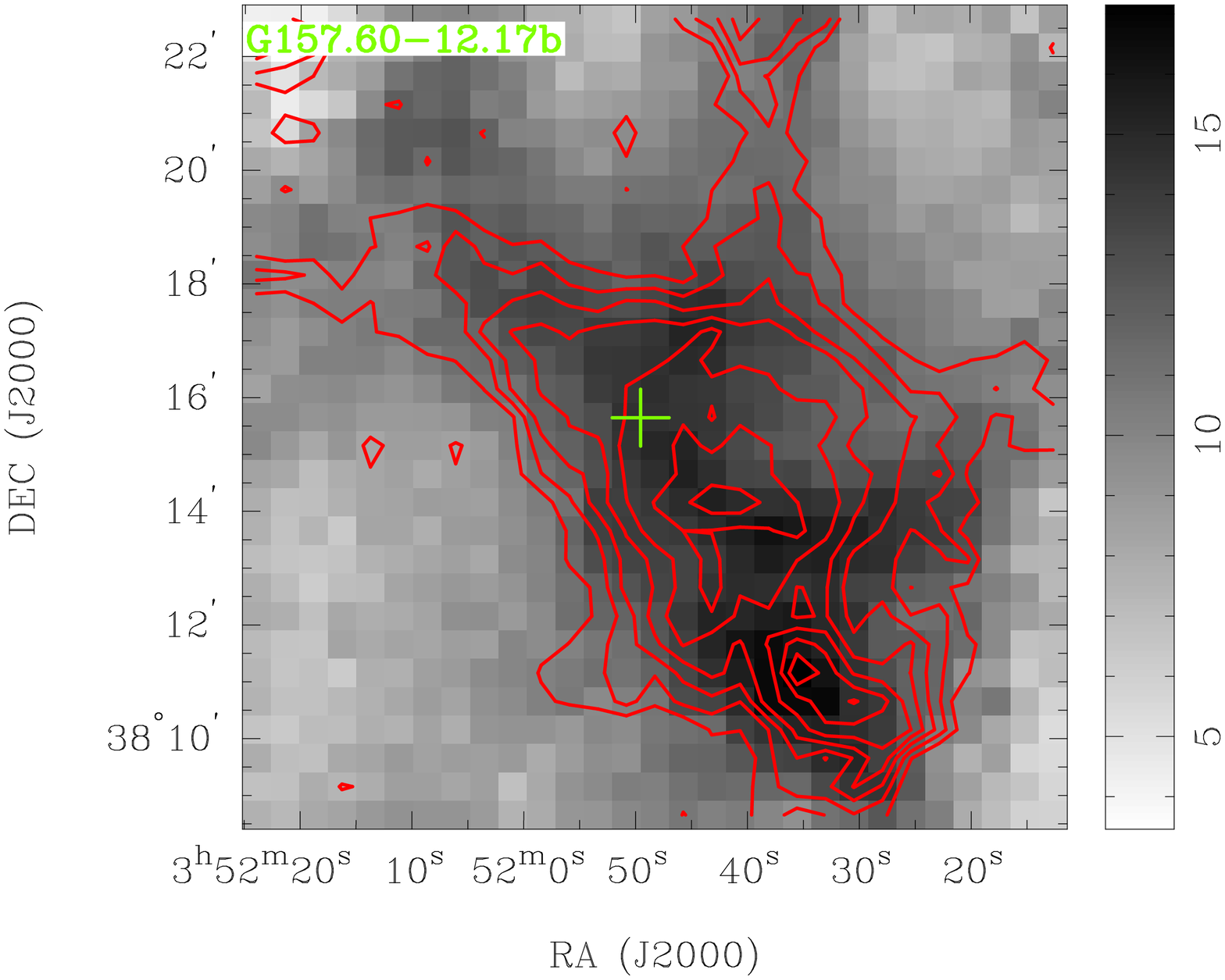}
\end{minipage}
\begin{minipage}[c]{0.33\textwidth}
  \centering
  \includegraphics[width=40mm,height=50mm,angle=-90]{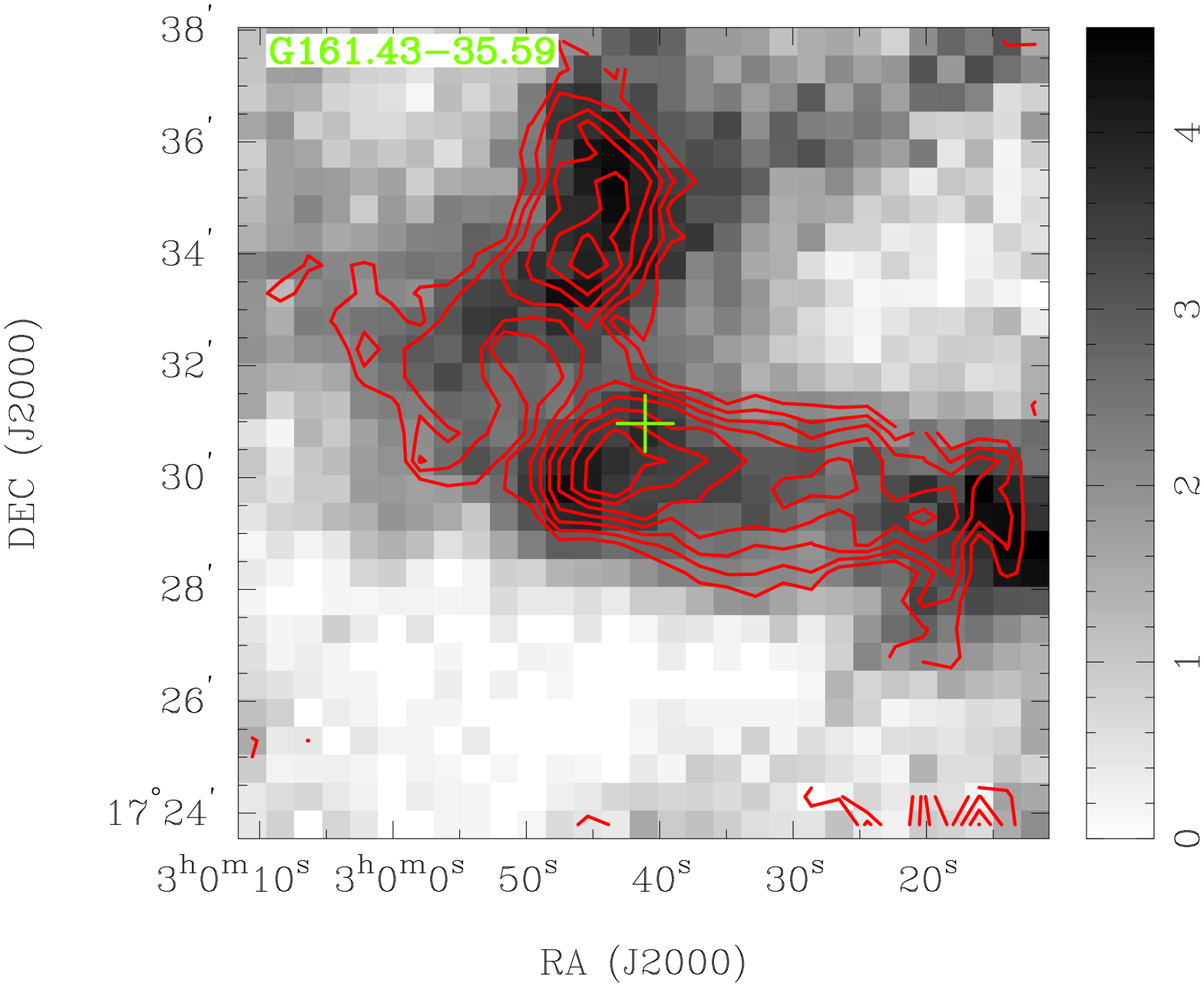}
\end{minipage}
\begin{minipage}[c]{0.33\textwidth}
  \centering
  \includegraphics[width=40mm,height=50mm,angle=-90]{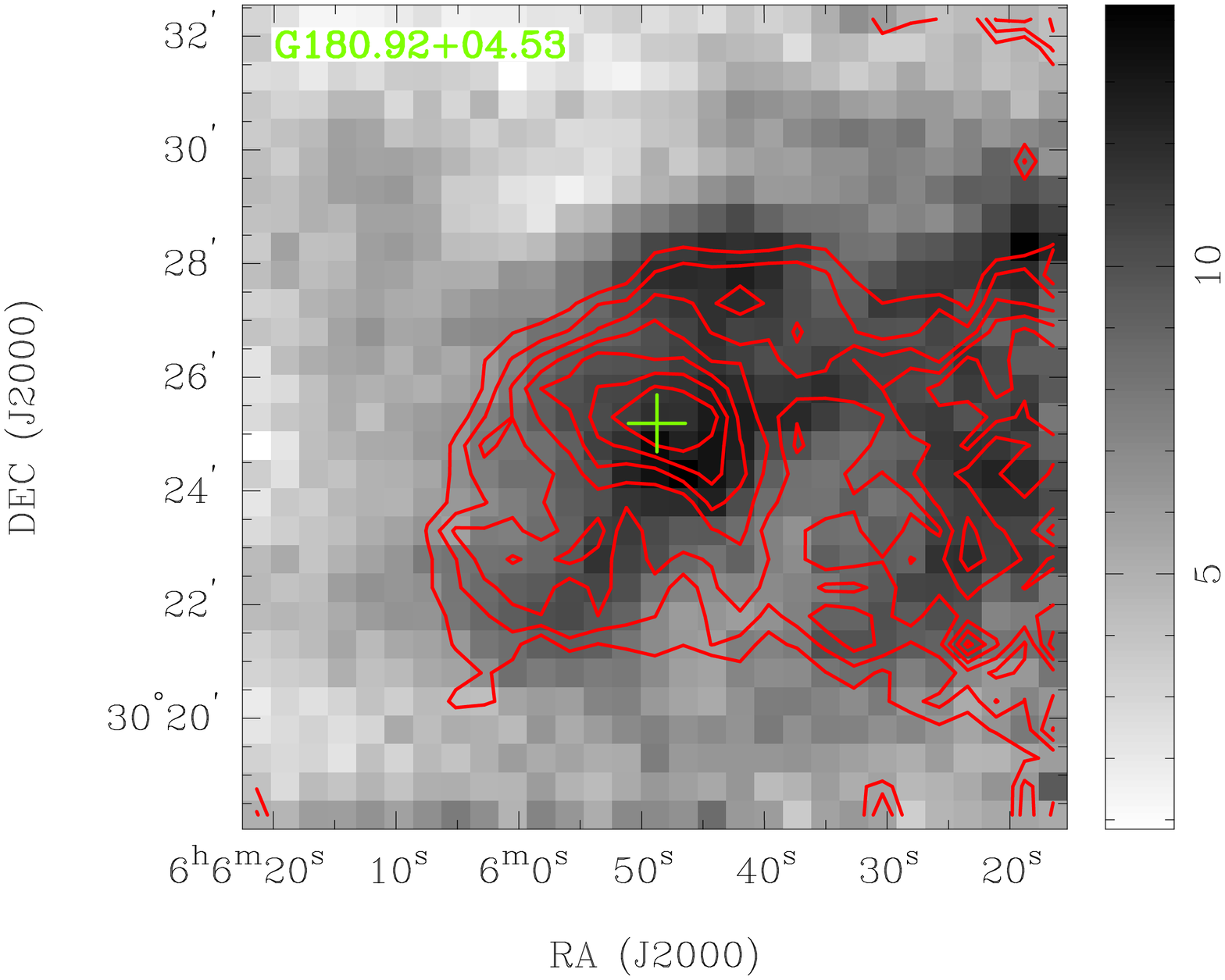}
\end{minipage}
\begin{minipage}[c]{0.33\textwidth}
  \centering
  \includegraphics[width=40mm,height=50mm,angle=-90]{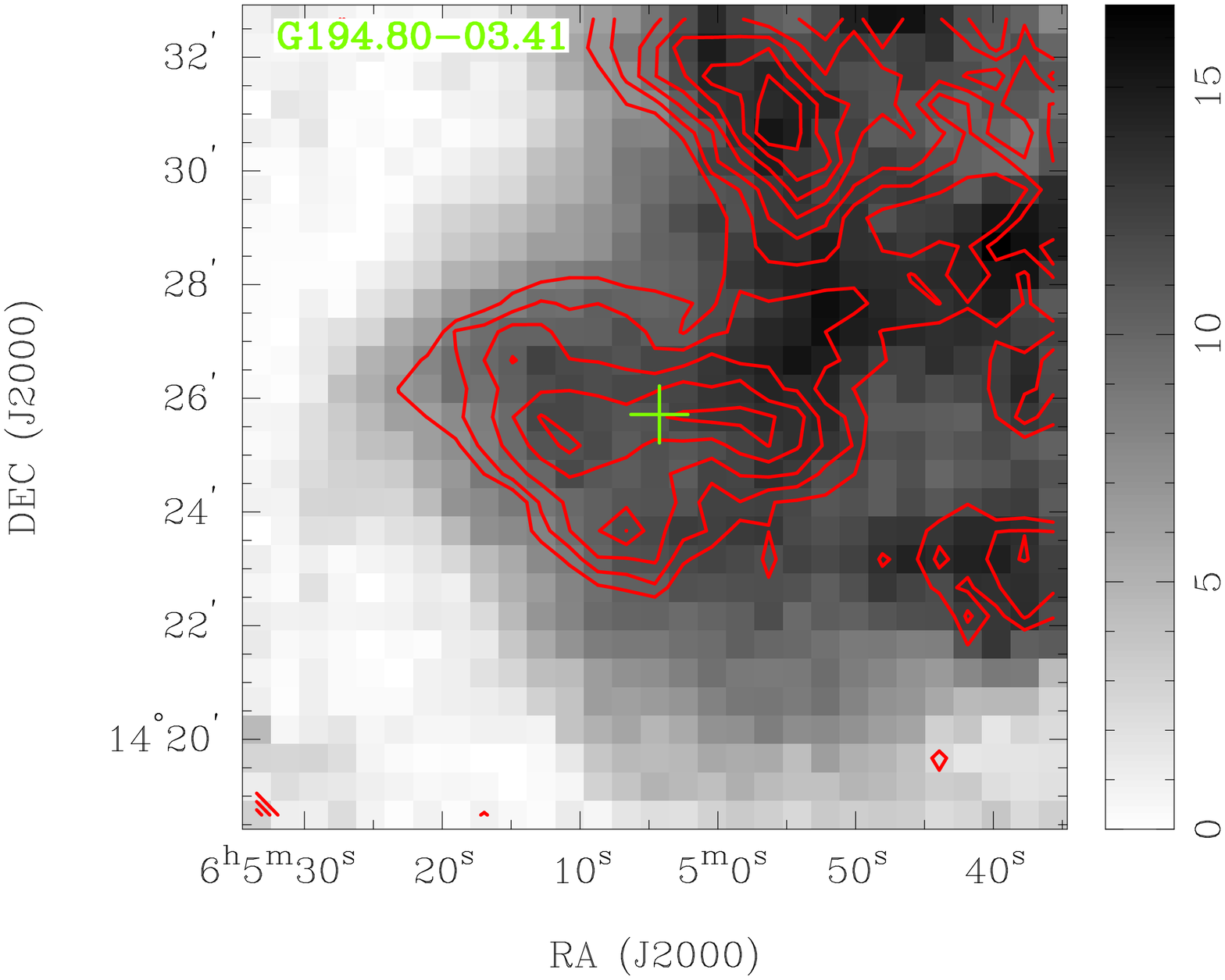}
\end{minipage}
\begin{minipage}[c]{0.33\textwidth}
  \centering
  \includegraphics[width=40mm,height=50mm,angle=-90]{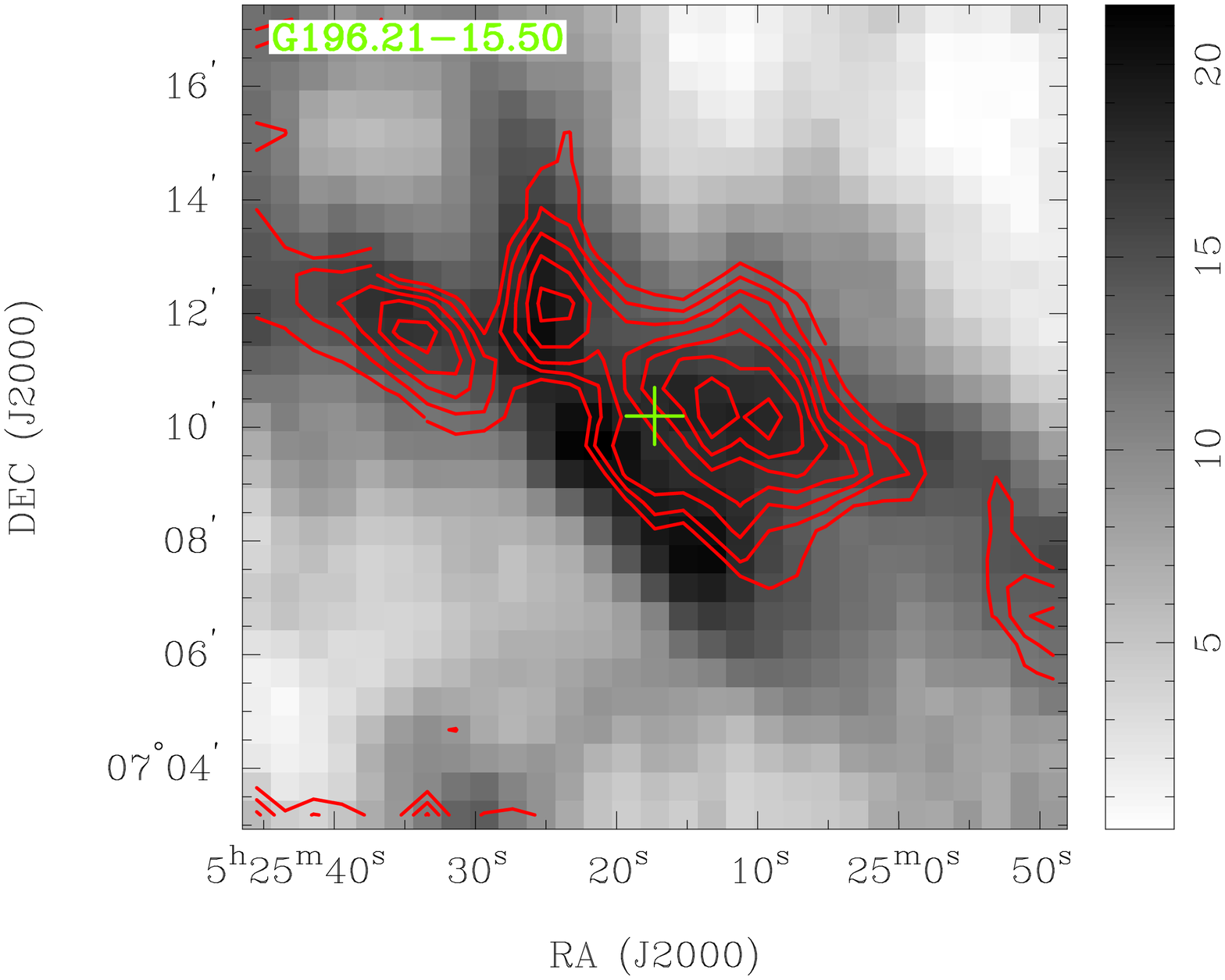}
\end{minipage}
\caption{Integration intensity images of the mapped clumps: contours for $^{13}$CO line and the grey scale for the $^{12}$CO lines. The contours are from 30\% to 90\% in a step of 10\% of the peak value.}
\end{figure}

\begin{figure}
\includegraphics[angle=0,scale=.50]{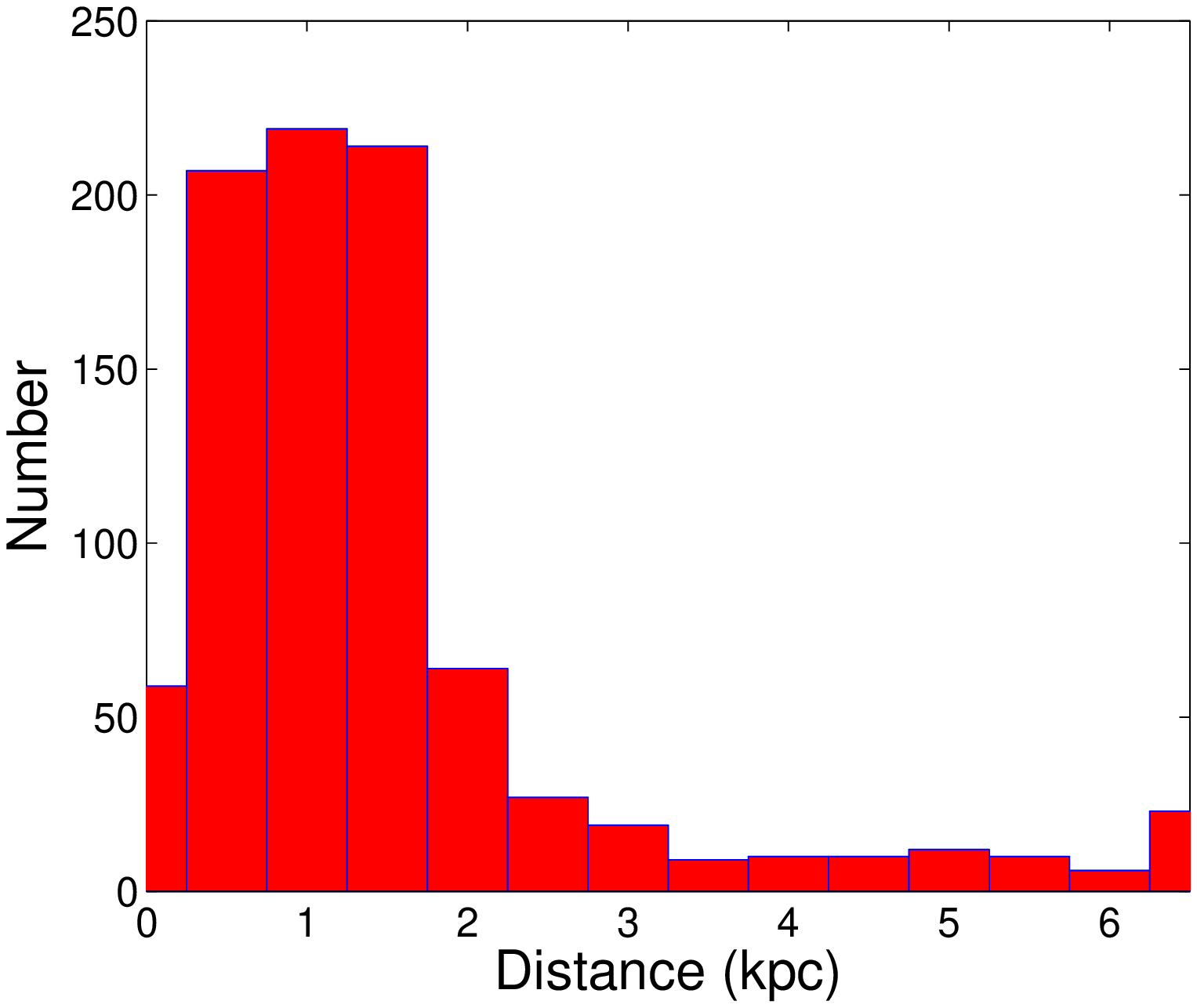}
\caption{The frequency distribution of the kinematic distances}
\end{figure}

\begin{figure}
\begin{minipage}[c]{0.5\textwidth}
  \centering
  \includegraphics[width=70mm,height=55mm,angle=0]{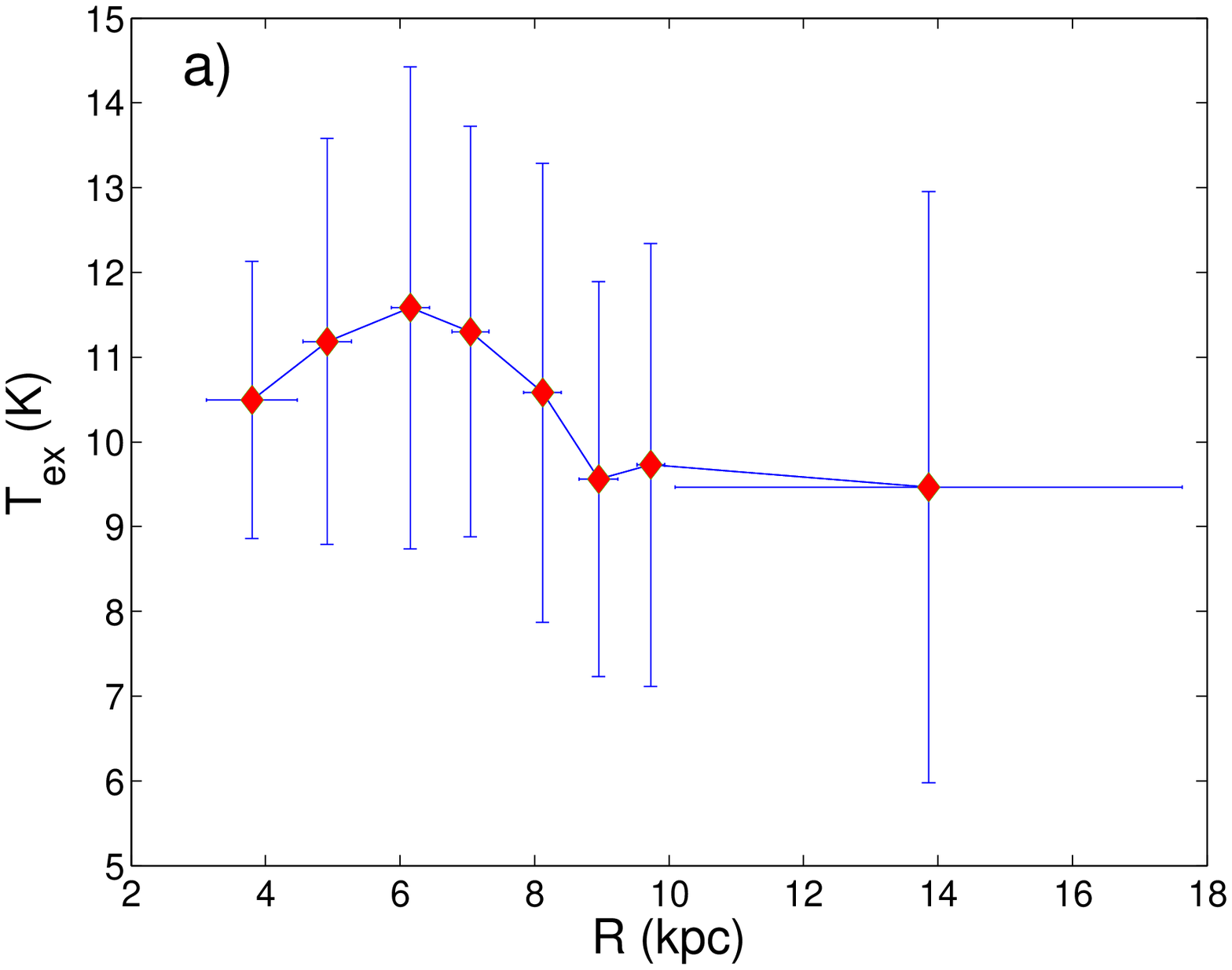}
\end{minipage}
\begin{minipage}[c]{0.5\textwidth}
  \centering
  \includegraphics[width=70mm,height=55mm,angle=0]{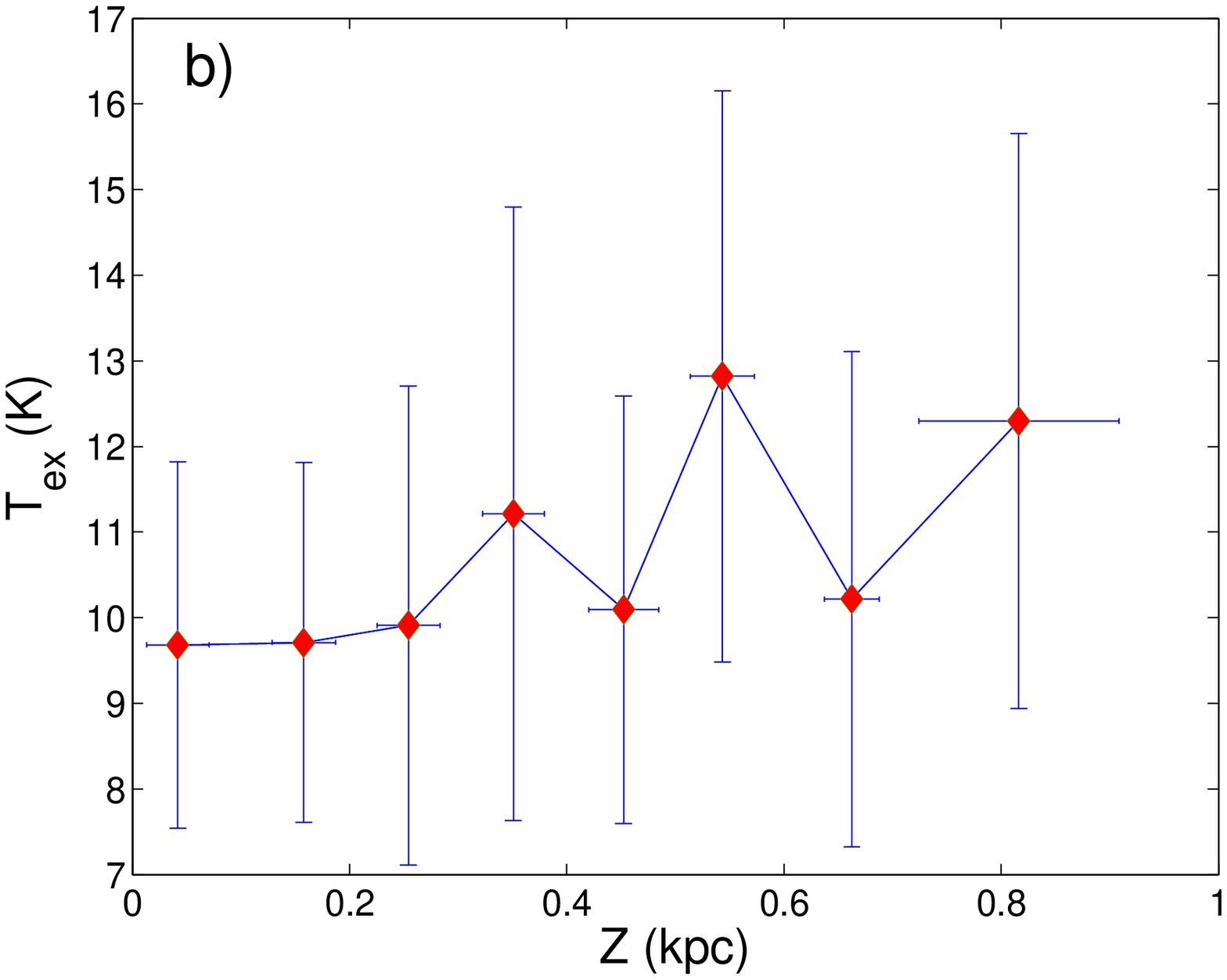}
\end{minipage}
\begin{minipage}[c]{0.5\textwidth}
  \centering
  \includegraphics[width=70mm,height=55mm,angle=0]{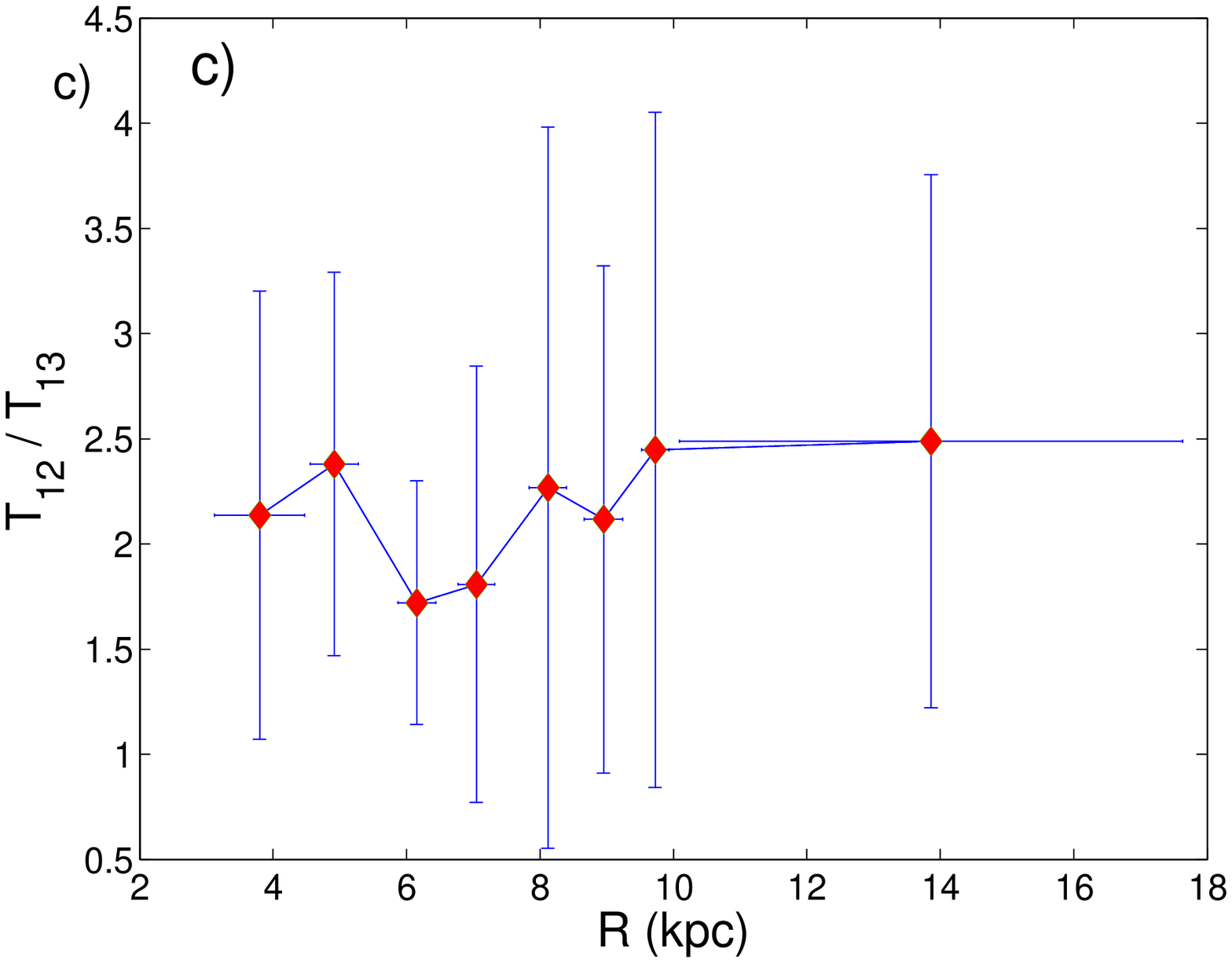}
\end{minipage}
\begin{minipage}[c]{0.5\textwidth}
  \centering
  \includegraphics[width=70mm,height=55mm,angle=0]{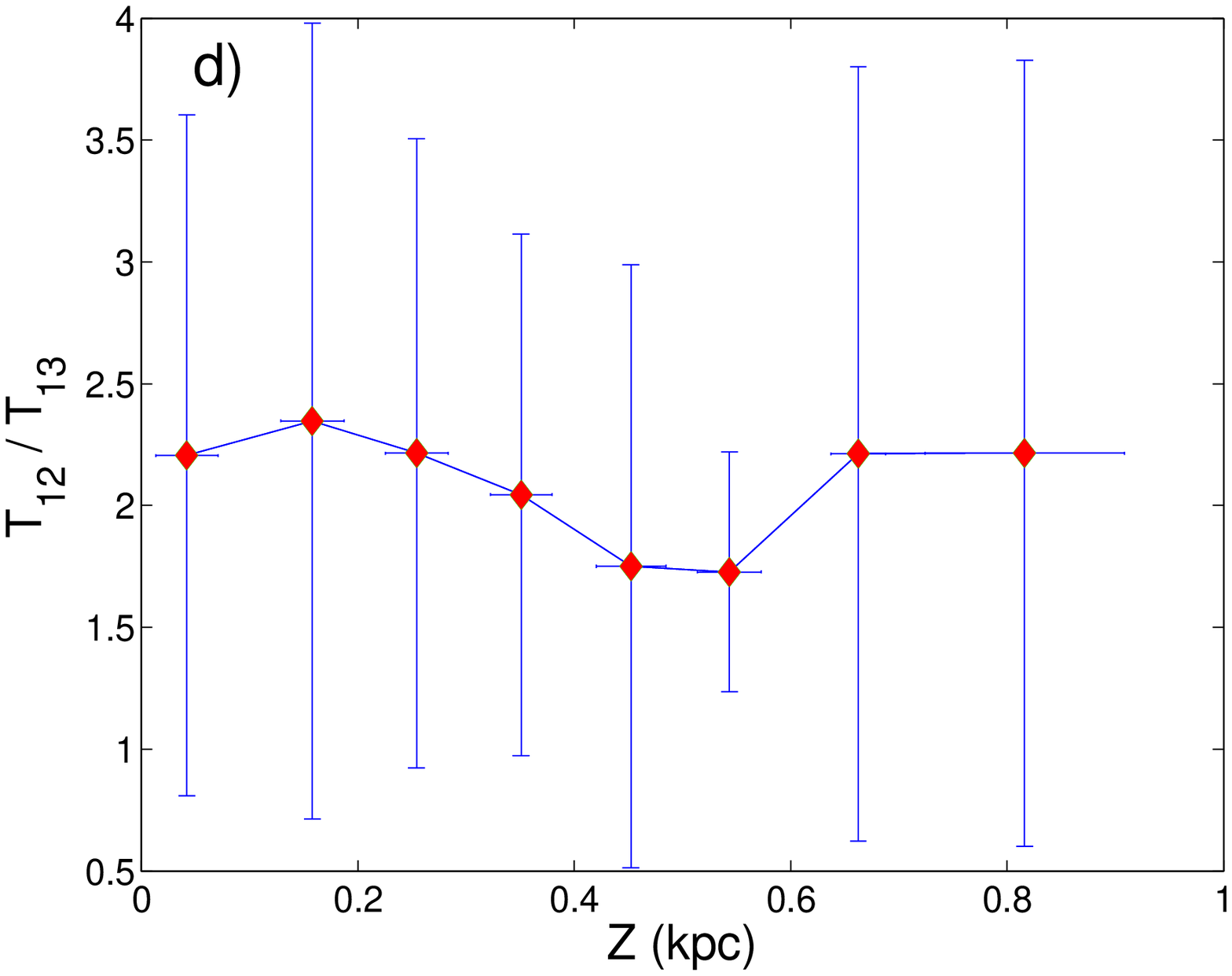}
\end{minipage}
\caption{Variations of bin-averaged T$_{ex}$ and T$_{12}/$T$_{13}$ with the distance from the Galactic center R and the altitude from the Galactic disk plane Z. The bin size in R is 1 kpc for those clumps with R$<$10 pc and the clumps with R$>$10 pc are put into a single bin. The bin size in Z is 0.1 pc. The clumps with Z$>$1 kpc are rare and are not included in analysis. The bin sizes in Figure 10 and 11 are the same as for figure 9.}
\end{figure}

\begin{figure}
\begin{minipage}[c]{0.5\textwidth}
  \centering
  \includegraphics[width=70mm,height=55mm,angle=0]{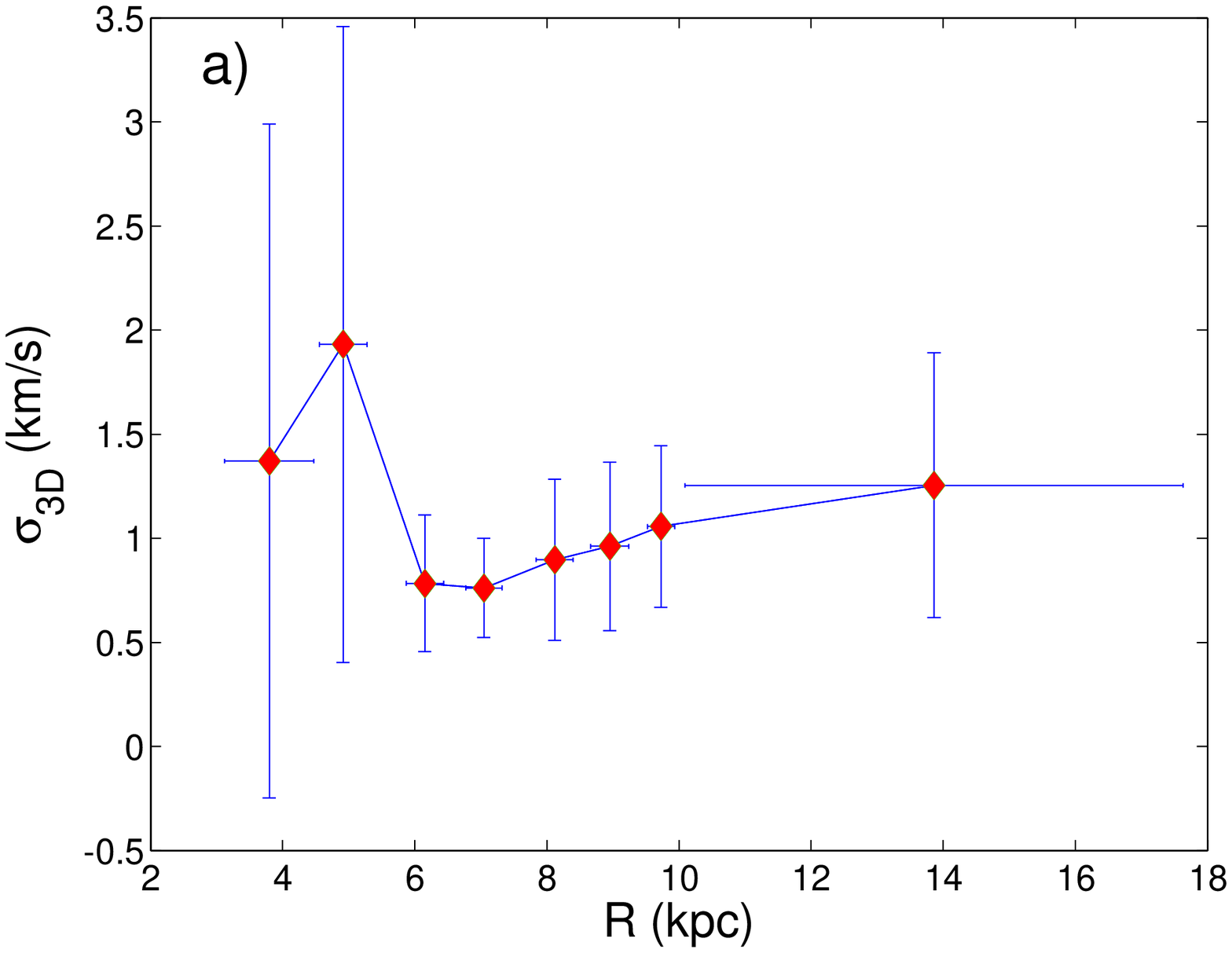}
\end{minipage}
\begin{minipage}[c]{0.5\textwidth}
  \centering
  \includegraphics[width=70mm,height=55mm,angle=0]{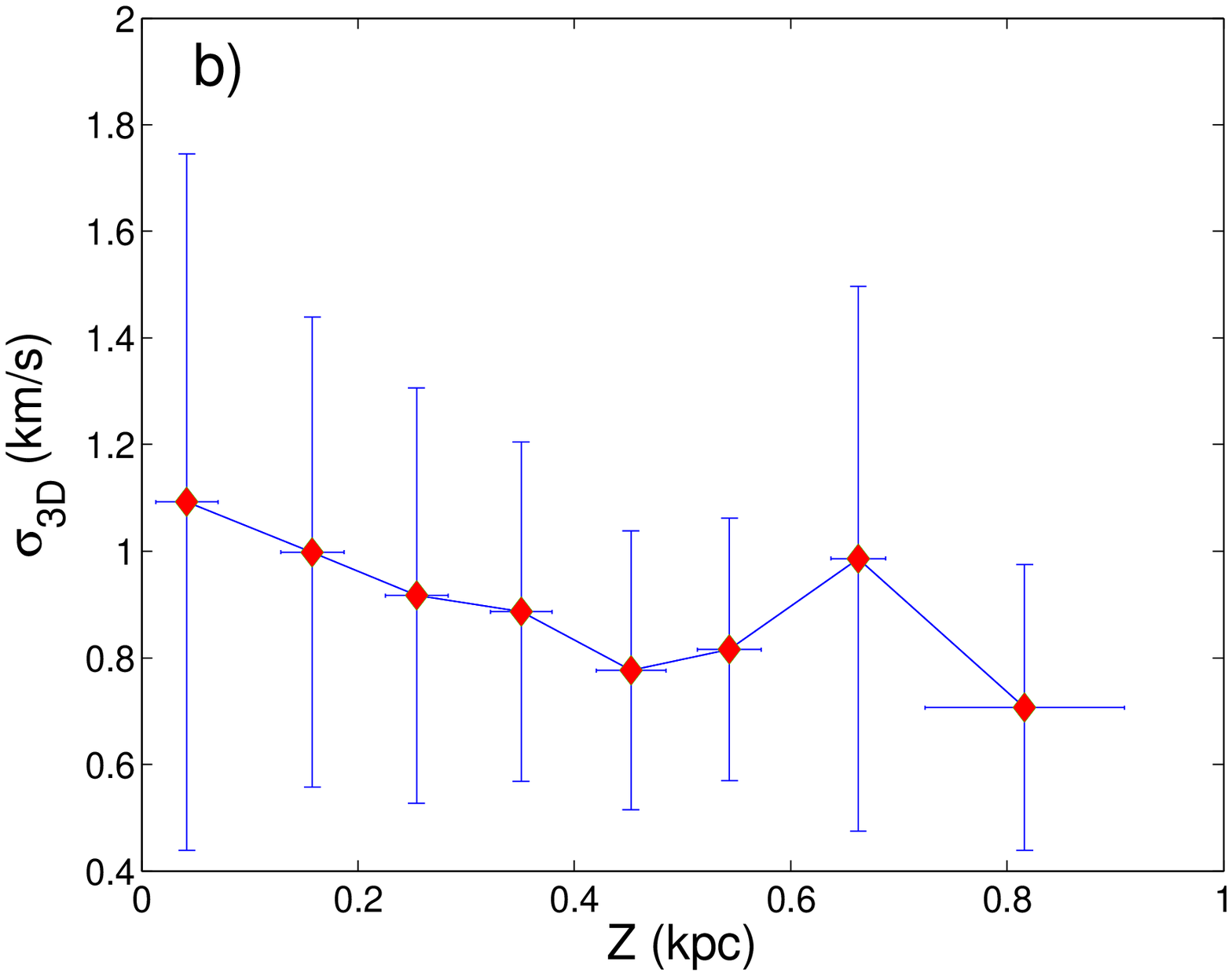}
\end{minipage}
\begin{minipage}[c]{0.5\textwidth}
  \centering
  \includegraphics[width=70mm,height=55mm,angle=0]{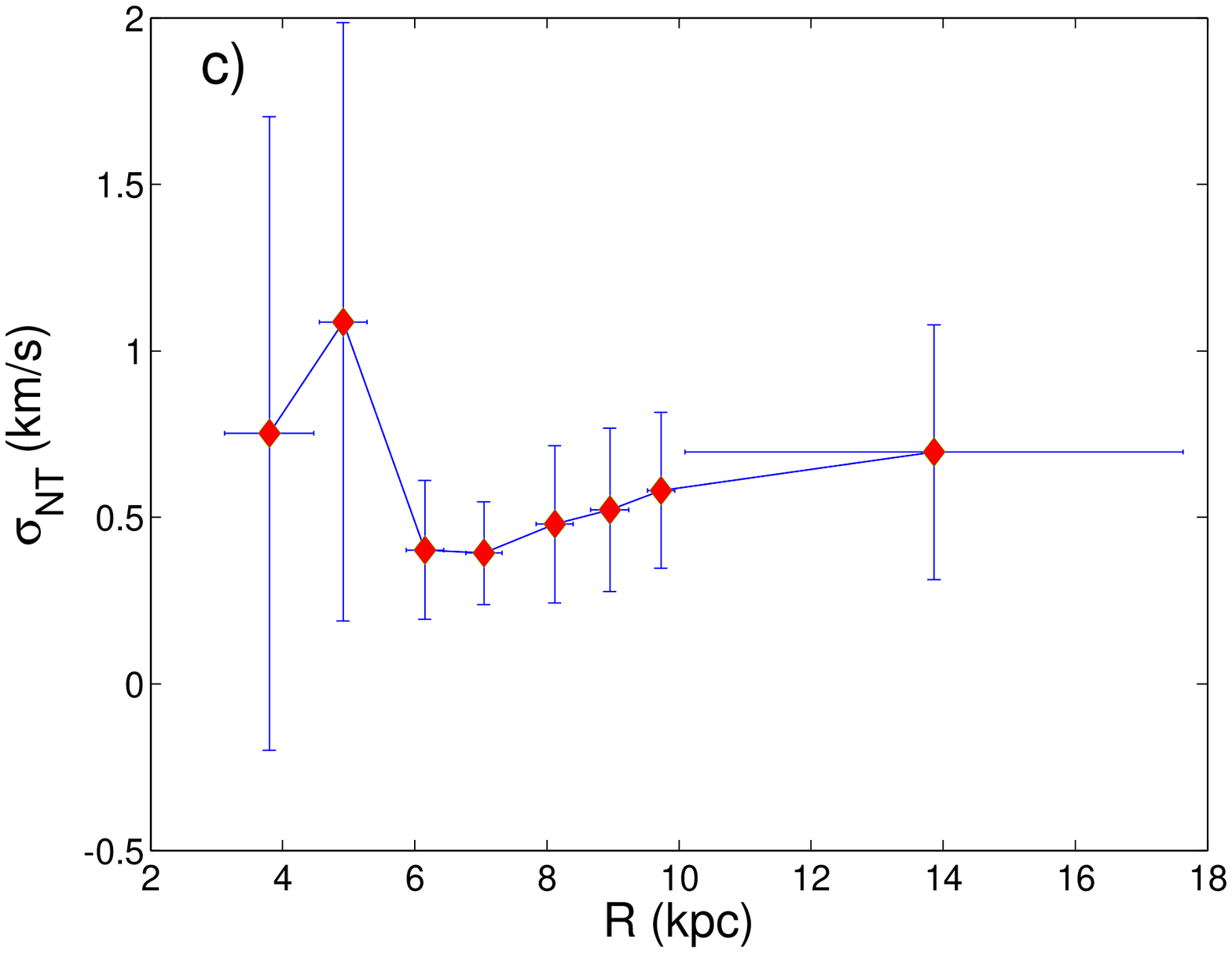}
\end{minipage}
\begin{minipage}[c]{0.5\textwidth}
  \centering
  \includegraphics[width=70mm,height=55mm,angle=0]{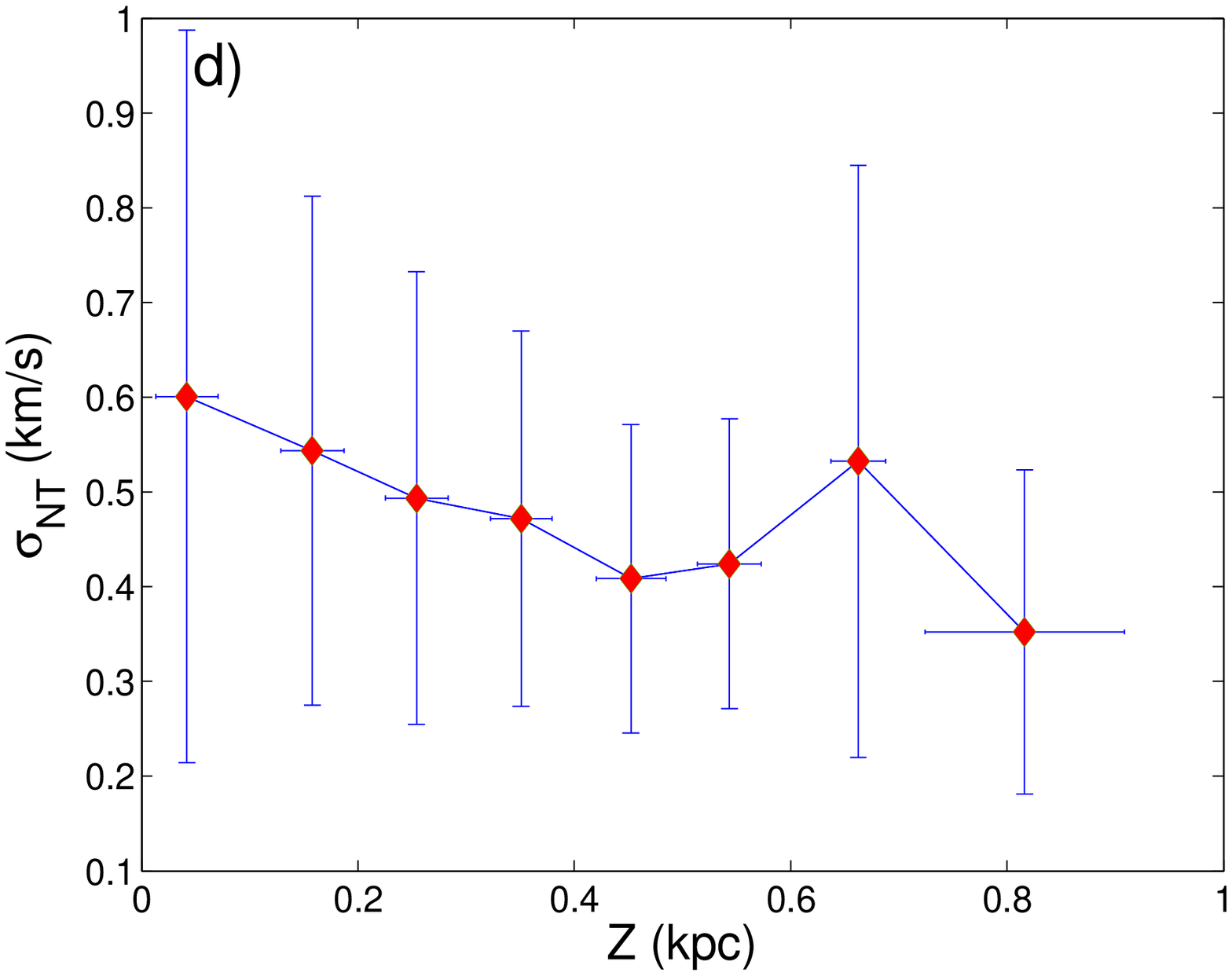}
\end{minipage}
\begin{minipage}[c]{0.5\textwidth}
  \centering
  \includegraphics[width=70mm,height=55mm,angle=0]{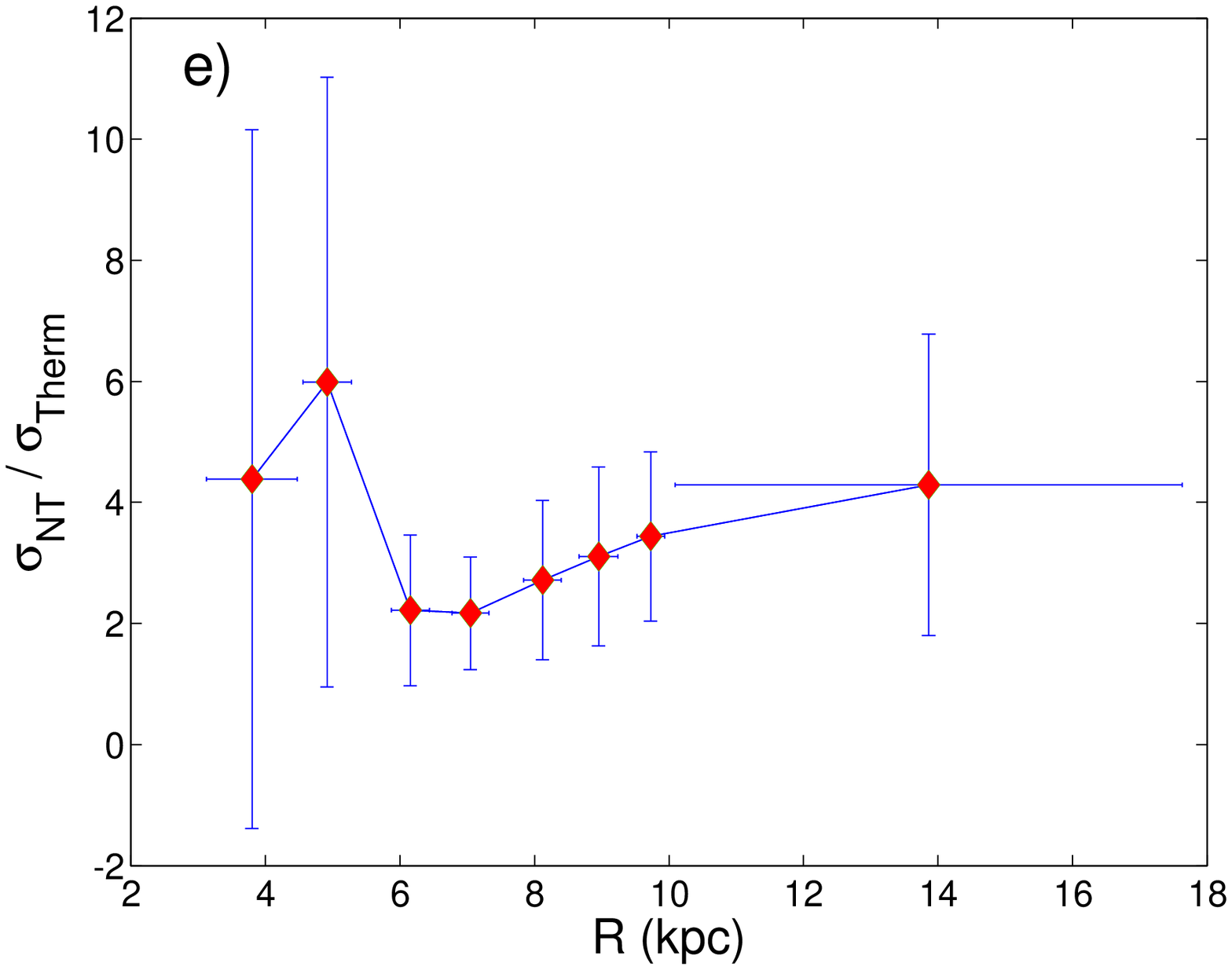}
\end{minipage}
\begin{minipage}[c]{0.5\textwidth}
  \centering
  \includegraphics[width=70mm,height=55mm,angle=0]{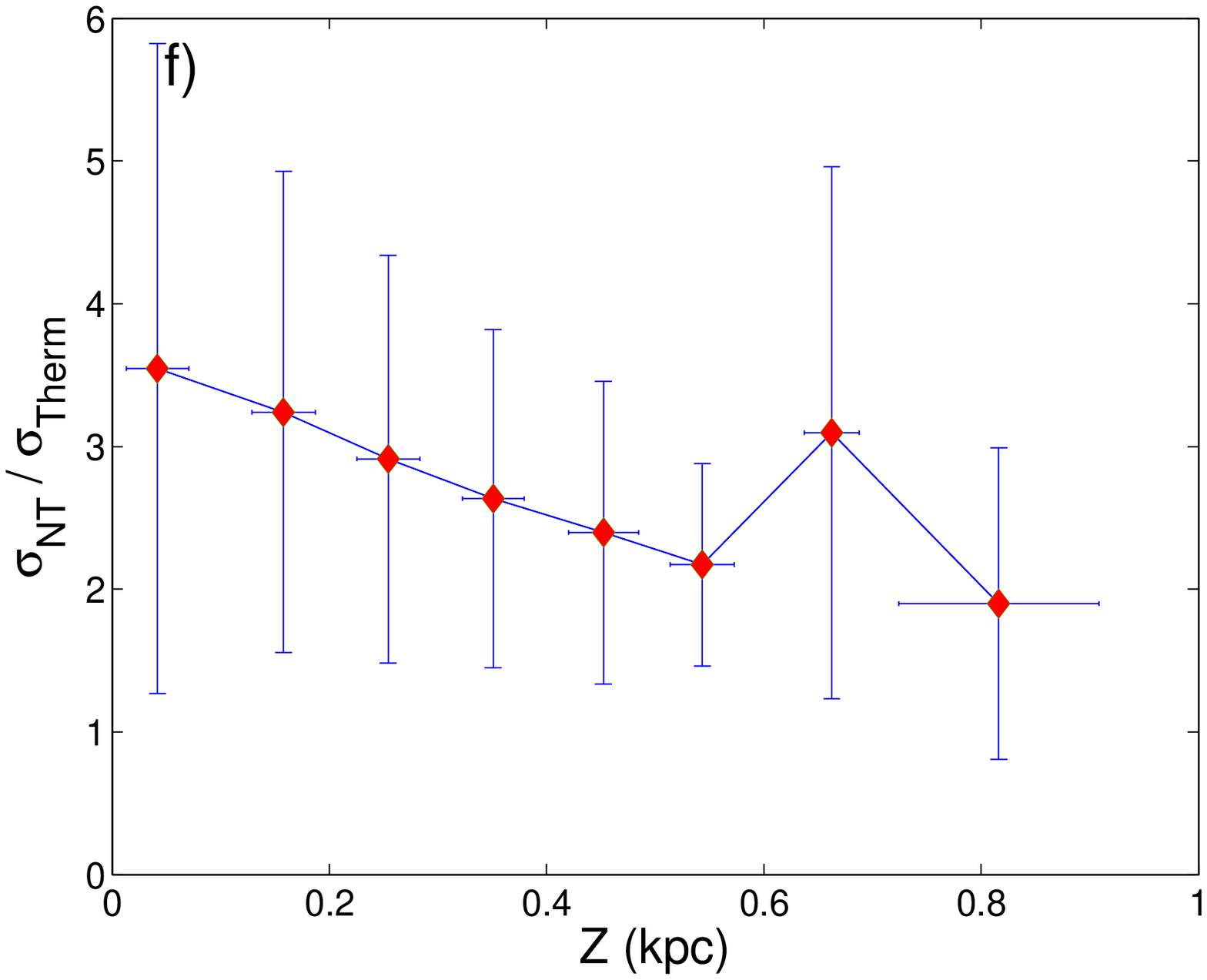}
\end{minipage}
\caption{Variations of bin-averaged $\sigma_{3D}$, $\sigma_{NT}$ and the ratio of $\sigma_{NT}/\sigma_{Therm}$ of $^{13}$CO lines with R and Z..}
\end{figure}

\begin{figure}
\begin{minipage}[c]{0.5\textwidth}
  \centering
  \includegraphics[width=70mm,height=55mm,angle=0]{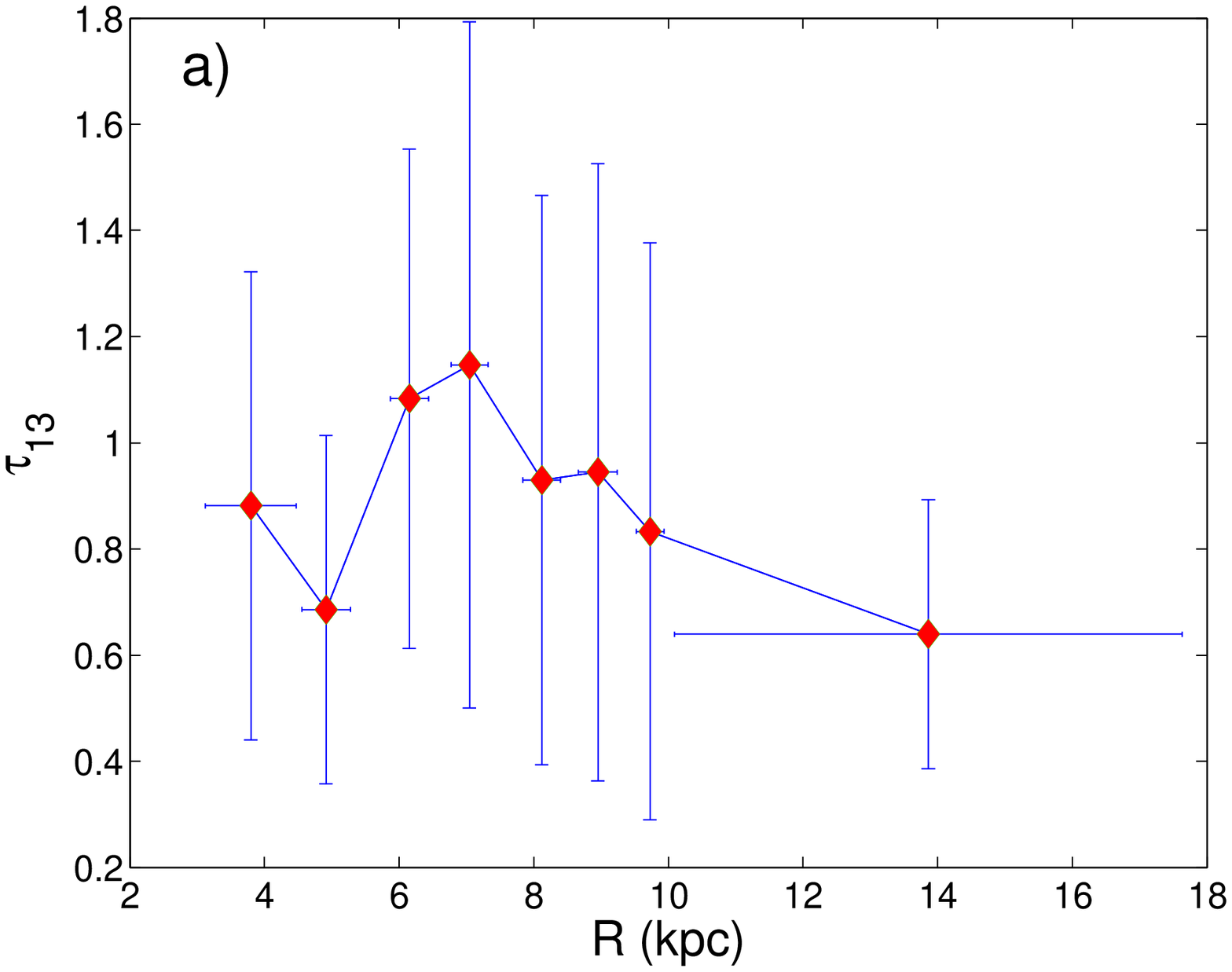}
\end{minipage}
\begin{minipage}[c]{0.5\textwidth}
  \centering
  \includegraphics[width=70mm,height=55mm,angle=0]{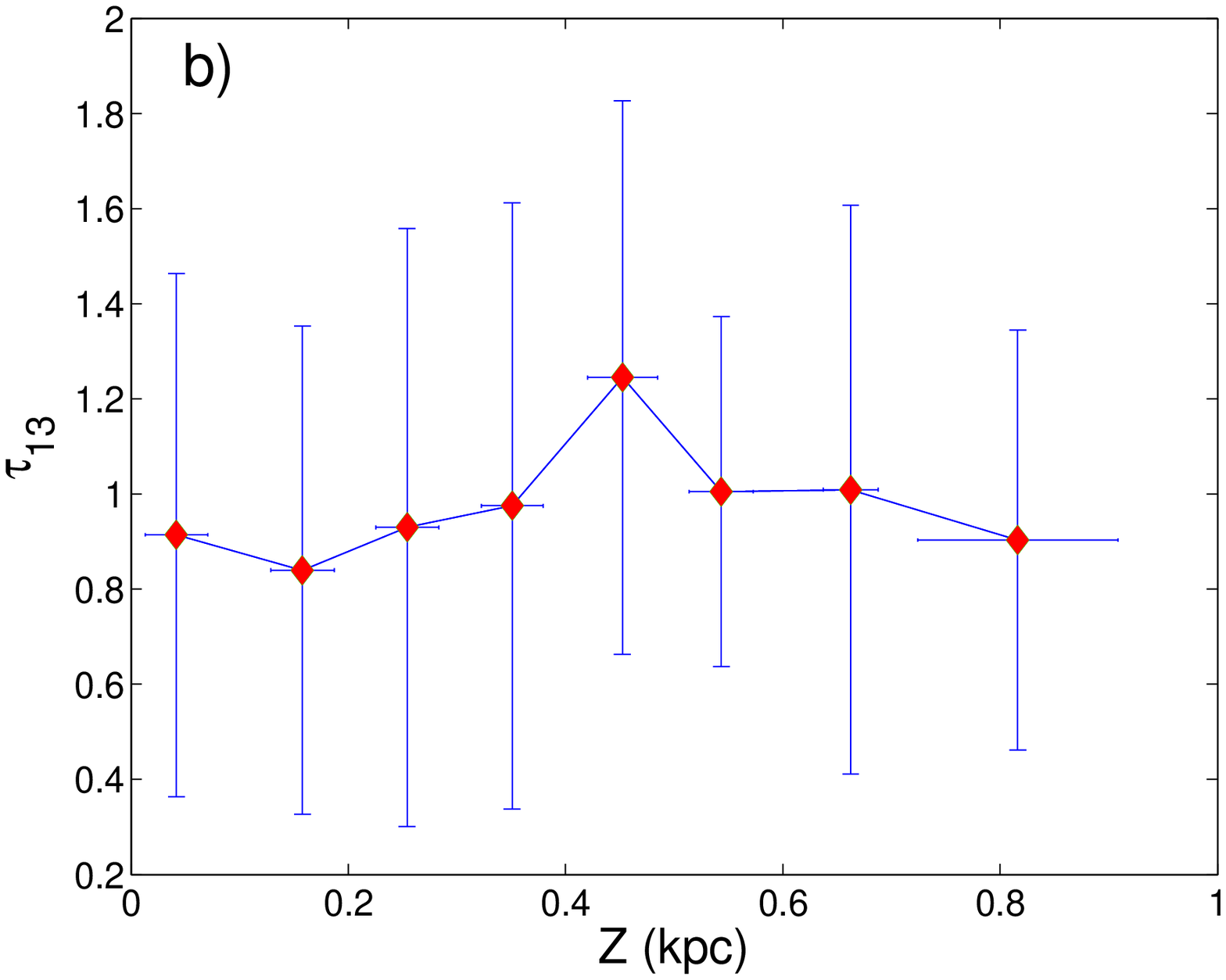}
\end{minipage}
\begin{minipage}[c]{0.5\textwidth}
  \centering
  \includegraphics[width=70mm,height=55mm,angle=0]{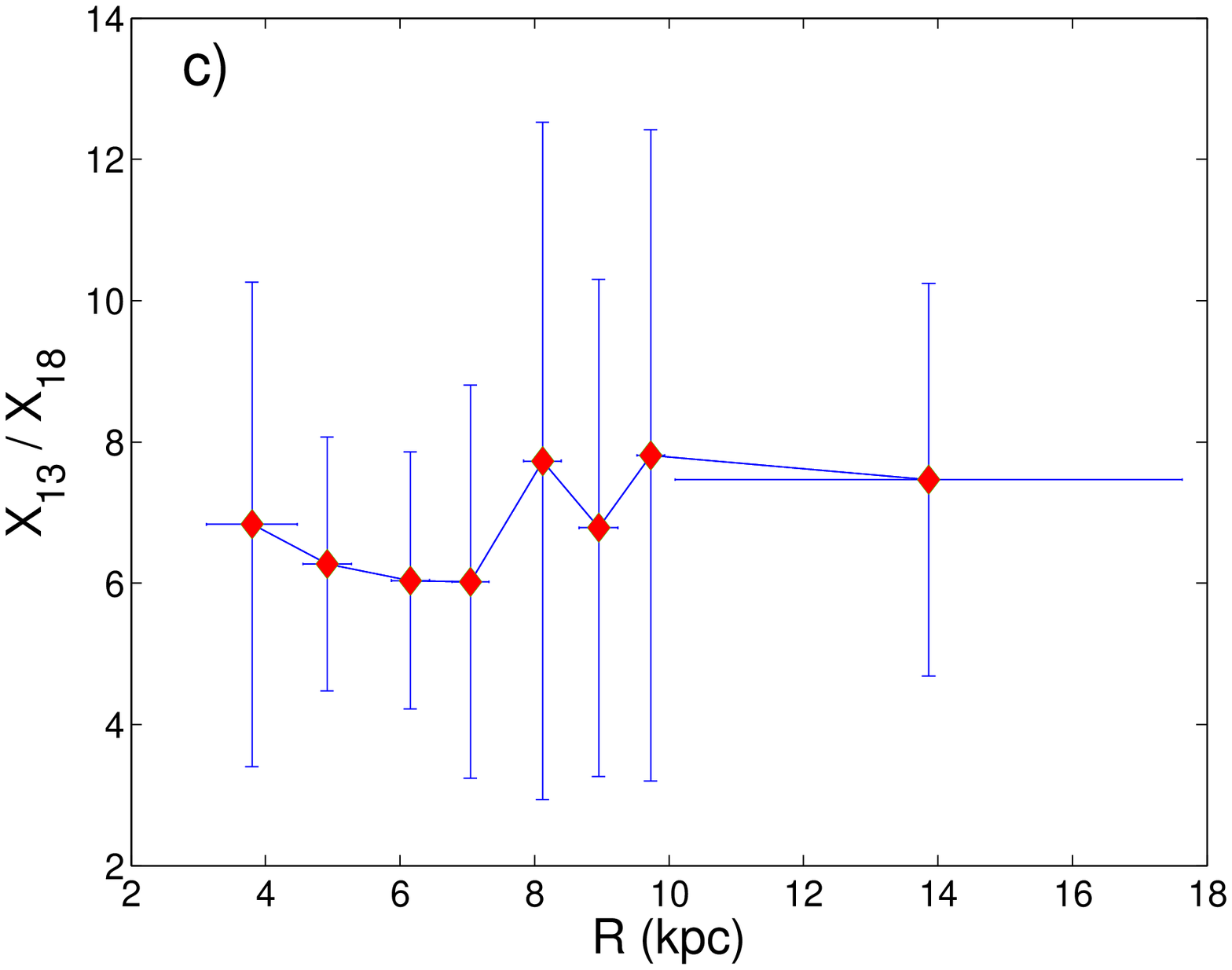}
\end{minipage}
\begin{minipage}[c]{0.5\textwidth}
  \centering
  \includegraphics[width=70mm,height=55mm,angle=0]{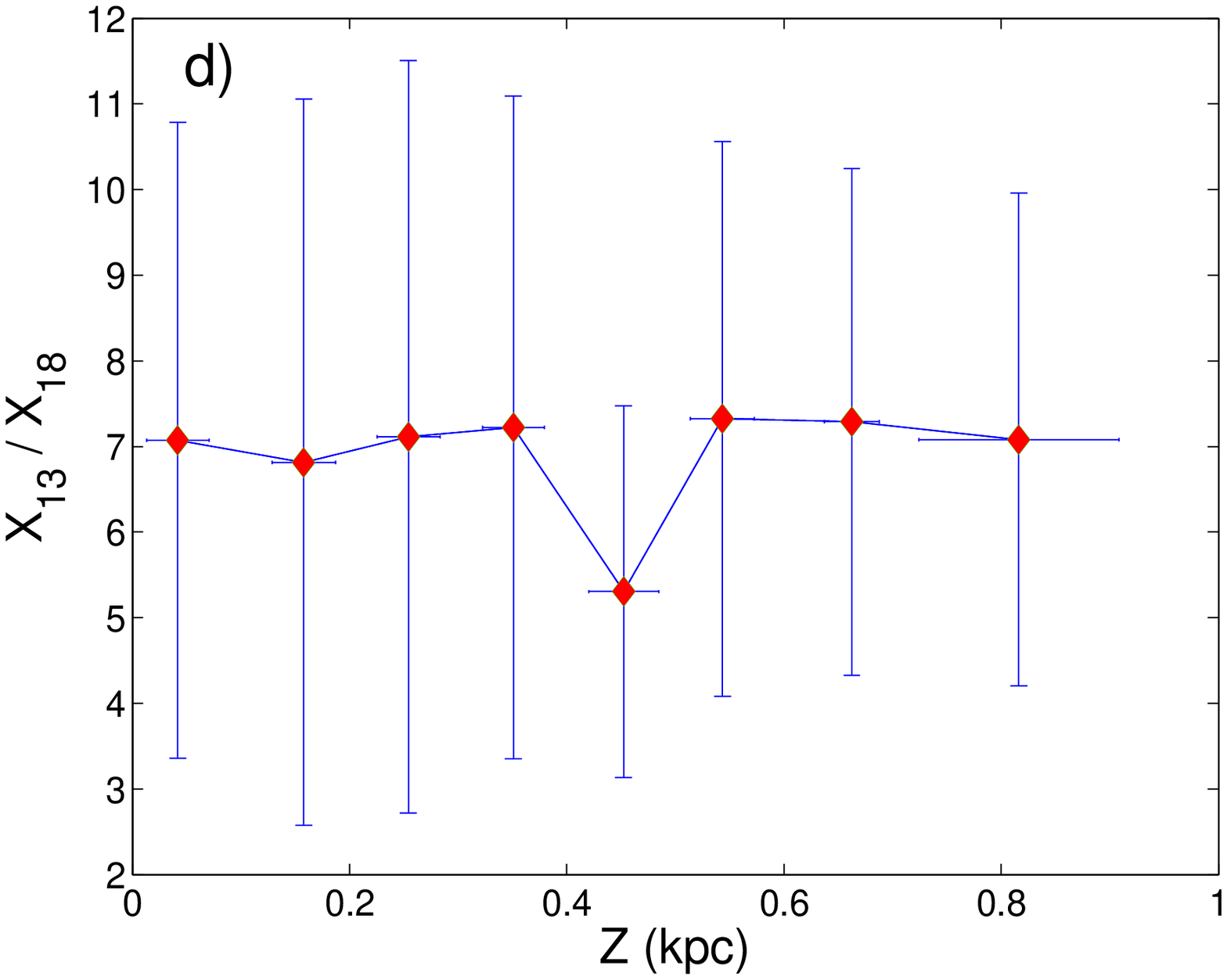}
\end{minipage}
\begin{minipage}[c]{0.5\textwidth}
  \centering
  \includegraphics[width=70mm,height=55mm,angle=0]{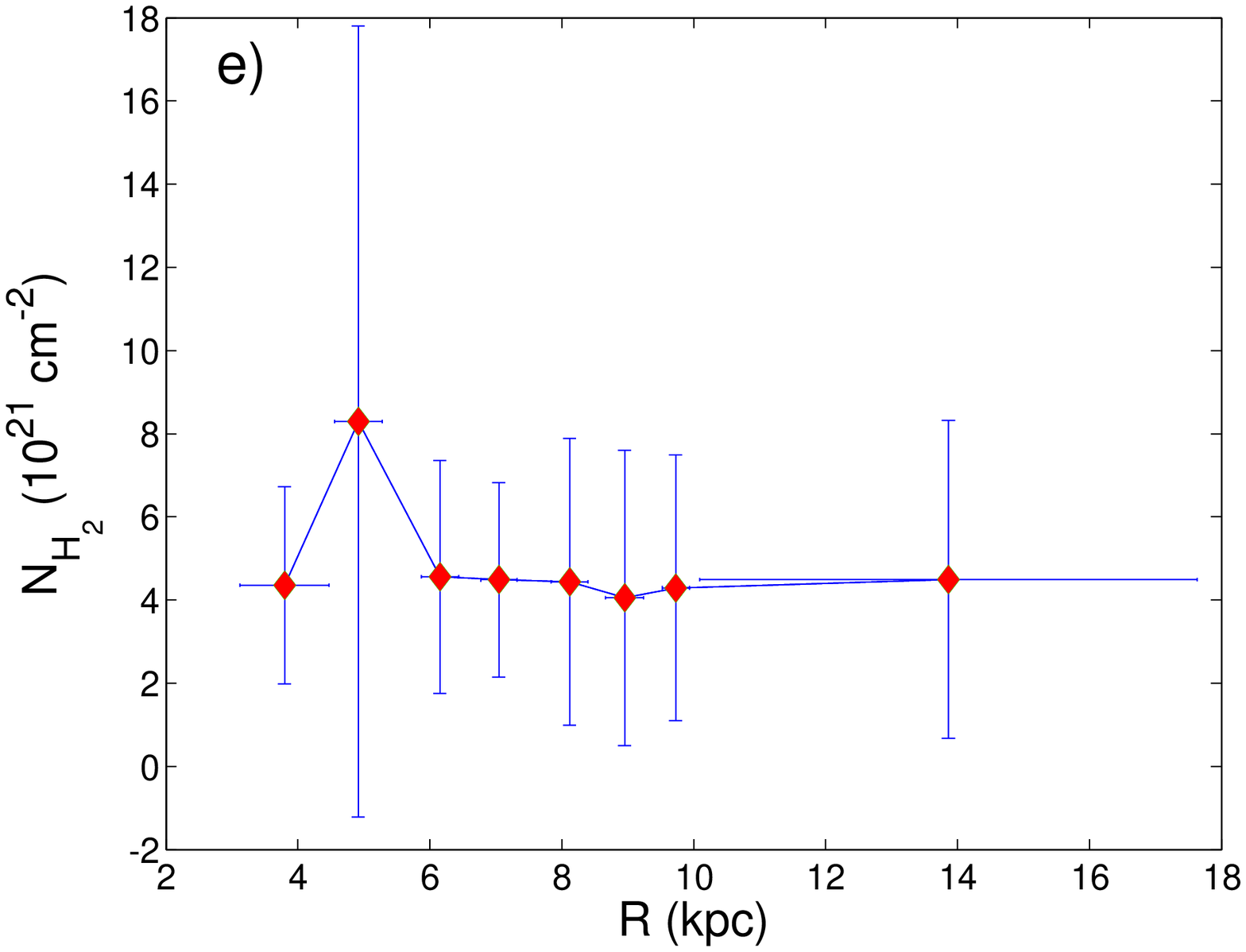}
\end{minipage}
\begin{minipage}[c]{0.5\textwidth}
  \centering
  \includegraphics[width=70mm,height=55mm,angle=0]{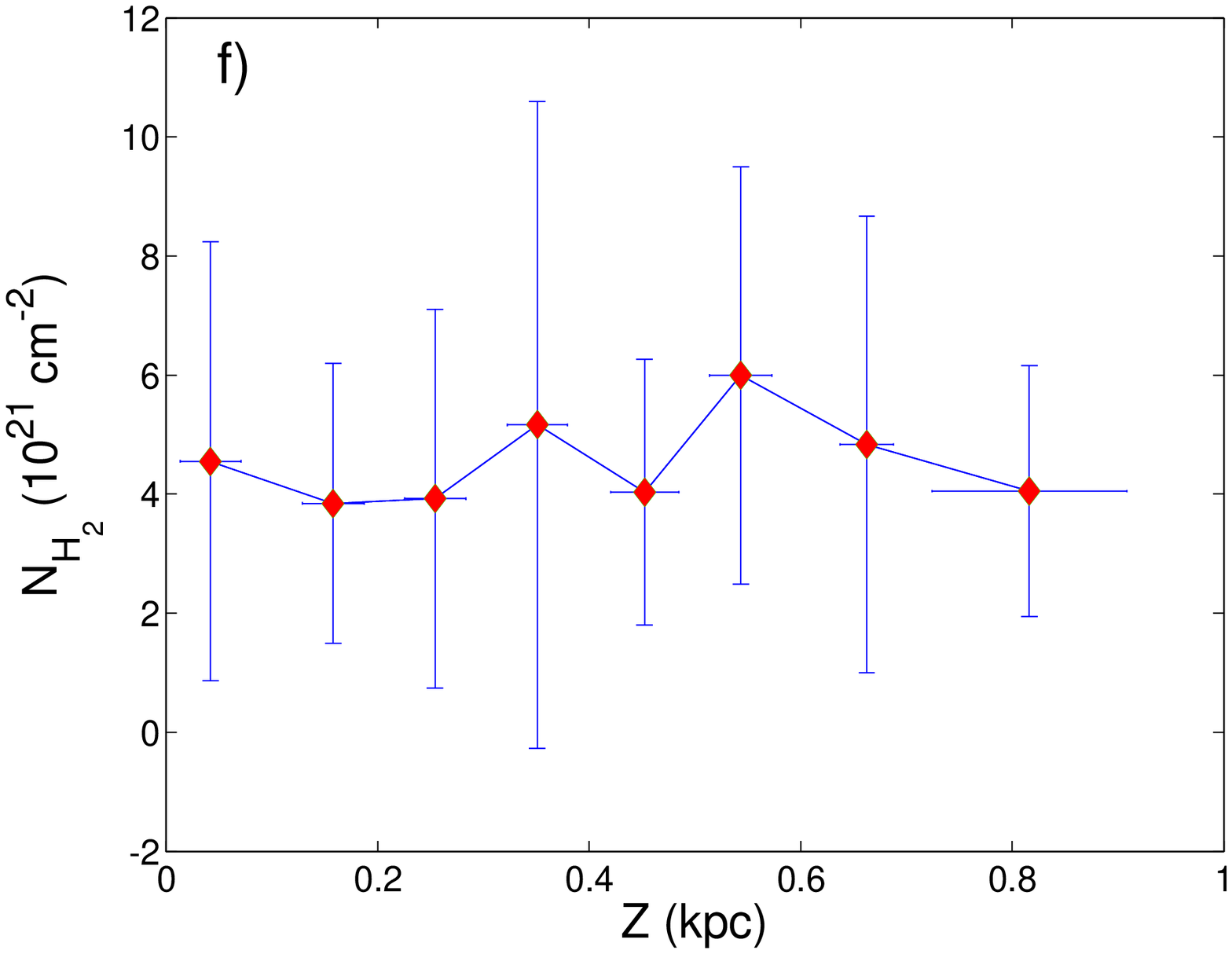}
\end{minipage}
\begin{minipage}[c]{0.5\textwidth}
  \centering
  \includegraphics[width=70mm,height=55mm,angle=0]{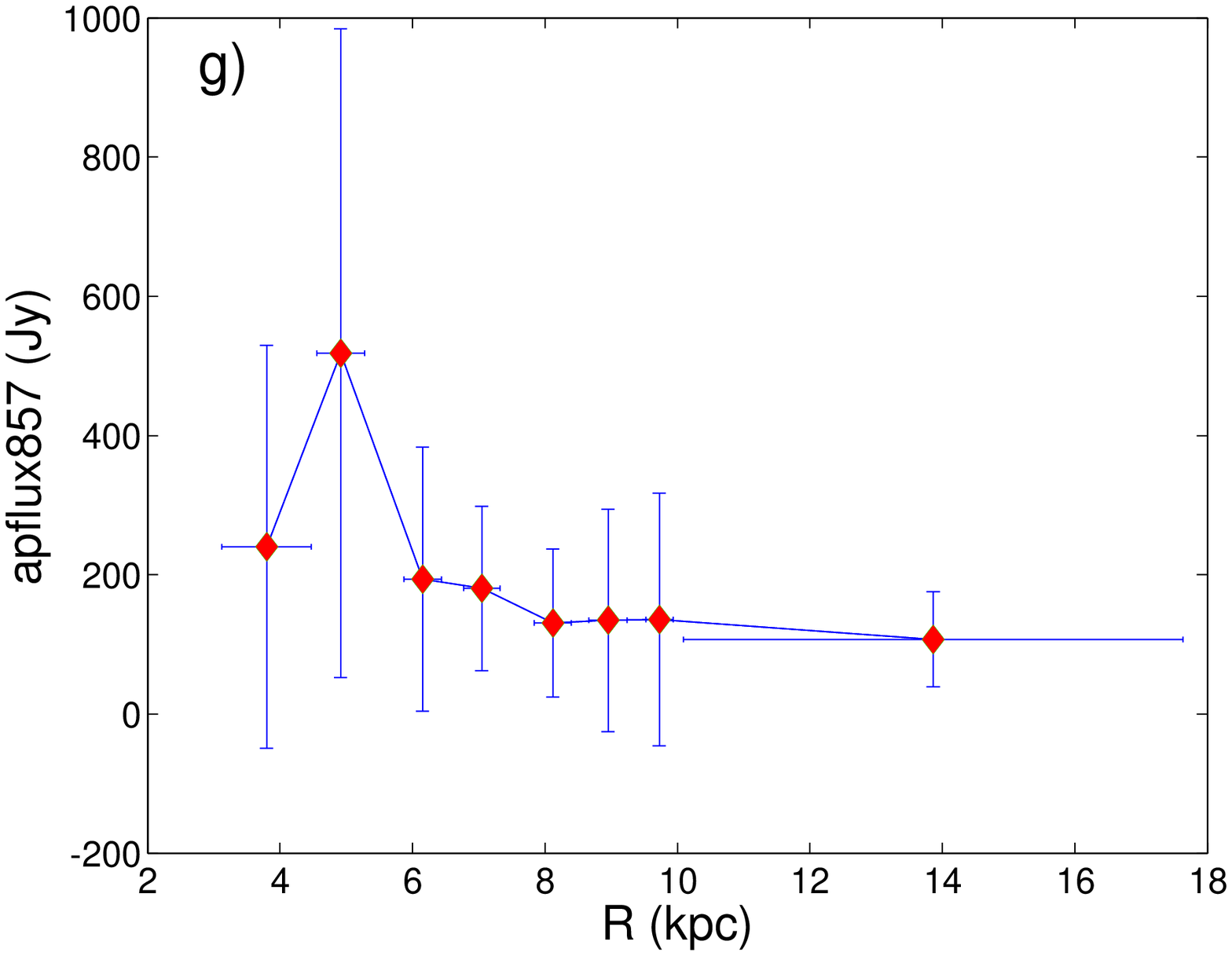}
\end{minipage}
\begin{minipage}[c]{0.5\textwidth}
  \centering
  \includegraphics[width=70mm,height=55mm,angle=0]{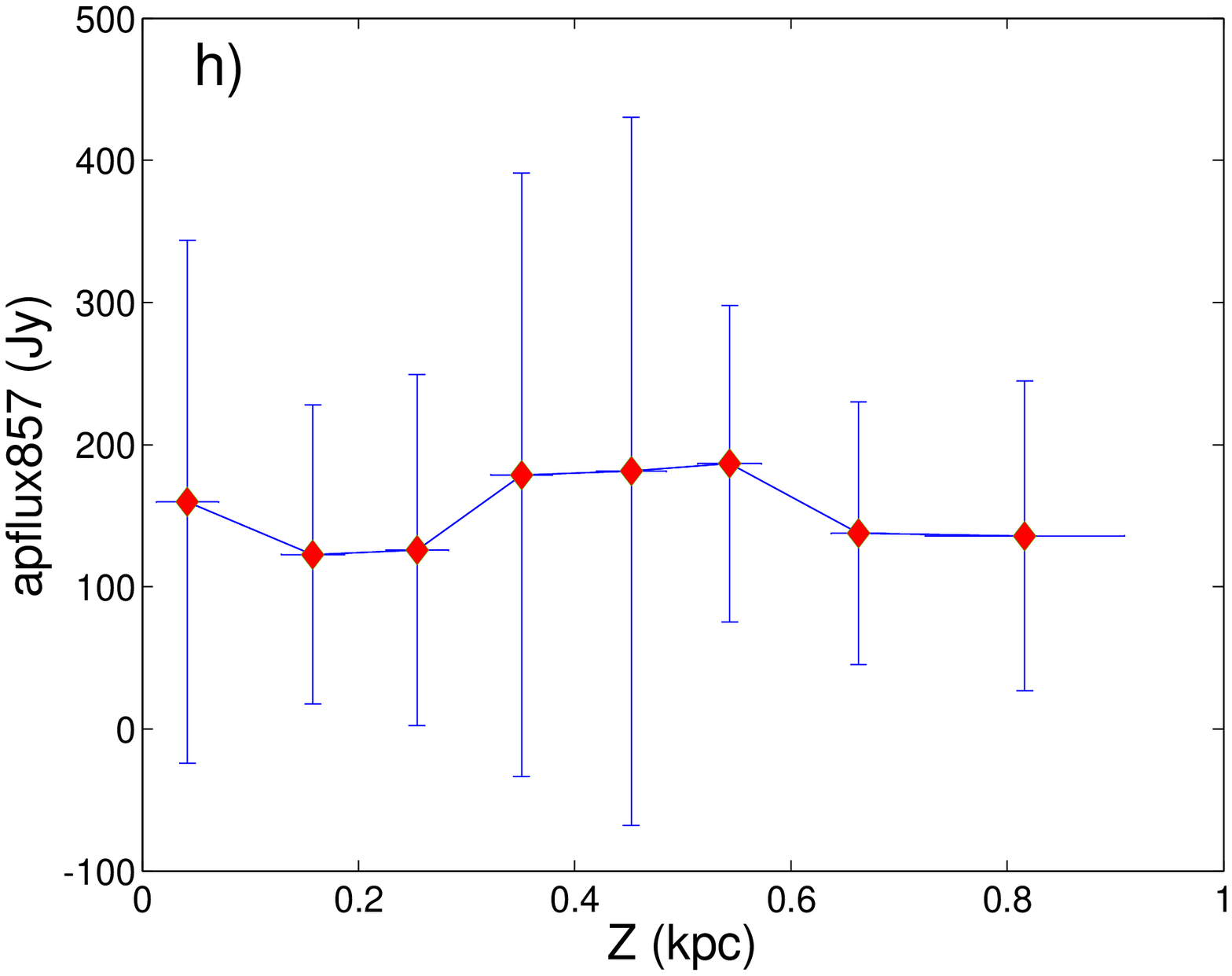}
\end{minipage}
\caption{Variations of bin-averaged $\tau_{13}$, [X13]/[X18], N$_{H_{2}}$ and the
flux at 857 GHz with R and Z.}
\end{figure}

\begin{figure}
\begin{minipage}[c]{0.5\textwidth}
  \centering
  \includegraphics[width=70mm,height=55mm,angle=0]{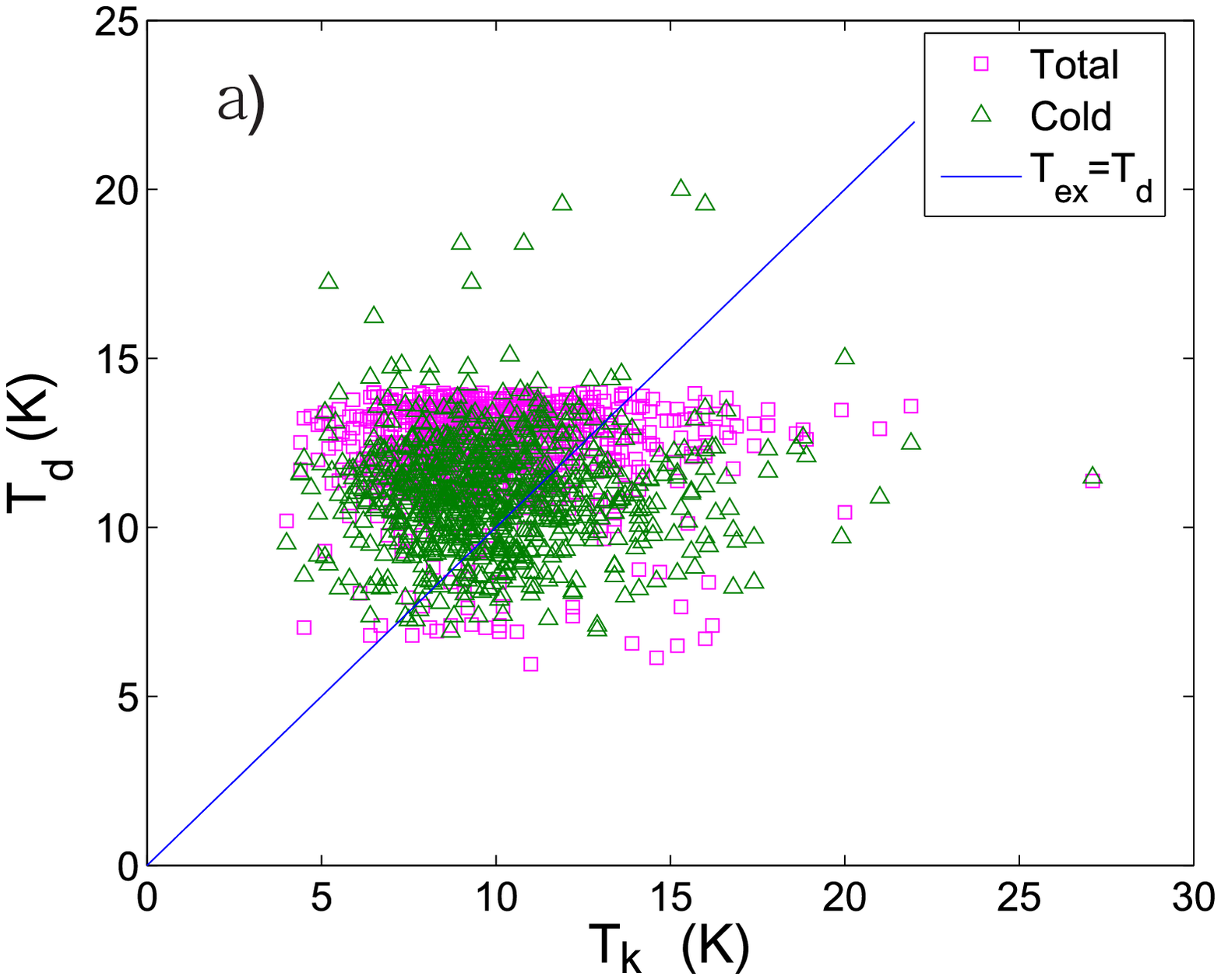}
\end{minipage}
\begin{minipage}[c]{0.5\textwidth}
  \centering
  \includegraphics[width=70mm,height=55mm,angle=0]{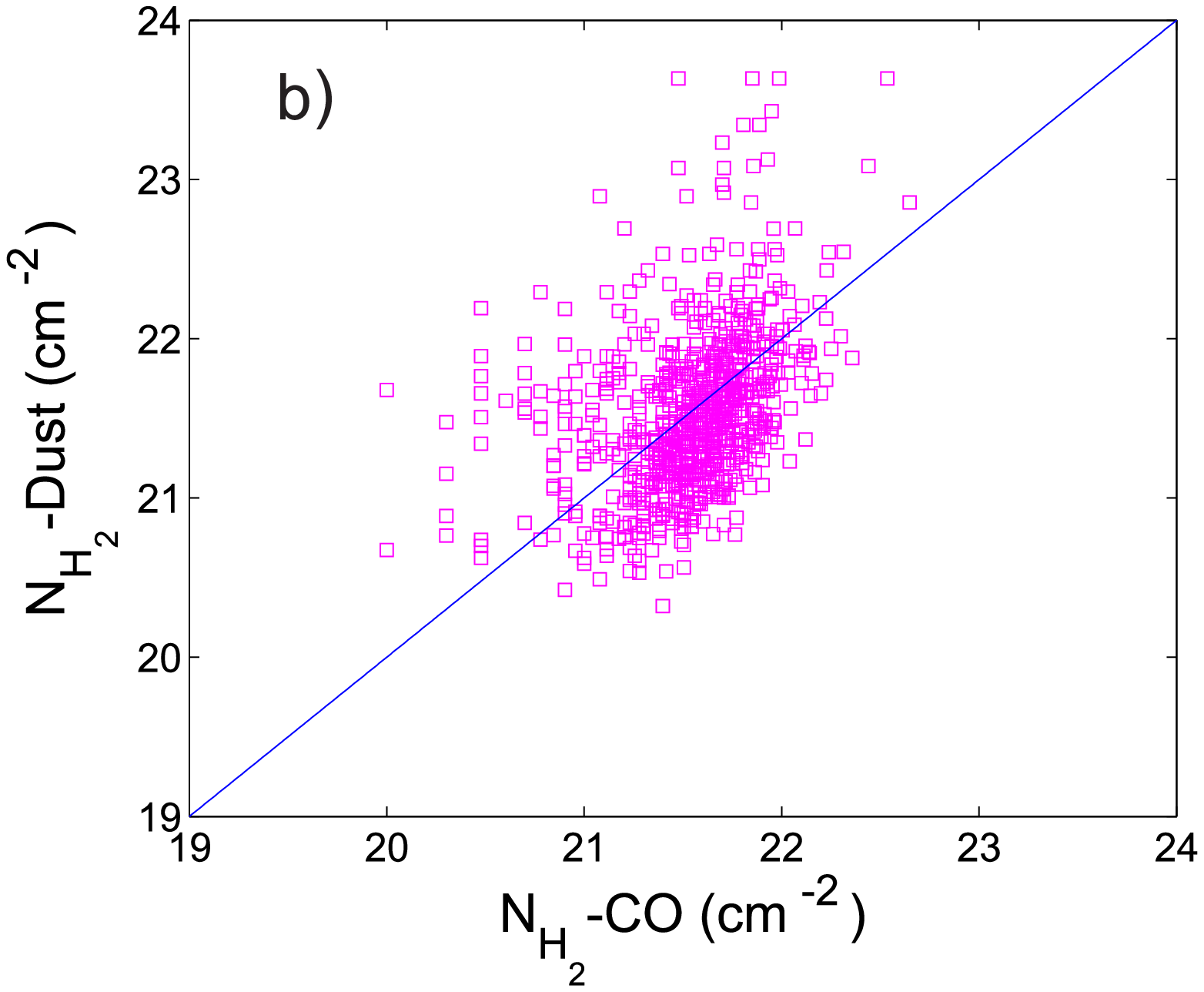}
\end{minipage}
\caption{Comparison T$_{k}$ to T$_{d}$ and N$_{H_{2}}$ to that
deduced from dust emission."Total" and "cold" indicate that Td is
calculated from the SED modeling with total fluxes and the SED of
the residual fluxes obtained by subtracting the modeled flux from
IRAS 100$ \mu$m respectively \citep{ade11a}.}
\end{figure}

\begin{figure}
\begin{minipage}[c]{0.5\textwidth}
  \centering
  \includegraphics[width=70mm,height=55mm,angle=0]{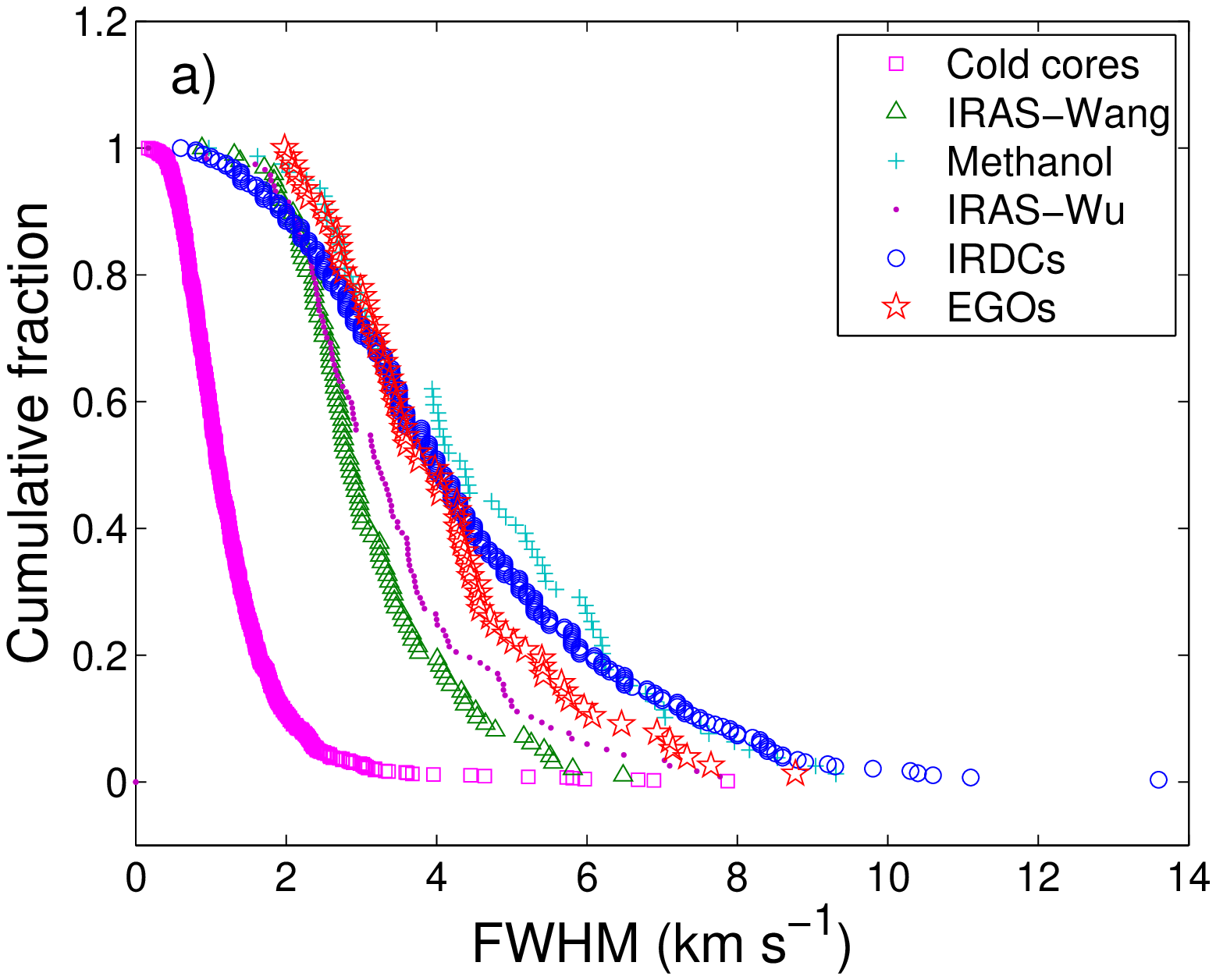}
\end{minipage}
\begin{minipage}[c]{0.5\textwidth}
  \centering
  \includegraphics[width=70mm,height=55mm,angle=0]{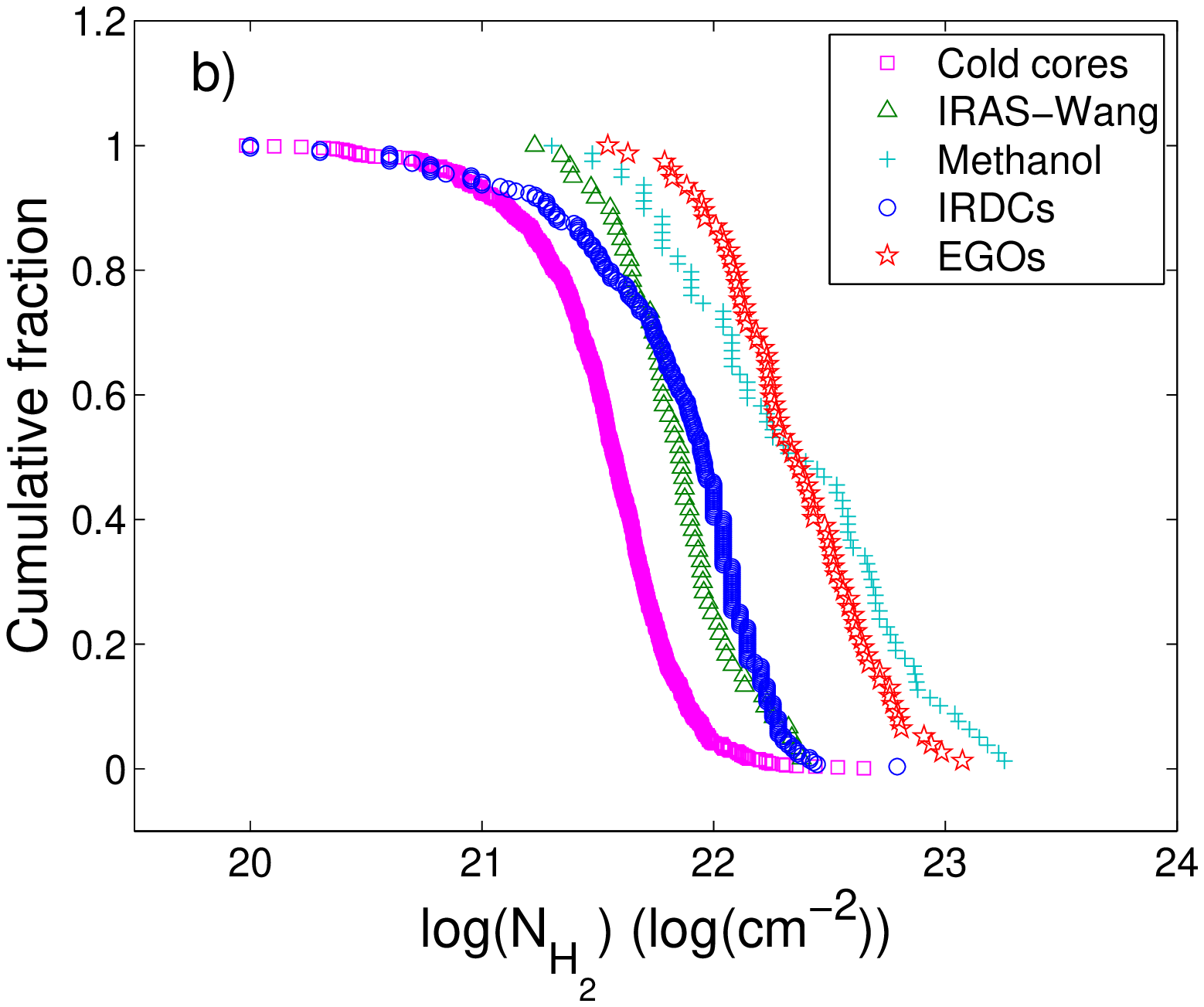}
\end{minipage}
\caption{Comparisons for Cumulative fraction of FWHM of $^{13}$CO lines and N$_{H_{2}}$  of different star formation samples.}
\end{figure}

\clearpage
\begin{deluxetable}{ccrrrrrrrrrrrrrrrrrrcrl}
\setlength{\tabcolsep}{0.05in}
\tabletypesize{\scriptsize}  \tablecaption{Surveyed ECC clump catalogue (the first page of 22 pages)}
 \tablewidth{0pt} \tablehead{
  Name& Glon & Glat &Ra(J2000)  &Dec(J2000) &Ra(B1950)  &Dec(B1950) & Region \\
      &($\arcdeg$) &($\arcdeg$) &(h m s) &(d m s) &(h m s) &(d m s)}
\startdata
G001.38+20.94     & 1.3842772    & 20.941952        & 16 34 38.06  & -15 46 40.71  & 16 31 46.85  &  -15 40 30.16 &Ophiuchus \\
G001.84+16.58     & 1.845703     & 16.587652        & 16 50 12.91  & -18 04 22.37  & 16 47 18.48  &  -17 59 15.82 &Ophiuchus \\
G003.73+16.39     & 3.7353513    & 16.393143        & 16 55 21.78  & -16 43 35.31  & 16 52 28.86  &  -16 38 50.32 &Ophiuchus \\
G003.73+18.30     & 3.7353513    & 18.308161        & 16 48 54.69  & -15 36 02.09  & 16 46 03.29  &  -15 30 50.23 &Ophiuchus \\
G004.02+16.64     & 4.0209961    & 16.646044        & 16 55 10.45  & -16 21 23.27  & 16 52 17.98  &  -16 16 37.51 &         \\
G004.19+18.09     & 4.1967773    & 18.092196        & 16 50 42.47  & -15 22 25.13  & 16 47 51.29  &  -15 17 20.75 &         \\
G004.15+35.77     & 4.152832     & 35.777241        & 15 53 29.82  & -04 38 52.39  & 15 50 51.56  &  -04 30 01.93 &High Glat  \\
G004.17+36.67     & 4.1748047    & 36.678967        & 15 50 42.66  & -04 04 20.84  & 15 48 05.00  &  -03 55 20.10 &High Glat  \\
G004.41+15.90     & 4.4165034    & 15.907708        & 16 58 35.99  & -16 28 36.07  & 16 55 43.28  &  -16 24 04.69 &         \\
G004.46+16.64     & 4.4604487    & 16.646044        & 16 56 11.73  & -16 00 51.08  & 16 53 19.65  &  -15 56 09.62 &         \\
G004.54+36.74     & 4.5483394    & 36.748764        & 15 51 14.27  & -03 47 40.14  & 15 48 36.88  &  -03 38 41.35 &High Glat  \\
G004.81+37.02     & 4.8120112    & 37.028599        & 15 50 52.79  & -03 27 20.38  & 15 48 15.74  &  -03 18 20.28 &High Glat  \\
G004.92+17.95     & 4.9218745    & 17.954901        & 16 52 50.41  & -14 53 40.51  & 16 49 59.75  &  -14 48 45.05 &         \\
G005.03+19.07     & 5.0317378    & 19.076056        & 16 49 19.74  & -14 09 20.80  & 16 46 30.04  &  -14 04 10.73 &         \\
G005.29+11.07     & 5.2954097    & 11.072874        & 17 17 19.97  & -18 30 57.71  & 17 14 24.31  &  -18 27 45.91 &         \\
G005.29+14.47     & 5.2954097    & 14.477516        & 17 05 31.09  & -16 36 07.96  & 17 02 38.08  &  -16 32 05.81 &         \\
G005.31+10.78     & 5.3173823    & 10.78794         & 17 18 23.02  & -18 39 22.30  & 17 15 27.16  &  -18 36 15.00 &         \\
G005.69+36.84*    & 5.6909175    & 36.84193         & 15 53 11.88  & -03 00 56.50  & 15 50 35.24  &  -02 52 04.98 &High Glat  \\
G005.80+19.92     & 5.8007808    & 19.926863        & 16 48 13.56  & -13 04 14.26  & 16 45 25.16  &  -12 58 59.65 &         \\
G006.08+20.26     & 6.0864253    & 20.264484        & 16 47 44.35  & -12 39 21.78  & 16 44 56.44  &  -12 34 05.17 &         \\
G006.04+36.74*    & 6.04248      & 36.748764        & 15 54 10.81  & -02 50 56.32  & 15 51 34.33  &  -02 42 08.46 &High Glat  \\
G006.32+20.44     & 6.3281245    & 20.443523        & 16 47 40.85  & -12 22 03.36  & 16 44 53.28  &  -12 16 46.52 &         \\
G006.41+20.56     & 6.4160151    & 20.562996        & 16 47 28.67  & -12 13 52.53  & 16 44 41.26  &  -12 08 34.85 &         \\
G006.70+20.66     & 6.7016597    & 20.66263         & 16 47 46.71  & -11 57 20.05  & 16 44 59.61  &  -11 52 03.63 &         \\
G006.96+00.89     & 6.965332     & 0.895288         & 17 57 59.28  & -22 29 20.42  & 17 54 57.90  &  -22 29 05.02 &4th Quad  \\
G006.94+05.84     & 6.9433589    & 5.8479061        & 17 39 41.49  & -19 58 05.07  & 17 36 43.62  &  -19 56 29.92 &4th Quad  \\
G006.98+20.72     & 6.9873042    & 20.722441        & 16 48 12.55  & -11 42 10.17  & 16 45 25.73  &  -11 36 55.55 &         \\
G007.14+05.94     & 7.1411128    & 5.9416566        & 17 39 47.47  & -19 45 05.14  & 17 36 49.87  &  -19 43 30.43 &4th Quad  \\
G007.53+21.10     & 7.5366206    & 21.101799        & 16 48 08.77  & -11 03 53.56  & 16 45 22.69  &  -10 58 38.70 &         \\
G007.80+21.10     & 7.8002925    & 21.101799        & 16 48 43.20  & -10 51 47.54  & 16 45 57.35  &  -10 46 35.08 &         \\
\enddata
\end{deluxetable}

\clearpage
\begin{deluxetable}{ccccccccccccccccccccccccc}
\setlength{\tabcolsep}{0.02in} \rotate
\tabletypesize{\scriptsize}  \tablecaption{Molecular complexes associated with the cold clumps.}
 \tablewidth{0pt} \tablehead{
region& l & b &Number & T$_{12}$ & T$_{13}$ &  T$_{18}$ & FWHM(12) & FWHM(13) & FWHM(18) & N$_{H_{2}}$ &  T$_{ex}$  & $\tau$(13)   & X13/X18  &$\sigma_{NT}$ &$\sigma_{Therm}$ &$\sigma_{3D}$ \\
      &[$^{\arcdeg}$,$^{\arcdeg}$]   &[$^{\arcdeg}$,$^{\arcdeg}$]   &       &(K)& (K)& (K) &(km~s$^{-1}$) &(km~s$^{-1}$)&(km~s$^{-1}$)  & (10$^{21}$ cm$^{-2}$) &(K) &   & & (km~s$^{-1}$) & (km~s$^{-1}$)& (km~s$^{-1}$)   }
\startdata
1st Quad       &[12,100] &[-10,10] & 84   & 2.69(1.20) & 1.85(0.77) & 0.96(0.72) & 2.70(2.23) & 1.73(1.34) & 0.96(0.72)  &5.7(5.1)  & 9.4(2.4 )   &1.1(0.9) & 5.6(3.3)   &0.75(0.60)    &   0.17(0.02)    &     1.34(1.02)     \\
2nd Quad       &[98,180] &[-4,10] & 70   & 2.57(1.23) & 1.67(0.77) & 0.85(0.51) & 2.09(1.18) & 1.42(0.68) & 0.85(0.51)  &4.5(3.4)  & 9.5(2.1 )   &0.9(0.5) & 7.6(3.3)   &0.62(0.29)    &    0.17(0.02)    &     1.13(0.49)     \\
3th Quad       &[180,279] &[-4,10] & 43   & 2.57(1.11) & 1.51(0.64) & 1.05(0.40) & 2.90(1.00) & 1.74(0.67) & 1.05(0.40)  &4.7(2.6)  & 9.1(2.0 )   &0.7(0.3) & 7.0(2.9)   &0.73(0.28)    &   0.17(0.02)    &     1.31(0.47)     \\
4th Quad and Ctr &[300,15] &[-8,8] & 6    & 3.09(0.44) & 1.54(0.59) & 2.81(5.01) & 3.03(2.57) & 1.83(1.96) & 2.81(5.01)  &4.2(2.6)  & 9.5(0.9 )   &0.8(0.4) & 4.4(2.3)   &0.77(0.83)    &   0.17(0.01)    &     1.39(1.42)     \\
Anticenter     &[175,210] &[-9,7] & 16   & 2.95(0.85) & 1.90(0.77) & 0.80(0.44) & 2.63(1.09) & 1.47(0.61) & 0.80(0.44)  &4.7(2.1)  & 9.3(1.7 )   &0.9(0.3) & 7.6(1.1)   &0.67(0.24)    &    0.17(0.02)    &     1.20(0.39)     \\
Aquila South   &[27,40] &[-21,-10] & 2    & 4.69(0.65) & 2.78(0.63) & 0.34(0.03) & 0.83(0.16) & 0.58(0.09) & 0.34(0.03)  &3.1(0.4)  & 12.8(1.3 )  &0.9(0.1) & 9.0(0.2)   &0.24(0.04)    &   0.20(0.01)    &     0.54(0.05)     \\
Cephens        &[99,143] &[8,22] & 87   & 2.60(0.93) & 1.66(0.64) & 0.69(0.29) & 1.94(0.87) & 1.13(0.43) & 0.69(0.29)  &3.5(1.9)  & 8.8(1.8 )   &1.0(0.5) & 5.4(2.1)   &0.48(0.18)    &     0.16(0.02)    &     0.88(0.30)     \\
High Glat      &         &$|$b$|\geq$25 & 41   & 3.30(1.12) & 1.84(0.97) & 0.46(0.18) & 1.47(0.66) & 0.89(0.40) & 0.46(0.18)  &3.0(2.0)  & 10.5(2.0 )  &0.8(0.5) &11.6(5.5)   &0.37(0.17)    &  0.18(0.02)    &     0.72(0.26)     \\
Oph-Sgr        &[8,40] &[9,24] & 9    & 3.23(1.49) & 2.11(1.08) & 0.42(0.16) & 1.30(0.46) & 0.81(0.29) & 0.42(0.16)  &3.0(1.6)  & 10.2(2.7 )  &0.9(0.5) & 6.1(2.9)   &0.34(0.13)    &   0.18(0.02)    &     0.67(0.20)     \\
Ophiuchs       &[344,4] &[7,25] & 6    & 5.84(0.56) & 3.89(0.40) & 0.42(0.11) & 1.10(0.23) & 0.70(0.16) & 0.42(0.11)  &5.9(2.0)  & 15.0(1.2 )  &1.1(0.1) & 7.0(5.6)   &0.29(0.07)    &  0.21(0.01)    &     0.63(0.10)     \\
Orion          &[180,225] &[-25,5] & 82   & 3.58(1.90) & 2.19(1.01) & 0.76(0.35) & 2.12(1.02) & 1.29(0.53) & 0.76(0.35)  &5.8(5.3)  & 11.1(4.0 )  &1.0(0.6) & 8.1(5.1)   &0.55(0.23)    &   0.18(0.03)    &     1.02(0.38)     \\
Taurus         &[152,180] &[-25,-3] & 153  & 3.15(1.32) & 2.08(0.86) & 0.62(0.26) & 1.67(0.78) & 1.07(0.42) & 0.62(0.26)  &4.2(2.9)  & 10.3(2.6 )  &1.0(0.7) & 6.2(3.4)   &0.45(0.18)    &  0.18(0.02)    &     0.85(0.29)     \\
other          & & & 75   & 3.45(1.61) & 1.97(1.22) & 0.52(0.21) & 1.51(0.68) & 0.92(0.45) & 0.52(0.21)  &3.4(2.4)  & 10.5(3.3 )  &0.9(0.7) & 7.4(4.4)   &0.38(0.20)    &    0.18(0.03)    &    0.75(0.31)     \\
\enddata
\end{deluxetable}

\clearpage

\begin{deluxetable}{cccccccccccccccccccccccccc}
 \setlength{\tabcolsep}{0.05in}
\tabletypesize{\scriptsize}  \tablecaption{Observed line parameters (the first page of 32 pages)}
 \tablewidth{0pt} \tablehead{
  Name &  V$_{lsr}$(12)  & FWHM(12)  & T$_{A}$(12) &  V$_{lsr}$(13)  & FWHM(13)  & T$_{A}$(13) &  V$_{lsr}$(18)  & FWHM(18)  & T$_{A}$(18)\\
      &(km~s$^{-1}$) &(km~s$^{-1}$) &(K)&(km~s$^{-1}$) &(km~s$^{-1}$) &(K)&(km~s$^{-1}$) &(km~s$^{-1}$) &(K)}
\startdata
G001.38+20.94&0.61(0.01)&1.16(0.02)&5.67(0.18)&0.7(0.01)&0.78(0.01)&3.98(0.09)&0.74(0.01)&0.54(0.03)&1.26(0.09)&\\
G001.84+16.58&6.05(0.01)&1.32(0.03)&5.58(0.17)&5.87(0.01)&0.69(0.01)&4(0.09)&5.87(0.01)&0.38(0.02)&1.67(0.09)&\\
G003.73+16.39&6.44(0.01)&1.11(0.02)&5.56(0.17)&6.17(0.01)&0.71(0.01)&3.41(0.09)&6.17(0.02)&0.53(0.07)&0.77(0.09)&RA\\
G003.73+18.30&4.44(0.01)&1.36(0.02)&6.68(0.21)&4.37(0.01)&0.97(0.01)&4.34(0.08)&4.32(0.01)&0.43(0.02)&2.45(0.08)&\\
G004.02+16.64&5.92(0.01)&0.69(0.02)&5.5(0.18)&5.87(0)&0.49(0.01)&3.81(0.09)&5.86(0.02)&0.34(0.05)&0.71(0.08)&\\
G004.19+18.09&3.64(0.01)&1.63(0.02)&5.94(0.19)&3.5(0.01)&1.23(0.01)&4.12(0.09)&3.73(0.02)&0.44(0.05)&0.94(0.08)&Dob\\
G004.19+18.09&6.68(0.01)&0.88(0.02)&6.63(0.19)&6.69(0.01)&0.62(0.01)&3.16(0.09)&&&&\\
G004.15+35.77&2.47(0.01)&1.92(0.03)&4.72(0.14)&2.61(0.01)&1.06(0.02)&3.17(0.09)&&&&\\
G004.17+36.67&2.55(0.01)&0.93(0.02)&5.08(0.17)&2.47(0.01)&0.71(0.01)&3.27(0.08)&&&&\\
G004.41+15.90&4.78(0.01)&0.57(0.01)&6.11(0.15)&4.73(0)&0.48(0.01)&3.69(0.09)&&&&\\
G004.46+16.64&5.16(0.01)&1.66(0.02)&6.1(0.15)&5.34(0.01)&1.19(0.02)&3.67(0.09)&5.55(0.01)&0.43(0.03)&1.42(0.09)&BA$^{+}$\\
G004.54+36.74&2.46(0.01)&1.17(0.02)&4.5(0.17)&2.53(0.01)&0.76(0.02)&2.94(0.09)&2.54(0.04)&0.44(0.1)&0.44(0.09)&\\
G004.81+37.02&3.6(0.01)&1.51(0.02)&4.47(0.13)&3.71(0.01)&1.04(0.02)&2.91(0.09)&3.82(0.04)&0.66(0.09)&0.46(0.08)&\\
G004.92+17.95&3.54(0.02)&1.49(0.03)&3.77(0.18)&3.72(0.01)&0.63(0.03)&3.26(0.17)&3.71(0.01)&0.32(0.03)&0.98(0.08)&\\
G005.03+19.07&3.83(0.01)&1.25(0.02)&7.17(0.19)&3.77(0.01)&0.99(0.01)&3.86(0.09)&3.81(0.03)&0.65(0.07)&0.73(0.09)&\\
G005.29+11.07&4.37(0.01)&0.92(0.01)&7.55(0.18)&4.22(0)&0.63(0.01)&4.07(0.09)&4.14(0.01)&0.34(0.04)&1.16(0.09)&\\
G005.29+14.47&3.76(0.01)&0.4(0.01)&5.07(0.18)&3.74(0.01)&0.3(0.02)&1.21(0.09)&&&&\\
G005.31+10.78&4.53(0.01)&1.24(0.02)&6.56(0.19)&4.38(0.01)&0.88(0.01)&4.1(0.1)&&&&\\
G005.69+36.84&0.56(0.03)&1.39(0.06)&3.84(0.17)&0.6(0.03)&0.96(0.07)&0.81(0.09)&&&&Dob\\
G005.69+36.84&2.25(0.04)&1.62(0.08)&3.4(0.17)&2.33(0.01)&0.9(0.02)&3.01(0.09)&2.27(0.01)&0.4(0.02)&1.46(0.09)&\\
G005.80+19.92&3.45(0.01)&1.41(0.02)&6.11(0.18)&3.06(0.01)&0.89(0.01)&3.87(0.09)&2.96(0.02)&0.47(0.05)&0.99(0.09)&\\
G006.08+20.26&4.03(0.01)&1.95(0.02)&5.23(0.16)&3.96(0.01)&0.91(0.01)&4.19(0.08)&3.97(0.01)&0.56(0.03)&1.16(0.08)&BA$^{+}$\\
G006.04+36.74&2.34(0.02)&1.99(0.04)&3.61(0.17)&2.61(0.01)&0.94(0.02)&2.73(0.1)&2.53(0.01)&0.47(0.03)&1.47(0.09)&\\
G006.32+20.44&4.48(0.01)&1.6(0.02)&4.98(0.16)&4.32(0.01)&0.72(0.01)&4.25(0.09)&4.27(0.01)&0.38(0.02)&1.88(0.09)&\\
G006.41+20.56&4.54(0.01)&1.38(0.02)&5.9(0.16)&4.39(0.01)&0.95(0.01)&4.43(0.08)&4.44(0.02)&0.67(0.04)&1.37(0.09)&\\
G006.70+20.66&3.49(0.01)&2(0.03)&4.44(0.15)&3.76(0.01)&1.06(0.01)&3.97(0.08)&3.79(0.01)&0.66(0.04)&1.35(0.09)&\\
G006.96+00.89&9.78(0.05)&4.1(0.12)&2.66(0.23)&9.33(0.05)&2.17(0.15)&0.88(0.1)&&&&Dob\\
G006.96+00.89&41.34(0.06)&9.42(0.15)&3.11(0.23)&41.53(0.1)&6.89(0.24)&0.77(0.1)&41.67(0.63)&13.03(2.05)&0.17(0.1)&\\
G006.94+05.84&5.22(0.02)&1.57(0.04)&3.26(0.19)&4.86(0.02)&0.63(0.06)&1.21(0.1)&&&&Dob\\
G006.94+05.84&10.59(0.02)&2.57(0.05)&3.38(0.19)&10.08(0.02)&1.39(0.03)&2.22(0.1)&10.05(0.03)&0.83(0.07)&0.77(0.09)&\\
\enddata
\end{deluxetable}

\clearpage

\begin{deluxetable}{ccrrrrrrrrrrrrrrrcrl}
\setlength{\tabcolsep}{0.05in} \rotate
\tabletypesize{\scriptsize}  \tablecaption{Derived parameters of Surveyed ECC clumps (The first page of 40 pages)}
 \tablewidth{0pt} \tablehead{
  Name &  V$_{lsr}$  & R  & D & Z &  T$_{ex}$  & $\tau$(13)  & N(13) &  $\tau$(18)  & N(18)  & N$_{H_{2}}$ & ratio(12/13) &X13/X18 &$\sigma_{NT}$  &$\sigma_{Therm}$ &$\sigma_{3D}$\\
       &  (km~s$^{-1}$) &(kpc)&(kpc) &(kpc) &(K)& & (10$^{15}$ cm$^{-2}$) & & (10$^{15}$ cm$^{-2}$) & (10$^{21}$ cm$^{-2}$) & &   (km~s$^{-1}$) &  (km~s$^{-1}$)&  (km~s$^{-1}$) }
\startdata
G001.38+20.94  & 0.74  &  7.47  &   1.1   &    0.39   & 14.8  & 1.2  & 7.3    & 0.2   & 1.6    & 6.5  & 2.1   & 4.6  &0.32   &0.21   & 0.67  \\
G001.84+16.58  & 5.87  &  4.51  &  4.16   &    1.19   & 14.6  & 1.2  & 6.5    & 0.4   & 1.5    & 5.7  & 2.7   & 4.3  &0.29   &0.21   & 0.61  \\
G003.73+16.39  & 6.17  &  5.87  &  2.75   &    0.78   & 14.6  & 0.9  & 5.7    & 0.1   & 1.0    & 5.0  & 2.5   & 5.9  &0.29   &0.21   & 0.63  \\
G003.73+18.30  & 4.32  &  6.48  &  2.14   &    0.67   & 16.8  & 1.0  & 10.6   & 0.5   & 2.7    & 9.4  & 2.2   & 4.0  &0.41   &0.23   & 0.80  \\
G004.02+16.64  & 5.86  &  6.10  &  2.52   &    0.72   & 14.4  & 1.2  & 4.3    & 0.1   & 0.6    & 3.9  & 2.0   & 7.7  &0.20   &0.21   & 0.50  \\
G004.19+18.09  & 3.73  &  6.88  &  1.71   &    0.53   & 15.3  & 1.2  & 12.1   & 0.2   & 1.0    & 10.8 & 1.9   & 12.2 &0.52   &0.22   & 0.97  \\
G004.19+18.09  & 6.69  &  5.92  &  2.73   &    0.85   & 16.7  & 0.6  & 4.9    &       &        & 4.4  & 3.0   &      &0.25   &0.23   & 0.59  \\
G004.15+35.77  & 2.61  &  7.13  &   1.7   &    0.99   & 12.9  & 1.1  & 7.4    &       &        & 6.6  & 2.7   &      &0.45   &0.20   & 0.85  \\
G004.17+36.67  & 2.47  &  7.19  &  1.64   &    0.98   & 13.6  & 1.0  & 5.2    &       &        & 4.7  & 2.0   &      &0.29   &0.20   & 0.62  \\
G004.41+15.90  & 4.73  &  6.62  &  1.96   &    0.54   & 15.7  & 0.9  & 4.3    &       &        & 3.8  & 2.0   &      &0.19   &0.22   & 0.50  \\
G004.46+16.64  & 5.55  &  6.37  &  2.23   &    0.64   & 15.6  & 0.9  & 10.6   & 0.3   & 1.5    & 9.4  & 2.3   & 7.1  &0.50   &0.22   & 0.95  \\
G004.54+36.74  & 2.54  &  7.25  &  1.56   &    0.93   & 12.4  & 1.0  & 4.9    & 0.1   & 0.4    & 4.3  & 2.4   & 11.5 &0.32   &0.19   & 0.64  \\
G004.81+37.02  & 3.82  &  6.79  &  2.15   &    1.30   & 12.3  & 1.0  & 6.6    & 0.1   & 0.7    & 5.9  & 2.2   & 9.9  &0.44   &0.19   & 0.83  \\
G004.92+17.95  & 3.71  &  7.10  &  1.48   &    0.46   & 10.9  & 2.0  & 4.3    & 0.3   & 0.7    & 3.8  & 2.7   & 6.5  &0.26   &0.18   & 0.55  \\
G005.03+19.07  & 3.81  &  7.08  &   1.5   &    0.49   & 17.8  & 0.8  & 9.9    & 0.1   & 1.2    & 8.8  & 2.3   & 8.0  &0.41   &0.23   & 0.82  \\
G005.29+11.07  & 4.14  &  7.09  &  1.44   &    0.28   & 18.6  & 0.8  & 6.8    & 0.2   & 1.1    & 6.1  & 2.7   & 6.5  &0.26   &0.24   & 0.61  \\
G005.29+14.47  & 3.74  &  7.20  &  1.35   &    0.34   & 13.6  & 0.3  & 0.8    &       &        & 0.7  & 5.6   &      &0.11   &0.20   & 0.40  \\
G005.31+10.78  & 4.38  &  7.03  &  1.51   &    0.28   & 16.6  & 1.0  & 9.0    &       &        & 8.0  & 2.3   &      &0.37   &0.22   & 0.75  \\
G005.69+36.84  & 0.6   &  8.33  &  0.21   &    0.13   & 11.1  & 0.2  & 1.6    &       &        & 1.4  & 6.9   &      &0.40   &0.18   & 0.77  \\
G005.69+36.84  & 2.27  &  7.60  &  1.13   &    0.68   & 10.2  & 2.1  & 5.5    & 0.6   & 1.2    & 4.9  & 2.0   & 4.6  &0.38   &0.18   & 0.72  \\
G005.80+19.92  & 2.96  &  7.52  &  1.05   &    0.36   & 15.7  & 1.0  & 8.3    & 0.2   & 1.1    & 7.4  & 2.5   & 7.4  &0.37   &0.22   & 0.75  \\
G006.08+20.26  & 3.97  &  7.26  &  1.33   &    0.46   & 13.9  & 1.6  & 8.7    & 0.2   & 1.5    & 7.7  & 2.7   & 5.9  &0.38   &0.21   & 0.75  \\
\enddata
\end{deluxetable}

\clearpage

\begin{deluxetable}{ccrrrrrrrrrrrrrrrrrrcrl}
\setlength{\tabcolsep}{0.05in}
\tabletypesize{\scriptsize}  \tablecaption{Sources with blue- or red- line profiles}
 \tablewidth{0pt} \tablehead{
   Name& V13 & FWHM13 &V18  &FWHM18 &V12\_peak  &$\delta V$\_13\tablenotemark{a}  &$\delta V$\_18\tablenotemark{b}  &T12\_B/T12\_R & $\Delta$T\tablenotemark{c}& Profile & region\\
      &(km~s$^{-1}$)  &(km~s$^{-1}$)  &(km~s$^{-1}$)  &(km~s$^{-1}$)  &(km~s$^{-1}$) }
\startdata
G004.46+16.64    &    5.34(0.01) & 1.19(0.02)  &   5.55(0.01) & 0.43(0.03) & 4.95    &  -0.33   &-1.40   &1.42   &1.88  & blue profile & others\\
G006.08+20.26    &    3.96(0.01) & 0.91(0.01)  &   3.97(0.01) & 0.56(0.03) & 3.25    &  -0.78   &-1.29   &1.09   &0.42  & blue profile & others\\
G089.03-41.28    &   -4.63(0.03) & 1.93(0.06)  &              &            & -5.37   &  -0.38   & 0.63  &1.23   &0.63  & blue profile &High Glat\\
G097.09+10.12    &   -2.71(0.03) & 1.09(0.05)  &  -2.67(0.04) & 0.63(0.09) & -3.33   &  -0.57   &-1.05   &1.57   &1.01  & blue profile &others\\
G111.33+19.94    &   -7.21(0.01) & 0.83(0.03)  &  -7.24(0.02) & 0.58(0.05) & -8.13   &  -1.11   &-1.53   &1.21   &0.5   & blue profile &Cephens\\
G111.77+13.78    &   -3.81(0.01) & 0.85(0.03)  &  -3.68(0.05) & 0.46(0.15) & -4.29   &  -0.56   &-1.33   &2.18   &1.69  & blue profile &Cephens\\
G111.77+20.26    &   -8.06(0.01) & 1.15(0.03)  &  -8.09(0.04) & 0.88(0.11) & -8.62   &  -0.49   &-0.60   &1.84   &1.55  & blue profile &Cephens\\
G111.97+20.52    &   -7.92(0.02) & 0.9 (0.05)  &  -7.91(0.03) & 0.48(0.08) & -8.62   &  -0.78   &-1.48   &1.69   &0.86  & blue profile &Cephens\\
G117.11+12.42    &    3.87(0.01) & 0.96(0.03)  &   3.92(0.02) & 0.53(0.06) &  3.32   &  -0.57   &-1.13   & 1.25  &0.78  & blue profile &Cephens\\
G126.49-01.30    &  -11.94(0.02) & 1.9 (0.04)  & -11.74(0.04) & 0.91(0.09) & -13     &  -0.56   &-1.38   &1.19   &0.6   & blue profile &2nd Quad\\
G130.14+13.78    &   -1.53(0.02) & 1.46(0.06)  &  -1.43(0.05) & 0.69(0.14) & -2.55   &  -0.70   &-1.62   &1.11   &0.23  & blue profile &Cephens\\
G155.45-14.59    &    1.97(0.02) & 1.39(0.05)  &   2.13(0.03) & 0.61(0.1)  & 0.83    &  -0.82   &-2.13   &2.50   &1.93  & blue profile &Taurus\\
G158.37-20.72    &    7.13(0.01) & 2.35(0.02)  &   7.05(0.03) & 1.4(0.08)  & 6.03    &  -0.47   &-0.73   &1.11   &0.80 & blue profile &Taurus\\
G164.94-08.57    &    -0.7(0.01) & 1.71(0.03)  &  -0.63(0.05) & 1.31(0.11) & -1.66   &  -0.56   &-0.79   &1.73   &1.89  & blue profile &Taurus\\
G181.84+00.31    &     3.3(0.02) & 1.49(0.04)  &   3.49(0.05) & 0.81(0.12) & 2.21    &  -0.73   &-1.58   &1.29   &0.74  & blue profile &3th Quad\\
G190.17-13.78    &    1.14(0.01) & 1.39(0.02)  &              &            & 0.74    &  -0.29   &       &1.12   &0.54  & blue profile &Orion \\
G210.01-20.16    &     8.1(0.02) & 1.83(0.04)  &   8.33(0.09) & 0.68(0.22) & 7.27    &  -0.45   &-1.56  &1.25   &0.99  & blue profile &Orion \\
G224.27-00.82    &   14.39(0.02) & 3.04(0.04)  &  14.45(0.02) & 1.61(0.05) & 13.09   &  -0.43   &-0.84   &1.28   &0.7   & blue profile &3th Quad\\
\cline{1-12}
G093.22-04.59    &    3.87(0.01) & 1.51(0.03)  &    3.9(0.06) & 1.14(0.15) &  4.41   &   0.36   & 0.45  &1.1    &0.29  & red profile  &1st Quad\\
G101.14-15.28    &   -9.58(0.01) & 0.98(0.03)  &              &            & -9.18   &  0.41        & 0.29  &0.92   &0.29  & red profile &others \\
G121.92-01.71    &  -14.07(0.01) & 1.39(0.03)  & -14.12(0.08) & 1.11(0.16) & -13.46  &  0.44      &0.59      &0.87   &0.5   & red profile &2nd Quad\\
G145.81+10.97    &  -14.94(0.02) & 1.06(0.04)  &              &              & -14.31  &  0.59      & 0.29  &0.90   &0.29  & red profile &others\\
G157.10-08.70    &   -7.48(0.02) & 1.43(0.04)  &  -7.48(0.03) & 0.9(0.08)  & -6.98   &  0.35      &-1.40     &0.55   &1.8   & red profile &Taurus\\
G169.84-07.61    &    6.46(0.01) & 0.67(0.02)  &   6.39(0.01) & 0.31(0.03) & 6.83    &  0.55      & 1.42     &0.90   &0.35  & red profile &Taurus\\
G172.85+02.27    &  -17.22(0.02) & 3.18(0.06)  & -17.49(0.07) & 1.47(0.17) & -15.83  &  0.44      & 1.13     &0.90   &0.45  & red profile &2nd Quad\\
G182.15-17.95    &     9.4(0.01) & 0.86(0.02)  & 9.35(0.03)   &0.53(0.06)  & 10.11   &  0.83      &-0.64     &0.71   &1.26  & red profile &Orion\\
G182.54-25.34    &    1.06(0.05) & 1.45(0.1)   &              &            & 1.78    &  0.50        & 0.87  &0.74   &0.87  & red profile &High Glat\\
G188.04-03.71    &    3.12(0.03) & 1.39(0.06)  & 3.15(0.08)   & 0.9(0.16)  & 4.03    &  0.65      &-1.02     &0.73   &0.52  & red profile &3th Quad\\
G192.28-11.33    &   10.19(0.01) & 1.47(0.01)  &  10.13(0.04) & 0.84(0.11  & 10.81   &  0.42      & 0.81     &0.75   &1.61  & red profile &Orion\\
G195.09-16.41    &   -0.08(0.01) & 1.84(0.02)  &   -0.1(0.05) &  1.3(0.13) & 0.47    &  0.30      &-0.47     &0.89   &0.77  & red profile &Orion\\
G210.67-36.77    &    0.47(0.01) &  0.6(0.02)  &   0.45(0.06) & 0.24(0.09) & 0.81    &  0.57      &1.50      &0.84   &0.93  & red profile &High Glat\\
G216.69-13.88    &    8.72(0.02) & 1.88(0.06)  &   8.81(0.07) &  0.8(0.13) & 9.71    &  0.53      &1.13      &0.91   &0.32  & red profile &Orion\\
G219.35-09.70    &    12.5(0.02) & 1.65(0.04)  &  12.45(0.03) & 0.93(0.07) & 13.41   &  0.55      &1.03      &0.67   &0.76  & red profile &Orion
\enddata
\tablenotetext{a}{$\delta V$\_13=(V12\_peak-V13)/FWHM13}\tablenotetext{b}{$\delta V$\_18=(V12\_peak-V18)/FWHM18}\tablenotetext{c}{$\Delta$T=abs(T12\_b-T12\_r)}
\end{deluxetable}

\clearpage
\begin{deluxetable}{ccrrrrrrrrrrrrrrrrrrcrl}
\setlength{\tabcolsep}{0.05in}
\tabletypesize{\scriptsize}  \tablecaption{Sources with blue- or red- line asymmetry}
 \tablewidth{0pt} \tablehead{
   Name & FWHM13  &FWHM18 & V12 &V12\_peak  &$\delta V(12)$\_13\tablenotemark{a}  &$\delta V(12)$\_18\tablenotemark{b}  & Profile & region\\
      &(km~s$^{-1}$)  &(km~s$^{-1}$)  &(km~s$^{-1}$)  &(km~s$^{-1}$)  }
\startdata
G018.39+19.39     & 0.52(0.01)     & 0.27(0.04)  & -0.46(0.01)  & -0.61    & -0.29   &  -0.56    & blue asymmetry &Oph-Sgr\\
G027.70-21.02     & 0.47(0.04)     &             & 4.95(0.02)   & 4.78     & -0.36   &           & blue asymmetry \\
G108.10+13.19     & 1.04(0.04)     & 0.43(0.03)  & -4.81(0.03)  & -5.17    & -0.35   &  -0.84    & blue asymmetry &Cephens\\
G111.66+20.20     & 1.34(0.03)     & 0.79(0.09)  & -8.17(0.03)  & -8.64    &-0.35    &-0.59      & blue asymmetry &Cephens\\
G143.85+11.49     & 1.63(0.08)     &             &-13.78(0.06)  & -14.3    &-0.32    &           & blue asymmetry &others\\
G147.96-08.02     & 0.63(0.1)      &             &-28.39(0.04)  & -28.59   &-0.32    &           & blue asymmetry &others\\
G156.90-08.49     & 1.92(0.06)     & 0.53(0.05)  & -7.72(0.04)  & -8.56    &-0.44    &-1.58      & blue asymmetry &Taurus\\
G158.24-21.80     & 1.63(0.03)     & 1.13(0.13)  &  4.09(0.02)  & 3.45     &-0.39    &-0.57      & blue asymmetry &Taurus\\
G158.40-21.86     & 1.21(0.03)     & 0.92(0.1)   &  4.54(0.01)  & 4.13     &-0.34    &-0.45      & blue asymmetry &Taurus\\
G159.65+11.39     & 1.11(0.02)     & 0.99(0.27)  &  2.61(0.01)  & 1.9      &-0.64    &-0.72      & blue asymmetry &others\\
G171.43-17.36     & 1.16(0.02)     &  0.4(0.03)  &  7.47(0.02)  & 6.99     &-0.41    &-1.20      & blue asymmetry &Taurus\\
G172.92-16.74a    &  0.7(0.02)     & 0.51(0.08)  &  5.06(0.02)  & 4.9      &-0.23    &-0.31      & blue asymmetry? &Taurus\\
G172.92-16.74b    & 0.87(0.02)     & 0.58(0.05)  &  6.96(0.02)  & 6.49     &-0.54    &-0.81      & blue asymmetry &Taurus\\
G173.18-09.12     &  0.5(0.02)     & 0.51(0.09)  &  6.23(0.02)  & 6.03     &-0.40    &-0.39      & blue asymmetry &Taurus\\
G179.14-06.27     & 0.52(0.02)     &             &  7.69(0.02)  & 7.45     &-0.46    &           & blue asymmetry &Anticenter\\
G182.02-00.16     & 1.28(0.05)     & 0.89(0.18)  &  3.91(0.02)  & 3.43     &-0.38    &-0.54      & blue asymmetry &3th Quad\\
G190.08-13.51     &  1.62(0.03)    &             &   1.11(0.02) & 0.64     & -0.29   &           & blue asymmetry &Orion \\
G196.21-15.50     &  2.29(0.08)    &             &   3.51(0.02) & 2.88     & -0.28   &           & blue asymmetry &Orion \\
G227.70+11.18     &  0.59(0.14)    &             &  14.98(0.02) & 14.78    & -0.34   &           & blue asymmetry &others \\
\cline{1-9}
G003.73+16.39     & 0.71(0.01)     & 0.53(0.07)  &  6.44(0.01)  & 6.67     &0.32     &0.43       & red asymmetry &Ophiuchus \\
G118.34+08.66     & 0.72(0.05)     & 0.41(0.14)  & -2.07(0.03)  &-1.79     &0.39     &0.68       & red asymmetry &Cephens\\
G118.36+21.74     & 0.86(0.06)     &             & -3.88(0.04)  &-3.5      &0.44     &           & red asymmetry &Cephens\\
G142.25+05.43     & 0.97(0.03)     &             &-10.18(0.01)  &-9.85     &0.34     &           & red asymmetry &2nd Quad\\
G142.62+07.29     & 0.66(0.03)     & 0.36(0.11)  &-11.42(0.02)  &-10.97    &0.68     &1.25       & red asymmetry &2nd Quad\\
G157.19-08.81     & 1.29(0.03)     & 0.74(0.06)  &    -7(0.02)  &-6.49     &0.40     &0.69       & red asymmetry &Taurus\\
G157.25-01.00     & 0.67(0.01)     & 0.36(0.02)  &  5.27(0.01)  &5.5       &0.34     &0.64       & red asymmetry &2nd Quad\\
G157.58-08.89     & 1.07(0.04)     & 0.41(0.03)  & -6.07(0.03)  &-5.37     &0.65     &1.71       & red asymmetry &Taurus \\
G159.58-32.84     & 1.38(0.07)     &             &  3.79(0.02)  &4.43      &0.46     &           & red asymmetry &High Glat\\
G174.50-19.88     & 0.81(0.02)     & 0.56(0.09)  &   7.7(0.01)  &7.96      &0.32     &0.46       & red asymmetry &Taurus\\
G175.16-16.74     & 1.11(0.03)     &  0.8(0.1)   &   5.7(0.03)  &6.23      &0.48     &0.66       & red asymmetry &2nd Quad\\
G195.00-16.95     &  1.1(0.02)     & 0.51(0.04)  & -1.99(0.01)  &-1.76     &0.21       &0.45       & red asymmetry &Orion \\
G221.46-17.89     & 0.46(0.03)     & 0.17(0.66)  &  7.99(0.03)  &8.12      &0.28     &0.76       & red asymmetry &Orion \\
\enddata
\tablenotetext{a}{$\delta V(12)$\_13=(V12\_peak-V12)/FWHM13}\tablenotetext{b}{$\delta V(12)$\_18=(V12\_peak-V12)/FWHM18}
\end{deluxetable}

\begin{deluxetable}{ccccccccccccccccccccccccc}
\setlength{\tabcolsep}{0.05in} \rotate
\tabletypesize{\scriptsize}  \tablecaption{A statistical analysis of parameters}
 \tablewidth{0pt} \tablehead{
 Stat & V$_{12}$-V$_{13}$ & V$_{13}$-V$_{18}$ &  & T$_{12}$& T$_{12} /$ T$_{13}$ & T$_{13} /$ T$_{18}$ & FWHM(12) & FWHM(13) & FWHM(18) & N$_{H_{2}}$ &  T$_{ex}$  & $\tau$(13)   & X13/X18  &$\sigma_{NT}$ &$\sigma_{Therm}$ &$\sigma_{3D}$ &$\sigma_{NT} / \sigma_{Therm}$\\
       &(km~s$^{-1}$) &(km~s$^{-1}$) & &(K)& & &(km~s$^{-1}$) &(km~s$^{-1}$)&(km~s$^{-1}$)  & (10$^{21}$ cm$^{-2}$) &(K) &   & & (km~s$^{-1}$) & (km~s$^{-1}$)& (km~s$^{-1}$)&   }
\startdata
 \multicolumn{18}{c}{Statistics} \\
\cline{1-18} \\
number\tablenotemark{1}       & 782  &437   &  &904 & 782 &437   &904   &782   &437   &782  &782  &782  &437 &782 &782  &782  &782  \\
mean         & 0.02 &0     &  &3.08&2.15 & 3.88 &2.03  &1.27  &0.76  &4.4  &10.1 &0.93 &7.0 &0.53&0.17 &0.98 &3.09 \\
std          & 0.29 &0.16  &  &1.37&1.35 & 1.71 &1.28  &0.77  &0.73  &3.6  &2.6  &0.56 &3.8 &0.31&0.02 &0.51 &1.83 \\
\cline{1-18} \\
&   \multicolumn{2}{c}{normal distribution fit} & & \multicolumn{10}{c}{lognormal distribution fit} \\
\cline{2-3} \cline{5-18}\\
$\mu$        &0.02 &0    &  &1.02&0.65 &1.26 &0.56 &0.10 &-0.44 &1.22 &2.29 &-0.24 &1.83 &-0.77 &-1.76 &-0.12 &0.99\\
$\sigma$     &0.29 &0.16 &  &0.48&0.43 &0.45 &0.55 &0.51 &0.53  &0.78 &0.24 &0.62  &0.46 &0.51  &-0.12 &0.43  &0.53\\
P            &0    &0    &  &0   &0    &0.234&0.636&0.830&0.515 &0    &0.227&0     &0.388&0.889 &0.227 &0.374 &0.822    \\
\enddata
\tablenotetext{1}{Only the parameters of $^{13}$CO components are analyzed. But we also include those $^{12}$CO components without $^{13}$CO emission in statistics of T$_{12}$ and FWHM(12). The 39 clumps without suitable reference positions in observations are excluded in statistics.}
\end{deluxetable}

\begin{deluxetable}{ccccccccccccccccccccccccccccccccccc}
\tabletypesize{\scriptsize} \tablecolumns{9} \rotate
\tablewidth{0pc} \setlength{\tabcolsep}{0.03in}
\tablecaption{Parameters of the ten mapped clumps} \tablehead{
 Name & V$_{lsr}$& d & \multicolumn{2}{c}{Offset}  & \multicolumn{3}{c}{Deconvolved Size} & R &\colhead{T$_{ex}$} &\colhead{N$_{H_{2}}$} &\colhead{$\sigma_{Therm}$} &\colhead{$\sigma_{NT}$} &\colhead{$\sigma_{3D}$} & n  &\colhead{M$_{LTE}$} &\colhead{M$_{vir}$} &\colhead{M$_{J}$}  & \colhead{Group} &\colhead{Region}\\
 & km~s$^{-1}$& (kpc) & \multicolumn{2}{c}{($\arcsec,\arcsec$)} & \multicolumn{3}{c}{($\arcsec\times\arcsec$($\arcdeg$))}& (pc) & (K) & (10$^{21}$ cm$^{-2}$) & (km~s$^{-1}$)& (km~s$^{-1}$)& (km~s$^{-1}$) &(10$^{3}$ cm$^{-3}$) &\colhead{(M$_{\sun}$)} &\colhead{(M$_{\sun}$)} &\colhead{(M$_{\sun}$)}\\
}\startdata
G001.38+20.94      &0.74   &1.1   &\multicolumn{2}{c}{(-110,-41)} &\multicolumn{3}{c}{677$\times$428(67.9)}  &1.4   &14.1(0.9)   & 5.3(1.7) &0.21(0.01) &0.33(0.09) &0.67(0.13) & 0.6  & 751  & 150   &43       & L     &Ophiuchus   \\
G006.96+00.89a\_1  &9.33   &2.21  &\multicolumn{2}{c}{(5,-91)}    &\multicolumn{3}{c}{187$\times$119(-8.1)}  &0.8   & 9.9(0.4)   & 3.2(0.7) &0.17(0.00) &0.88(0.19) &1.56(0.32) & 0.6  & 141  & 453   &496      &H      &4th Quad     \\
G006.96+00.89a\_2  &9.33   &2.21  &\multicolumn{2}{c}{(-90,30)}   &\multicolumn{3}{c}{149$\times$102(-42.5)} &0.7   & 9.7(0.5)   & 2.6(0.9) &0.17(0.00) &0.70(0.20) &1.25(0.34) & 0.6  & 78   & 240   &260      &      &4th Quad    \\
G006.96+00.89a\_3  &9.33   &2.21  &\multicolumn{2}{c}{(-207,39)}  &\multicolumn{3}{c}{301$\times$64(-70.9)}  &0.7   & 9.4(0.3)   & 3.2(1.0) &0.17(0.00) &0.92(0.01) &1.61(0.17) & 0.7  & 122  & 449   &544      &      &4th Quad    \\
G006.96+00.89a\_4  &9.33   &2.21  &\multicolumn{2}{c}{(-334,66)}  &\multicolumn{3}{c}{268$\times$143(89.1)}  &1.0   & 8.9(0.7)   & 2.7(0.7) &0.16(0.01) &1.02(0.30) &1.79(0.51) & 0.4  & 204  & 783   &945      &      &4th Quad    \\
G006.96+00.89b     &41.67  &5.36  &\multicolumn{2}{c}{(-64,-1)}   &\multicolumn{3}{c}{203$\times$161(38.2)}  &2.3   &10.7(1.1)   & 6.8(2.6) &0.18(0.01) &1.52(0.22) &2.66(0.37) & 0.5  & 2579 & 3873  &2905      &H      &4th Quad    \\
G049.06-04.18      &9.93   &0.6   &\multicolumn{2}{c}{(16,41)   } &\multicolumn{3}{c}{239$\times$194(-51.9)} &0.3   & 8.9(1.2)   & 1.3(0.5) &0.16(0.01) &0.15(0.04) &0.39(0.05) & 0.7  & 9    & 11    &7         &  L    &1st Quad    \\
G089.64-06.59      &12.51  &0.6   &\multicolumn{2}{c}{(81,-20)  } &\multicolumn{3}{c}{320$\times$187(-0.9)}  &0.4   &10.2(0.9)   & 3.4(1.5) &0.18(0.01) &0.40(0.09) &0.74(0.15) & 1.5  & 30   & 45    &38       &  L    &1st Quad    \\
G108.85-00.80      &-49.51 &5.4   &\multicolumn{2}{c}{(-20,-12) } &\multicolumn{3}{c}{776$\times$213(45.1)}  &5.3   &11.5(3.7)   & 7.7(5.4) &0.18(0.03) &0.89(0.28) &1.58(0.46) & 0.2  & 14993& 3096  &858      & H     &2nd Quad    \\
G157.60-12.17a     &-7.75  &1.17  &\multicolumn{2}{c}{(-38,-44) } &\multicolumn{3}{c}{432$\times$283(16.4)}  &1.0   &10.4(0.9)   & 3.6(1.2) &0.18(0.01) &0.40(0.01) &0.76(0.12) & 0.6  & 243  & 133   &61       &L      &Taurus     \\
G157.60-12.17b\_1  &-2.58  &0.47  &\multicolumn{2}{c}{(-83,-63) } &\multicolumn{3}{c}{535$\times$421(41.7)}  &0.5   &14.7(1.6)   & 4.1(1.5) &0.21(0.01) &0.43(0.11) &0.83(0.17) & 1.2  & 82   & 87    &55       &L      &Taurus     \\
G157.60-12.17b\_2  &-2.58  &0.47  &\multicolumn{2}{c}{(-197,-278)}&\multicolumn{3}{c}{301$\times$196(52.2)}  &0.3   &14.6(1.3)   & 4.1(1.4) &0.21(0.01) &0.60(0.13) &1.11(0.21) & 2.4  & 22   & 79    &92       &      &Taurus     \\
G161.43-35.59\_1   &-5.83  &1.49  &\multicolumn{2}{c}{(60,212)}   &\multicolumn{3}{c}{269$\times$143(-12.5)} &0.7   & 9.8(0.4)   & 2.3(0.5) &0.17(0.00) &0.29(0.06) &0.59(0.09) & 0.5  & 79   & 57    &29        &L     &High Glat   \\
G161.43-35.59\_2   &-5.83  &1.49  &\multicolumn{2}{c}{(-3,-49)}   &\multicolumn{3}{c}{261$\times$150(82)}    &0.7   &12.0(1.2)   & 2.6(0.8) &0.19(0.01) &0.18(0.03) &0.46(0.04) & 0.6  & 91   & 35    &13        &     &High Glat   \\
G161.43-35.59\_3   &-5.83  &1.49  &\multicolumn{2}{c}{(-166,-71)} &\multicolumn{3}{c}{387$\times$168(87.6)}  &0.9   &10.5(0.9)   & 2.2(0.5) &0.18(0.01) &0.20(0.03) &0.47(0.04) & 0.4  & 128  & 47    &17        &     &High Glat   \\
G161.43-35.59\_4   &-5.83  &1.49  &\multicolumn{2}{c}{(-303,-86)} &\multicolumn{3}{c}{141$\times$140(-21.1)} &0.5   &10.7(0.9)   & 1.9(0.6) &0.18(0.01) &0.25(0.06) &0.53(0.09) & 0.6  & 34   & 33    &21        &     &High Glat   \\
G180.92+04.53      &0.98   &3.62  &\multicolumn{2}{c}{(0,-12)}    &\multicolumn{3}{c}{359$\times$339(62.2)}  &3.1   & 9.1(0.6)   & 3.4(1.1) &0.17(0.01) &0.59(0.11) &1.07(0.19) & 0.2  & 2191 & 817   &303       &H     &3th Quad    \\
G194.80-03.41\_1   &12.84  &2.89  &\multicolumn{2}{c}{(-4,-19)  } &\multicolumn{3}{c}{690$\times$341(81.3)}  &3.4   & 9.8(0.5)   & 4.5(1.7) &0.18(0.01) &0.86(0.31) &1.52(0.52) & 0.2  & 3573 & 1829  &812      & H     &3th Quad    \\
G194.80-03.41\_2   &12.84  &2.89  &\multicolumn{2}{c}{(-132,331)} &\multicolumn{3}{c}{355$\times$242(61.3)}  &2.1   &10.3(0.9)   & 5.3(2.4) &0.17(0.01) &0.78(0.20) &1.28(0.34) & 0.4  & 1536 & 784   &436      &      &3th Quad \\
G196.21-15.50\_1   &3.76   &0.8   &\multicolumn{2}{c}{(-79,5)}    &\multicolumn{3}{c}{276$\times$203(54.2)}  &0.5   &14.5(0.8)   & 3.1(0.9) &0.21(0.01) &0.32(0.12) &0.68(0.16) & 1.1  & 45   & 49    &30       & H  &Orion       \\
G196.21-15.50\_2   &3.76   &0.8   &\multicolumn{2}{c}{(104,126)}  &\multicolumn{3}{c}{228$\times$ 84(3.5)}   &0.3   &15.1(0.8)   & 2.8(0.7) &0.21(0.01) &0.31(0.07) &0.65(0.11) & 1.7  & 14   & 26    &22       &   &Orion       \\
G196.21-15.50\_3   &3.76   &0.8   &\multicolumn{2}{c}{(265,85)}   &\multicolumn{3}{c}{256$\times$ 89(60.8)}  &0.3   &14.6(0.6)   & 2.5(0.9) &0.21(0.00) &0.30(0.12) &0.66(0.15) & 1.4  & 15   & 30    &23       &   &Orion
\enddata
\end{deluxetable}


\begin{thebibliography}{}
{\small

\bibitem[Bally \& Lada(1983)]{bal83}Bally, J. \& Lada, C. J., 1983, \apj, 265, 824

\bibitem[Benson \& Myers(1989)]{ben89}Benson, P. J. \& Myers, P. C., 1989, \apjs, 71, 89

\bibitem[Bergin et al.(1997)]{Ber97}Bergin, E. A., Ungerechts, H., Goldsmith, P., F. et al. 1997, \apj, 482, 267

\bibitem[Beuther et al.(2002)]{beu02}Beuther, H., Schilke, P., Menten, K. M., et al. 2002, \apj, 566, 945

\bibitem[Cesaroni, Walmsley \& Churchwell(1992)]{ce92}Cesaroni, R., Walmsley, C. M., \& Churchwell, E., 1992, \aap, 256, 618

\bibitem[Chen et al.(2010)]{chen10}Chen, Xi., Shen, Z.-Q., Li, J.-J., Xu, Y., \& He J.-H., 2010, \apj, 710, 150

\bibitem[Clemens(1985)]{cle85}Clemens, D. P. 1985, \apj, 295, 422

\bibitem[Clark \& Johnson(1981)]{clar81}Clark, F. O., \& Johnson, D. R., 1981, \apj, 247, 104

\bibitem[Dame et al.(1987)]{dame87}Dame, T. M., Ungerechts, H., Cohen, R. S. et al. 1987, \apj, 322, 706

\bibitem[Dame et al.(2001)]{dame01}Dame, T. M., Hartmann, D., \& Thaddeus, P., 2001, \apj, 547, 792

\bibitem[Dobashi et al.(2005)]{do05}Dobashi, K., Uehara, H., Kandori, R. et al. 2005, \pasj, 57, 1

\bibitem[Du \& Yang(2008)]{du08}Du, F. J., \& Yang, J., 2008, \apj, 686, 384

\bibitem[Egan et al.(1998)]{egan98}Egan, M. P., Shipman, R. F., Price, S. D., Carey, S. J., \& Clark, F. O. 1998, \apj, 494, L199

\bibitem[Evans, Beckwith \& Blair (1977)]{evans77}Evans, N. J., II., Beckwith, S., Blair, G. N., 1977, \apj, 217, 448

\bibitem[Evans(2003)]{ev03}Evans, N. J., II. 2003, in Chemistry as a Diagnostic of Star Formation, ed. C. L. Curry \& M. Fich (Ottawa: NRC Press), 157

\bibitem[Froebrich et al.(2007)]{fro07}Froebrich D., Murphy G. C., Smith M. D., Walsh J., Del Burgo C., 2007. \mnras, 378, 1447

\bibitem[Fuller, Williams, \& Sridharan(2005)]{fu05}Fuller, G. A., Williams, S. J., \& Sridharan, T. K. 2005, \aap, 442, 949

\bibitem[Garden et al.(1991)]{gar91}Garden, R. P., Hayashi, M., Hasegawa, T., Gatley, I., Kaifu,
N., 1991, \apj, 374, 540

\bibitem[G\'{o}mez et al.(2006)]{Go06}G\'{o}mez, Jos\'{e}F., de Gregorio-Monsalvo, I., Su\'{a}rez, O., Kuiper, T. B. H., 2006, 132, 1322

\bibitem[Goldreich \& Kwan(1974)]{gol74}Goldreich, P., \& Kwan, J., 1974, \apj, 189, 441

\bibitem[Goodman, Pineda \& Schnee(2009)]{goo09}Goodman A. A., Pineda J. E., Schnee S. L., 2009, \apj, 692, 91

\bibitem[Guilloteau \& Lucas(2000)]{gui00}Guilloteau, S. \& Lucas, R., 2000, in Astronomical Society of the Pacific Conference Series, Vol. 217, Imaging at Radio through Submillimeter
Wavelengths, ed. J. G. Mangum \& S. J. E. Radford, 299

\bibitem[Hartquist \& Williams(1998)]{har98}Hartquist, T., Caselli P., Rawlings J., Ruffle D., \& Williams, D., 1998, The Chemistry of Star Formation Regions, in the Molecular Astrophysics of Stars and Galaxies, T.Hartquist, D. Hartquist, eds., Oxford:Clarendon,p.101

\bibitem[Harvey, Campbell \& Hoffmann (1977)]{Har77}Harvey, P. M., Campbell, M. F., Hoffmann, W. F., 1977, \apj, 211, 786

\bibitem[Harvey et al.(1998)]{harvey98}Harvey, P. M., Smith, B. J., di Francesco, J., Colome, C., 1998, \apj, 499, 294

\bibitem[Hauser et al.(1984)]{hau84}Hauser, M. G., Gillett, F. C., Low, F. J. et al. 1984, \apj, 278, L15

\bibitem[Heithausen et al.(1993)]{hei93}Heithausen, A., Stacy, J. G., de Vries, H. W., Mebold, U., Thaddeus, P., 1993, \aap, 268, 265

\bibitem[Hennebelle \& Chabrier(2008)]{hen08}Hennebelle, P., \& Chabrier, G., 2008, \apj, 684, 395

\bibitem[Huard, Sandell, \& Weintraub(1999)]{hua99}Huard, Tracy L., Sandell, G\"{o}ran., \& Weintraub, David A., 1999, \apj ,526 ,833

\bibitem[Juvela et al.(2010)]{ju10}Juvela, M., Ristorcelli, I., Montier, L. A. et al. 2010, \aap, 518, L93

\bibitem[Juvela et al.(2012)]{ju12}Juvela, M., Ristorcelli, I., Pagani, L. et al. 2012, arXiv: 1202.1672

\bibitem[Kainulainen et al.(2009)]{ka09}Kainulainen J., Beuther H., Henning T., Plume R., 2009, \aap, 508, L35

\bibitem[Lee \& Myers(1999)]{lee99}Lee, Chang Won., \& Myers, Philip C., 1999, \apjs, 123, 233

\bibitem[Liu, Wu \& Wang(2010)]{liu10}Liu, T., Wu, Y., \& Wang, K., 2010, RAA, 10, 67

\bibitem[Low et al.(1984)]{low84}Low, F. J., Young, E., Beintema, D. A. et al. 1984, \apj, 278, L19

\bibitem[Maddalena et al.(1986)]{ma86}Maddalena, R. J., Morris, M., Moscowitz, J., Thaddeus, P., 1986, \apj, 303, 375

\bibitem[Mardones et al.(1997)]{mar97}Mardones, D., et al. 1997, \apj, 489, 719

\bibitem[Molinari et al.(1996)]{mol96}Molinari, S., Brand, J., Cesaroni, R., \& Palla, F. 1996, \aap, 308, 573

\bibitem[Muller et al.(2011)]{mu11}Muller, S., Beelen, A., Gu\'{e}lin, M. et al. 2011, \aap, 535, 103

\bibitem[Myers, Linke \& Benson(1983)]{my83a}Myers, P. C., Linke, R. A., \& Benson, P. J., 1983, \apj, 264

\bibitem[Myers \& Benson(1983)]{my83b}Myers, P. C. \& Benson, P. J., 1983, \apj, 266, 309

\bibitem[Planck Collaboration. et al.(2011a)]{ade11a}Planck Collaboration., Ade, P. A. R., Aghanim, N., Arnaud, M. et al. 2011a, arXiv:1101.2035

\bibitem[Planck Collaboration. et al.(2011b)]{ade11b}Planck Collaboration., Ade, P. A. R., Aghanim, N., Arnaud, M. et al. 2011b, arXiv:1101.2034

\bibitem[Planck Collaboration. et al.(2011c)]{ade11c}Planck Collaboration., Ade, P. A. R., Aghanim, N., Arnaud, M. et al. 2011c, \aap, 536, 19


\bibitem[Planck Collaboration. et al.(2011d)]{ade11d}Planck Collaboration., Ade, P. A. R., Aghanim, N.,
Arnaud, M. et al. 2011d, \aap, 536, 23

\bibitem[Qin et al.(2008)]{qin08}Qin, S.-L., Huang, M., Wu, Y., Xue, R., Chen, S., 2008, \apj, 686, L21

\bibitem[Rathborne, Jackson, \& Simon(2006)]{ra06}Rathborne, J. M., Jackson, J. M., \& Simon, R., 2006, \apj, 641, 389

\bibitem[Ridge et al.(2006)]{rid06}Ridge N. A. et al., 2006, \aj, 131, 2921

\bibitem[Sharpless(1959)]{shar59}Sharpless, S., 1959, \apjs, 4, 257

\bibitem[Simon et al.(2006)]{si06}Simon, R., Rathborne, J. M., Shah, R. Y., Jackson, J. M., Chambers, E. T., 2006, \apj, 653, 1325

\bibitem[Snell, Loren \& Plambeck.(1980)]{sne80}Snell, R. L., Loren, R. B., Plambeck, R. L., 1980, \apj, 239, L17

\bibitem[Sridharan et al.(2002)]{sr02}Sridharan, T. K., Beuther, H., Schilke, P., Menten, K. M., \& Wyrowski, F. 2002, \apj, 566, 931

\bibitem[Sridharan et al.(2005)]{sr05}Sridharan, T. K., Beuther, H., Saito, M., Wyrowski, F., Schilke, P., 2005, \apj, 634, L57

\bibitem[Strom(1975)]{str75}Strom, K. M., Strom, S. E., Carrasco, L., Vrba, F. J., 1975, \apj, 196, 489

\bibitem[Strom, Vrba \& Strom (1976)]{str76}Strom, S. E., Vrba, F. J., \& Strom, K. M., 1976, \aj, 81, 314

\bibitem[Ungerechts et al.(2000)]{ung00}Ungerechts H., Umbanhowar P., \& Thaddeus P., 2000, \apj, 537, 221

\bibitem[V\'{a}zquez-Semadeni (1994)]{va94}V\'{a}zquez-Semadeni E., 1994, \apj, 423, 681

\bibitem[Velusamy et al.(2008)]{ve08}Velusamy, T., Peng, R., Li, D., Goldsmith, P. F., Langer, William D., 2008, \apj, 688, L87

\bibitem[Wang et al.(2009)]{wang09}Wang, K., Wu, Y, Ran, L., Yu, W. T., Miller, M., 2009, \aap, 507, 369

\bibitem[Watson et al.(1997)]{wat97}Watson, Alan M., Coil, Alison L., Shepherd, Debra S., Hofner, Peter, Churchwell, Ed., 1997, \apj, 487, 818

\bibitem[Winnewisser, Churchwell, \& Walmsley(1979)]{wi79}Winnewisser, G., Churchwell, E., Walmsley, C. M., 1979, \aap, 72, 215

\bibitem[Wood \& Churchwell(1989)]{wood89}Wood, Douglas O. S. \& Churchwell, Ed., 1989, \apj, 340, 265

\bibitem[Wu, Wu \& Wang(2001)]{wu01}Wu, Y., Wu, J., Wang, J., 2001, \aap, 380, 665

\bibitem[Wu et al.(2006)]{wu06}Wu, Y., Zhang, Q., Yu, W. et al. 2006, \aap, 450, 607

\bibitem[Wu et al.(2007)]{wu07}Wu, Y., Henkel, C., Xue, R., Guan, X., Miller, M., 2007, 669, L37

\bibitem[Yamamoto et al.(2003)]{ya03}Yamamoto, H., Onishi, T., Mizuno, A., Fukui, Y., 2003, \apj, 592, 217

\bibitem[Zhang et al.(2011)]{zhang11}Zhang, S. B., Yang, J., Xu, Y. et al. \apjs, 193, 10

\bibitem[Zhou et al.(1993)]{zhou93}Zhou, S., Evans, N. J. II, Koempe, C., \& Walmsley, C. M. 1993, \apj, 404, 232
}
\end{thebibliography}
\end{document}